\documentclass[11pt]{article}

\usepackage{geometry}
\usepackage{fullpage}

\usepackage{amsmath}
\usepackage{amsfonts}
\usepackage{amsthm}
\usepackage{epsfig}     

\usepackage[authoryear, sort]{natbib}

\usepackage{graphicx}
\usepackage{caption}
\usepackage{color, colortbl, xcolor}
\usepackage{multirow}
\usepackage{authblk}
\usepackage[hidelinks]{hyperref}
\usepackage{lipsum}     
\usepackage{enumitem}

\def\R{\mathbb{R}}
\def\E{\mathbb{E}} 
\def\P{\mathbb{P}} 

\def\BIC{\textsc{bic}}

\newtheorem{theorem}{\textsc{\bf Theorem}}
\newtheorem{remark}{\underline{\bf Remark}}

\newtheorem{definition}{\textsc{\bf Definition}}
\newtheorem{proposition}{\bf Proposition}

\def\sign{\textup{sign}}

\DeclareMathOperator*{\maximize}{maximize}
\DeclareMathOperator*{\minimize}{minimize}
\DeclareMathOperator*{\argmin}{argmin}

\def\cov{\hbox{cov}}

\def\NPN{{\textup{NPN}}}
\def\LNPN{{\textup{LNPN}}}
\def\N{{\textup{N}}}

\def\R{\mathbb{R}}
\def\E{\mathbb{E}} 
\def\P{\mathbb{P}} 
\newcommand{\bpm}{\begin{pmatrix}}
	\newcommand{\epm}{\end{pmatrix}}

\title{\LARGE \bf Sparse semiparametric canonical correlation analysis for data of mixed types}
\author{Grace Yoon, Raymond J. Carroll and Irina Gaynanova\thanks{Corresponding author. E-mail: irinag@stat.tamu.edu}}
\affil{Department of Statistics, Texas A\&M University\\ 3143 TAMU, College Station, TX 77843}
\date{}

\begin{document}

\maketitle

\vspace{-.3cm}
\begin{abstract}
Canonical correlation analysis investigates linear relationships between two sets of variables, but often works poorly on modern data sets due to high-dimensionality and mixed data types such as continuous, binary and zero-inflated. To overcome these challenges, we propose a semiparametric approach for sparse canonical correlation analysis based on Gaussian copula. Our main contribution is a truncated latent Gaussian copula model for data with excess zeros, which allows us to derive a rank-based estimator of the latent correlation matrix for mixed variable types without the estimation of marginal transformation functions. The resulting canonical correlation analysis method works well in high-dimensional settings as demonstrated via numerical studies, as well as in application to the analysis of association between gene expression and micro RNA data of breast cancer patients.
\end{abstract}


\section{Introduction}\label{sec:intro}
Canonical correlation analysis investigates linear associations between two sets of variables, and is widely used in various fields including biomedical sciences, imaging and genomics \citep{Hardoon:2004, Chi:2013gj, Safo:2018he}. However, sample canonical correlation analysis often performs poorly due to two main challenges: high-dimensionality and non-normality of the data.

In high-dimensional settings, sample canonical correlation analysis is known to overfit the data due to the singularity of sample covariance matrices \citep{Hardoon:2004, SuffCCA}. Additional regularization is often used to address this challenge. \citet{RCCA:2008} focus on ridge regularization of sample covariance matrices to avoid singularity, while more recent methods focus on sparsity regularization of canonical vectors \citep{parkhomenko2009sparse, Witten:2009PMD, ChenLiu, Chi:2013gj, FastRegCCA, Wilms:2015wna, Gao:2015_minimax, Safo:2018he}.  At the same time, with the advancement in technology, it is common to collect data of different types. For example, the Cancer Genome Atlas Project contains matched data of mixed types such as gene expression (continuous), mutation (binary) and micro RNA (count) data. While regularized canonical correlation methods work well for Gaussian data, they still are based on the sample covariance matrix, and therefore are not appropriate for the analysis in the presence of binary data or data with excess zero values.

Several approaches have been proposed to address the non-normality of the data. There are completely nonparametric approaches such as kernel canonical correlation analysis \citep{Hardoon:2004}. Alternatively, there are parametric approaches building upon a probabilistic interpretation of \citet{BachJordan}. For example, \citet{ZohPCAN2016} develop probabilistic canonical correlation analysis for count data based on Poisson distribution. More recently, \citet{semiCCA} utilize a normal semiparametric transformation model for the analysis of mixed types of variables; however, the method requires estimation of marginal transformation functions via nonparametric maximum likelihood.

In summary, significant progress has been made in developing regularized variants of sample canonical correlation analysis that work well in high-dimensional settings. However, these approaches are not suited for mixed data types. At the same time, several methods have been proposed to account for non-normality of the data, however they are not designed for high-dimensional settings. More importantly, to our knowledge none of the existing methods explicitly address the case of zero-inflated measurements, which, for example, is common for micro RNA and microbiome abundance data.

To bridge this major gap, we propose a semiparametric approach for sparse canonical correlation analysis, which allows us to handle high-dimensional data of mixed types via a common latent Gaussian copula framework. Our work has three main contributions.

First, we model the zeros in the data as observed due to truncation of an underlying latent continuous variable, and define a corresponding truncated Gaussian copula model. We derive explicit formulas for the bridge functions that connect the Kendall's $\tau$ of the observed data to the latent correlation matrix for different combinations of continuous, binary and truncated data types, and use these formulas to construct a rank-based estimator of the latent correlation matrix for the mixed data. \citet{Fan:2016um} use a similar bridge function approach in the context of graphical models, however the authors do not consider the truncated variable type. The latter requires derivation of new bridge functions, and those derivations are considerably more involved than the corresponding derivations for the continuous/binary case. The significant advantage of the bridge function technique is that it allows us to estimate the latent correlation structure of a Gaussian copula without estimating marginal transformation functions, in contrast to \citet{semiCCA}.  

Secondly, we use the derived rank-based estimator instead of the sample correlation matrix within the sparse canonical correlation analysis framework that is motivated by \citet{Chi:2013gj} and \citet{Wilms:2015wna}. This allows us to take into account the dataset-specific correlation structure in addition to the cross-correlation structure.  In contrast, \citet{parkhomenko2009sparse} and \citet{Witten:2009PMD} model the variables within each data set as uncorrelated. We develop an efficient optimization algorithm to solve the corresponding problem. 

Finally, we propose two types of Bayesian Information Criteria (\BIC) for tuning parameter selection, which leads to significant computational savings compared to commonly used cross-validation and permutation techniques \citep{Witten:SAGMB}. \citet{Wilms:2015wna} also use {\BIC} in the canonical correlation analysis context, however only one criterion is proposed. Our two criteria correspond to the cases of the error variance being either known or unknown.  We found that both are competitive in our numerical studies, however one criterion works best for variable selection, whereas the other works best for prediction.


\section{Background}


\subsection{Canonical correlation analysis}
In this section we review both the classical canonical correlation analysis, and its sparse alternatives. Given two random vectors $\mathbf{X}_1\in \R^{p_1}$ and $\mathbf{X}_2\in \R^{p_2}$, let $\Sigma_1 = \cov (\mathbf{X}_1)$, $\Sigma_2 = \cov(\mathbf{X}_2)$ and $\Sigma_{12} = \cov (\mathbf{X}_1, \mathbf{X}_2)$. Population canonical correlation analysis \citep{Hotelling:1936} seeks linear combinations $w_1^{\top}\mathbf{X}_1$ and $w_2^{\top}\mathbf{X}_2$ with maximal correlation, that is
\begin{equation}\label{eq:pCCA}
\maximize_{w_1,w_2} \Big\{ w_1^{\top}\Sigma_{12}w_2 \Big\} \quad \mbox{subject to}\quad w_1^{\top}\Sigma_1 w_1=1, \quad w_2^{\top}\Sigma_2 w_2=1.
\end{equation}
Problem \eqref{eq:pCCA} has a closed form solution via the singular value decomposition of $\Sigma_{1}^{-1/2} \Sigma_{12} \Sigma_{2}^{-1/2}$. Given the first pair of singular vectors $(u,v)$, the solutions to~\eqref{eq:pCCA} can be expressed as $w_1 = \Sigma_1^{-1/2}u$ and $w_2 = \Sigma_2^{-1/2}v$.

Sample canonical correlation analysis replaces $\Sigma_1$, $\Sigma_2$ and $\Sigma_{12}$ in~\eqref{eq:pCCA} by corresponding sample covariance matrices $S_1$, $S_2$ and $S_{12}$. In high-dimensional settings when sample size is small compared to the number of variables, $S_1$ and $S_2$ are singular, thus leading to non-uniqueness of solution and poor performance due to overfitting. A common approach to circumvent this challenge is to consider sparse regularization of $w_1$ and $w_2$ via the addition of a $\ell_1$ penalty in the objective function of~\eqref{eq:pCCA} \citep{Witten:2009PMD, parkhomenko2009sparse, Chi:2013gj, Wilms:2015wna}. Sparse canonical correlation analysis is then formulated as
\begin{equation}\label{eq:sparseCCA}
\maximize_{w_1,w_2} \Big\{ w_1^{\top}S_{12}w_2-\lambda_1\|w_1\|_1-\lambda_2\|w_2\|_1 \Big\} \quad \mbox{subject to}\quad w_1^{\top}S_1 w_1\le1, \quad w_2^{\top}S_2 w_2\le1.
\end{equation}
In addition to $\ell_1$ penalties, the equality constraints in~\eqref{eq:pCCA} are replaced with inequality constraints which define convex sets. This generalization is possible since nonzero solutions to~\eqref{eq:sparseCCA} satisfy the constraints with equality, see Proposition~\ref{prop:kkt} below.

While problem~\eqref{eq:sparseCCA} works well in high-dimensional settings, it still relies on sample covariance matrices, and therefore is not well-suited for skewed or non-continuous data, such as binary or zero-inflated. We next review the Gaussian copula models that we propose to use to address these challenges.

\subsection{Latent Gaussian copula model for mixed data}

In this section we review the Gaussian copula model in~\citet{Liu2009:NPN}, and its extension to mixed continuous and binary data in~\cite{Fan:2016um}.

\begin{definition}[Gaussian copula model] A random vector $\mathbf{X}=(X_1,\ldots,X_p)^{\top}$ satisfies a Gaussian copula model if there exists a set of monotonically increasing transformations $f=(f_j)^p_{j=1}$ satisfying $f(\mathbf{X}) = \left\{f_1(X_1),\ldots,f_p(X_p)\right\}^{\top}\sim \N_p(0,\Sigma)$ with $\Sigma_{jj}=1$ for all $j$. We denote $\mathbf{X} \sim \NPN(0, \Sigma, f)$.
\end{definition}
\begin{definition}[Latent Gaussian copula model for mixed data] Let $\mathbf{X}_1\in \R^{p_1}$ be continuous and $\mathbf{X}_2\in \R^{p_2}$ be binary random vectors with $\mathbf{X}=(\mathbf{X}_1, \mathbf{X}_2)$. Then $\mathbf{X}$ satisfies the latent Gaussian copula model if there exists a $p_2$-dimensional random vector $\mathbf{U}_2=(U_{p_1+1},\ldots,U_{p_1+p_2})^{\top}$ such that $\mathbf{U}:=(\mathbf{X}_1,\mathbf{U}_2)\sim \NPN(0,\Sigma,f)$ and
$X_j = I(U_j >C_j)$ for all $j=p_1+1,\ldots,p_1+p_2$, where $I(\cdot)$ is the indicator function and $\mathbf{C}=(C_1,\ldots,C_{p_2})$ is a vector of constants. We denote this as $\mathbf{X} \sim \LNPN(0,\Sigma, f, \mathbf{C})$, where $\Sigma$ is the latent correlation matrix. 
\end{definition}

\citet{Fan:2016um} consider the problem of estimating $\Sigma$ for the latent Gaussian copula model based on the Kendall's $\tau$. Given the observed data $(X_{1j}, X_{1k}), \ldots, (X_{nj}, X_{nk}) $ for variables $X_j$ and $X_k$, Kendall's $\tau$ is defined as
\[
\widehat\tau_{jk} = \dfrac{2}{n(n-1)}\sum_{1\le i<i'\le n}\sign(X_{ij}-X_{i'j})\sign(X_{ik}-X_{i'k}).
\]
Since $\widehat \tau_{jk}$ is invariant under monotone transformation of the data, it is well-suited to capture associations in copula models. Let $\tau_{jk} = \E(\widehat \tau_{jk})$ be the population Kendall's $\tau$. The latent correlation matrix $\Sigma$ is connected to Kendall's $\tau$ via the so-called bridge function $F$ such that $\Sigma_{jk} = F^{-1}(\tau_{jk})$ for all variables $j$ and $k$. \citet{Fan:2016um} derive an explicit form of the bridge function for continuous, binary and mixed variable pairs, which allows to estimate the latent correlation matrix via the method of moments. We summarize these results below.

\begin{theorem}[\citet{Fan:2016um}]\label{d:hatR}
Let $\mathbf{X}=(\mathbf{X}_1, \mathbf{X}_2)\sim\LNPN(0,\Sigma, f, \mathbf{C})$ with $p_1$-dimensional continuous $\mathbf{X}_1$ and $p_2$-dimensional binary $\mathbf{X}_2$. The rank-based estimator of $\Sigma$ is the symmetric matrix $\widehat R$ with $\widehat R_{jj}=1$ and $\widehat R_{jk} = \widehat R_{kj}=F_{jk}^{-1}(\widehat \tau_{jk})$, where for $r\in (0,1)$,
$$
F_{jk}(r) = \begin{cases}
2\sin^{-1}(r)/\pi& \mbox{if}\quad 1\leq j<k\leq p_1;\\
2\left\{\Phi_2( \Delta_j,  \Delta_{k}; r)-\Phi( \Delta_j)\Phi( \Delta_{k})\right\}& \mbox{if}\quad p_1+1 \leq j<k\leq p_1+p_2;\\
4\Phi_2( \Delta_k, 0; r/\surd{2})-2\Phi( \Delta_k)&\mbox{if}\quad 1\leq j\leq p_1, p_1+1\leq k\leq p_1+ p_2.
\end{cases}
$$
Here $\Delta_j = f_j(C_j)$, $\Phi(\cdot)$ is the cdf of the standard normal distribution, and $\Phi_2(\cdot, \cdot; r)$ is the cdf of the standard bivariate normal distribution with correlation $r$.
\end{theorem}
\begin{remark}
Since $\Delta_j = f_j(C_j)$ is unknown in practice, \citet{Fan:2016um} propose to use a plug-in estimator from the moment equation $\E(X_{ij})= 1- \Phi(\Delta_j)$, leading to $\widehat \Delta_j = \Phi^{-1}(1-\bar X_{j})$, where $\bar X_{j} = \sum_{i=1}^{n}X_{ij}/n$.
\end{remark}

\citet{Fan:2016um} use these results in the context of Gaussian graphical models, and replace the sample covariance matrix with a rank-based estimator $\widehat R$, which allows one to use Gaussian models with skewed continuous and binary data. However, \citet{Fan:2016um} do not consider the case of zero-inflated data, which requires formulation of a new model, and derivation of new bridge functions.


\section{Methodology}

\subsection{Truncated latent Gaussian copula model}\label{sec:trunc}

Our goal is to model the zero-inflated data through latent Gaussian copula models. Two motivating examples are micro RNA and microbiome data, where it is common to encounter a large number of zero counts. In both examples it is reasonable to assume that zeros are observed due to truncation of underlying latent continuous variables. More generally, one can think of zeros as representing the measurement error due to truncation of values below a certain positive threshold. This intuition leads us to consider the following model.

\begin{definition}[Truncated latent Gaussian copula model]\label{d:trunc}A random vector $\mathbf{X}=(X_1, \ldots, X_p)^{\top}$ satisfies the truncated Gaussian copula model if there exists a $p$-dimensional random vector $\mathbf{U} = (U_1,\ldots, U_p)^{\top} \sim \NPN(0, \Sigma, f)$ such that
$$
X_j = I(U_j>C_j)U_j \quad (j=1,\ldots,p),
$$
where $I(\cdot)$ is the indicator function and $\mathbf{C}=(C_1,\ldots,C_p)$ is a vector of positive constants. We denote $X \sim \textup{TLNPN}(0,\Sigma, f, \mathbf{C})$, where $\Sigma$ is the latent correlation matrix. 
\end{definition}

The methodology in \citet{Fan:2016um} allows them to estimate the latent correlation matrix in the presence of mixed continuous and binary data. Our Definition~\ref{d:trunc} adds a third type, which we denote as \textit{truncated} for short. To construct a rank-based estimator for $\Sigma$ as in Theorem~\ref{d:hatR} in the presence of truncated variables, below we derive an explicit form of the bridge function for all possible combinations of the data types. Throughout, we use $\Phi(\cdot)$ for the cdf of a standard normal distribution and $\Phi_d(\cdots; \Sigma_d)$ for the cdf of a standard $d$-variate normal distribution with correlation matrix $\Sigma_d$. All the proofs are deferred to the Supplementary Material.

\begin{theorem}\label{bridge1}
Let $X_j$ be truncated and $X_k$ be binary. Then $\E (\widehat \tau_{jk}) = F_{\rm TB}(\Sigma_{jk};\Delta_j, \Delta_k)$, where
\[
    F_{\rm TB}(\Sigma_{jk};\Delta_j, \Delta_k) =
    2 \{1-\Phi(\Delta_j)\}\Phi(\Delta_k)
    -2 \Phi_3\left(-\Delta_j, \Delta_k, 0; \Sigma_{3a} \right)
    -2 \Phi_3\left(-\Delta_j, \Delta_k, 0; \Sigma_{3b} \right),
\]
$\Delta_j = f_j(C_j)$, $\Delta_k = f_k(C_k)$,
\[
\Sigma_{3a}=\bpm
1 & -\Sigma_{jk} & 1/\surd{2} \\
-\Sigma_{jk} & 1 & -\Sigma_{jk}/\surd{2}\\
1/\surd{2} & -\Sigma_{jk}/\surd{2} & 1
\epm, \quad
 \Sigma_{3b}=\bpm
1 & 0 & -1/\surd{2} \\
0 & 1 & -\Sigma_{jk}/\surd{2}\\
-1/\surd{2} & -\Sigma_{jk}/\surd{2} & 1
\epm.
\]
\end{theorem}

\begin{theorem}\label{bridge2}
Let $X_j$ be truncated and $X_k$ be continuous. Then $\E (\widehat \tau_{jk}) = F_{\rm TC}(\Sigma_{jk};\Delta_j)$, where
\[
F_{\rm TC}(\Sigma_{jk};\Delta_j) = -2 \Phi_2 (-\Delta_j,0; 1/\surd{2} ) +4\Phi_3 \left(-\Delta_j,0,0; \Sigma_3\right),
\]
$\Delta_j = f_j(C_j)$ and
\[
\Sigma_3 = \bpm
        1 & 1/\surd{2} & \Sigma_{jk}/\surd{2}\\
        1/\surd{2} & 1 & \Sigma_{jk}\\
        \Sigma_{jk}/\surd{2} & \Sigma_{jk} & 1
    \epm.
\]
\end{theorem}

\begin{theorem}\label{bridge3}
Let both $X_j$ and $X_k$ be truncated. Then $\E (\widehat \tau_{jk}) = F_{\rm TT}(\Sigma_{jk};\Delta_j, \Delta_k)$, where
\[
F_{\rm TT}(\Sigma_{jk};\Delta_j, \Delta_k)  = ~-2 \Phi_4 (-\Delta_j, -\Delta_k, 0,0; \Sigma_{4a}) + 2 \Phi_4 (-\Delta_j, -\Delta_k, 0,0; \Sigma_{4b}),
\]
$\Delta_j = f_j(C_j)$, $\Delta_k = f_k(C_k)$ and
\[
\Sigma_{4a} = \bpm
1 & 0 & 1/\surd{2} & -\Sigma_{jk}/\surd{2}\\
0& 1 & -\Sigma_{jk}/\surd{2}& 1/\surd{2}\\
1/\surd{2}& -\Sigma_{jk}/\surd{2} & 1& -\Sigma_{jk}\\
-\Sigma_{jk}/\surd{2}& 1/\surd{2}& -\Sigma_{jk} & 1
\epm
\]
and
\[
\Sigma_{4b} = \bpm
1 & \Sigma_{jk} & 1/\surd{2} & \Sigma_{jk}/\surd{2}\\
\Sigma_{jk} & 1 & \Sigma_{jk}/\surd{2}& 1/\surd{2}\\
1/\surd{2}& \Sigma_{jk}/\surd{2} & 1& \Sigma_{jk}\\ \Sigma_{jk}/\surd{2}& 1/\surd{2}& \Sigma_{jk} & 1
\epm.
\]
\end{theorem}

We also show that the inverse bridge function exists for all of the cases.
\begin{theorem}\label{t:monotone}
For any constants $\Delta_j$, $\Delta_k$, the bridge functions $F(\Sigma_{jk})$ in Theorems \ref{bridge1}--\ref{bridge3} are strictly increasing in $\Sigma_{jk} \in(-1,1)$, and thus the corresponding inverse functions $F^{-1}(\tau_{jk})$ exist.
\end{theorem}
\begin{remark}
While the inverse functions exist, they do not have the closed form. In practice we estimate $\widehat R$ element-wise by solving $\widehat R_{jk}=\argmin_{r}\{F(r)-\widehat \tau_{jk}\}^2$. This leads to $O(p^2)$ computations which can be done in parallel to alleviate the computational burden.
\end{remark}

Theorems~\ref{bridge1}--\ref{t:monotone} complement the results of \citet{Fan:2016um} summarized in Theorem~\ref{d:hatR} by adding three more cases: continuous/truncated, binary/truncated and truncated/truncated. This allows us to construct a rank-based estimator $\widehat R$ for $\Sigma$ in the presence of mixed variables.

\begin{remark}\label{rem:Rtilde} Since $\widehat R$ is not guaranteed to be positive semidefinite, \citet{Fan:2016um} regularize $\widehat R$ by projecting it onto the cone of positive semidefinite matrices. We follow this approach using the \texttt{nearPD} function in the \texttt{Matrix} R package leading to estimator $\widehat R_p$. Furthermore, we consider
\begin{equation}\label{d:Rtilde}
\widetilde R = (1-\nu)\widehat R_p + \nu I
\end{equation}
with a small value of $\nu>0$, so that $\widetilde R$ is strictly positive definite. Throughout, we fix $\nu = 0.01$.
\end{remark}

\begin{remark} As in the binary case, $\Delta_j = f_j(C_j)$ is unknown for truncated variables. Similar to \cite{Fan:2016um}, we use a plug-in estimator $\widehat \Delta_j$ based on the moment equation $\E \left\{I(X_{ij}>0) \right\} = \P (X_{j}>0) = \P \left\{f_j(U_{j})>\Delta_j\right\} = 1-\Phi(\Delta_j)$. Let $n_{\text{zero}}=\sum_{i=1}^{n} I(X_{ij}=0)$, then we use $\widehat{\Delta}_j = \Phi^{-1} \left({n_{\text{zero}}}/{n}\right)$.
\end{remark}

For clarity, we summarize below all the steps in the construction of our rank-based estimator $\widetilde R$ based on the observed data matrix $\mathbf{X}\in \R^{n \times p}$.
\begin{enumerate}
	\item Calculate $\widehat \tau_{jk}$ for all pairs of variables $1\le j < k \le p$.
	\item Estimate $\widehat{\Delta}_j = \Phi^{-1} \{\sum_{i=1}^{n} I( X_{ij} \neq 0)/n \}$ for all $j$ of truncated or binary type.
	\item Compute $\widehat R_{jk} = F^{-1} (\widehat \tau_{jk})$, where $F$ is the bridge function chosen according to the type of variables $j$ and $k$ (with possible dependence on $\widehat \Delta_j$, $\widehat \Delta_k$).
	\item Project $\widehat R$ onto the cone of positive semidefinite matrices to form $\widehat R_p$.
	\item Set $\widetilde R = (1-\nu)\widehat R_p + \nu I$ for small $\nu > 0$.
\end{enumerate}

\subsection{Consistency of rank-based estimator for latent correlation matrix}
We next show that our proposed estimator $\widetilde R$ is consistent for $\Sigma$. Similar to \citet{Fan:2016um}, we use the following two assumptions:
\begin{itemize}[leftmargin=1in]
    \item[Assumtion 1.] All the elements of $\Sigma$ satisfy $|\Sigma_{jk}|\leq 1-\delta$ for some $\delta > 0$.
    \item[Assumtion 2.] All the thresholds $\Delta_j$ satisfy $|\Delta_j|\leq M$ for some constant $M>0$.
\end{itemize}



We first prove Lipschitz continuity of the inverse of the bridge function, $F^{-1}(\tau_{jk})$. 

\begin{theorem}\label{t:lipschitz}Under Assumptions~1--2,
for any constants $\Delta_j$ and $\Delta_k$, the inverses of the bridge functions in Theorems \ref{bridge1}--\ref{bridge3}, $F^{-1}(\cdot)$, satisfy for any $\tau_1$, $\tau_2$
\[
|F^{-1}(\tau_1) - F^{-1}(\tau_2) | \le L |\tau_1 - \tau_2|,
\]
where $L>0$ is a constant independent of $\tau_1$, $\tau_2$, $\Delta_j$ and $\Delta_k$.
\end{theorem}

\citet{Fan:2016um} also prove Lipschitz continuity in the continuous/binary case, however their proof technique cannot be directly used for the truncated case considered here due to a more complex form of the bridge functions. Instead, we develop a new proof technique based on the multivariate chain rule, which also leads to simplified proofs in the continuous/binary case. The full proof is given in the Supplementary Material Section S.1. The Lipschitz continuity of the inverse bridge functions is then used to prove consistency of $\widehat R$.

\begin{theorem}\label{t:consistency} Let a random $\mathbf{X}=(\mathbf{X}_1,\mathbf{X}_2, \mathbf{X}_3) \in \R^p$ satisfy the latent Gaussian copula model with correlation matrix $\Sigma$, with $\mathbf{X}_1\in \R^{p_1}$ being continuous, $\mathbf{X}_2\in \R^{p_2}$ being binary, and $\mathbf{X}_3\in \R^{p_3}$ being truncated with $p=p_1 + p_2 + p_3$. Let $\widehat R$ be the rank-based estimator for the correlation matrix $\Sigma$ from Section~\ref{sec:trunc} constructed by inverting corresponding bridge functions element-wise. Under Assumptions~1--2, with probability at least $1-p^{-1}$, for some $C>0$ independent of $n$, $p$
$$
\|\widehat R - \Sigma\|_{\max} = \max_{j,k}|\widehat R_{jk} - \Sigma_{jk}| \leq C(\log p/n)^{1/2}.
$$
\end{theorem}

Theorem~\ref{t:consistency} states that $\widehat R$ is consistent in estimating $\Sigma$ with respect to sup norm, and the consistency rate coincides up to constants with the rate obtained by the sample covariance matrix in the Gaussian case. In practice, we further regularize $\widehat R$ by forming $\widetilde R = (1-\nu)\widehat R_p + \nu I$. By Corollary~2 in \citet{Fan:2016um}, $\widehat R_p$ has the same consistency rate as $\widehat R$, hence Theorem~\ref{t:consistency} implies the consistency of $\widetilde R$ with the same rate as long as $\nu \le (\log p / n)^{1/2}$.

\subsection{Semiparametric sparse canonical correlation analysis}\label{sec:optim}
Our proposal is based on formulating sparse canonical correlation analysis using a latent correlation matrix from the Gaussian copula model for mixed data. At a population level, let $\Sigma$ be the latent correlation matrix for $(\mathbf{X}_1,\mathbf{X}_2) \sim \LNPN(0,\Sigma, f, \mathbf{C})$ where each $\mathbf{X}_1$ and $\mathbf{X}_2$ follows one of the three data types: continuous, binary or truncated. In Section~\ref{sec:trunc} we derived a rank-based estimator for $\Sigma$, which we propose to use within the sparse canonical correlation analysis framework~\eqref{eq:sparseCCA}.

Given the semiparametric estimator $\widetilde R$ in \eqref{d:Rtilde}, we propose to find canonical vectors by solving
\begin{equation}\label{eq:sparseCCAwithR}
\minimize_{w_1,w_2} \Big\{-w_1^{\top}\widetilde{R}_{12}w_2+\lambda_1\|w_1\|_1+\lambda_2\|w_2\|_1\Big\}\quad\mbox{subject to}\quad w_1^{\top}\widetilde{R}_1 w_1\le1,\quad w_2^{\top}\widetilde{R}_2 w_2\le1.
\end{equation}

\begin{remark} \citet{qingmaiBiometrics} establish the consistency of estimated canonical vectors from the sparse canonical correlation analysis problem~\eqref{eq:sparseCCA} in the Gaussian case. Their proof relies on the sup norm bound for the sample covariance matrix. Since Theorem~\ref{t:consistency} establishes such a bound for our rank-based estimator, these results can be directly extended to~\eqref{eq:sparseCCAwithR}.
\end{remark}

While we focus only on the estimation of the first canonical pair, the subsequent canonical pairs can be found sequentially by using a deflation scheme as follows. Let $\widetilde R_{12}^{(1)}= \widetilde{R}_{12}$ and let $\widehat{w}_{1}$, $\widehat{w}_2$ be the $(k-1)$th estimated canonical pair. To estimate the $k$th pair for $k>1$, form
$$\widetilde{R}_{12}^{(k)}=\widetilde R_{12}^{(k-1)} - (\widehat w_1^{\top}\tilde R_{12}^{(k-1)}\widehat w_2)\widetilde R_1\widehat w_{1}\widehat w_{2}^{\top}\widetilde R_2,$$
and solve~\eqref{eq:sparseCCAwithR} using $\widetilde R_{12}^{(k)}$ instead of $\widetilde R_{12}$.

While problem~\eqref{eq:sparseCCAwithR} is not jointly convex in $w_1$ and $w_2$, it is biconvex. Therefore, we propose to iteratively optimize over $w_1$ and $w_2$. First, consider optimizing over $w_1$ with $w_2$ fixed.

\begin{proposition}\label{prop:kkt} For a fixed $w_2\in \R^{p_2}$, let
\begin{equation}\label{eq:prop1_1}
\begin{split}
\widehat w_1 =\argmin_{w_1} & \Big\{ - w_1^{\top} \widetilde R_{12} w_2 + \lambda_1\|w_1\|_1 \Big\}\quad \mbox{subject to}\quad w_1^{\top}\widetilde R_1 w_1\le 1.
\end{split}
\end{equation}
This problem is equivalent to finding
\begin{equation}\label{eq:prop1_2}
\begin{split}
\widetilde{w}_1 = \argmin_{w_1} & \Big\{(1/2) w_1^{\top} \widetilde R_1 w_1 - w_1^{\top} \widetilde R_{12} w_2 + \lambda_1\|w_1\|_1 \Big\},
\end{split}
\end{equation}
and then setting $\widehat{w}_1 =0$ if $\widetilde{w}_1=0$, and $\widehat{w}_1 =\widetilde{w}_1/(\widetilde{w}_1^{\top}\widetilde R_1\widetilde{w}_1)^{1/2}$ if $\widetilde{w}_1\neq 0$.
\end{proposition}

Both problems~\eqref{eq:prop1_1} and~\eqref{eq:prop1_2} are convex, but unlike~\eqref{eq:prop1_1}, problem~\eqref{eq:prop1_2} is unconstrained. Furthermore, problem~\eqref{eq:prop1_2} is of the same form as the well-studied penalized LASSO problem \citep{tibshirani1996regression}, which can be solved efficiently using for example the coordinate-descent algorithm. Hence, the proposed optimization algorithm for~\eqref{eq:sparseCCAwithR} can be viewed as a sequence of LASSO problems with rescaling. Given the value of $w_2$ at iteration $t$, the updates at iteration $t+1$ have the form
\begin{align*}
&\widetilde w_{1}=\argmin_{w_{1}}\Big\{(1/2){w_1^{\top} \widetilde R_1w_1}-w_1^{\top} \widetilde R_{12}w_2^{(t)}+\lambda_{1}\|w_{1}\|_{1}\Big\};\\
&\widehat w_{1}^{(t+1)}=\widetilde w_{1}/(\widetilde w_{1}^{\top} \widetilde R_1\widetilde w_{1})^{1/2};\\
&\widetilde w_{2}=\argmin_{w_{2}}\Big\{(1/2){w_2^{\top} \widetilde R_2w_2}-w_2^{\top} \widetilde R_{12}^{\top}w_1^{(t+1)}+\lambda_{2}\|w_{2}\|_{1}\Big\};\\
&\widehat w_{2}^{(t+1)}=\widetilde w_{2}/(\widetilde w_{2}^{\top} \widetilde R_2\widetilde w_{2})^{1/2}.
\end{align*}

If a zero solution is obtained at any of the steps, the optimization algorithm stops, and both $w_1$ and $w_2$ are returned as zeros. Otherwise, the algorithm proceeds until convergence, which is guaranteed due to biconvexity of~\eqref{eq:sparseCCAwithR} \citep{Gorski:2007uv}.

We further describe a coordinate-descent algorithm for~\eqref{eq:prop1_2}. Consider the KKT conditions \citep{Boyd:2004uz}
\[
\widetilde R_1 w_1   - \widetilde R_{12}w_2 + \lambda_1 s_1 = 0,
\]
where $s_1$ is the subgradient of $\|w_1\|_1$. If $\lambda_1 \geq \|\widetilde R_{12}w_2\|_{\infty}$, it follows that $\widetilde w_1 = 0$. Otherwise, the $i$th element of $w_1$ can be expressed through the other coordinates as
\[
w_{1i} = S_{\lambda_1}\Big\{(\widetilde R_{12})_i w_2^{(t)} - (\widetilde R_{1})_{i, -i} (w_{1})_{-i}\Big\},
\]
where $S_{\lambda} (t) = \sign(t) \left(| t | - \lambda \right)_{+}$ is the soft-thresholding operator, $(R_{12})_i$ denotes the $i$th row of matrix $R_{12}$ and $(R_1)_{i, -i}$ denotes $i$th row of matrix $R_1$ without the $i$th component that is $(R)_{i, -i} = (R_{i1}, \ldots, R_{i,i-1}, R_{i,i+1}, \ldots, R_{ip})$. The coordinate-descent algorithm proceeds by using the above formula to update one coordinate at a time until the convergence to a global optimum is achieved. This convergence is guaranteed due to convexity of the objective function and separability of the penalty with respect to coordinates \citep{Tseng:1988}.

\subsection{Selection of tuning parameters}\label{sec:tuning}
Cross-validation is a popular approach to select the tuning parameter in LASSO. In our context, however, it amounts to performing a grid search over both $\lambda_1$ and $\lambda_2$.  Moreover, splitting the data as in cross-validation may lead to too small a number of testing samples to construct the rank-based estimator of the latent correlation matrix. Instead, motivated by \cite{Wilms:2015wna},  we propose to adapt the Bayesian information criterion to the canonical correlation analysis to avoid splitting the data and decrease computational costs.

For the Gaussian linear regression model, the Bayesian information criterion (\BIC) has the form
\[
\text{\BIC} = -2\ell + \text{df} \log n,
\]
where df indicates the number of parameters in the model, and $\ell$ is the log-likelihood
\[
\ell = \log L = -(n/2)\log{\sigma^2} -\sum_{i=1}^{n} \left(y_i - X_i \boldsymbol{\beta} \right)^2/({2\sigma^2}).
\]
Two cases can be considered depending on whether the variance $\sigma^2$ is known or unknown.
\begin{enumerate}
    \item If $\sigma^2$ is known, and the data are scaled so that $\sigma^2 = 1$, then
    \[
    \text{\BIC} = n^{-1}\sum_{i=1}^{n} \Big(y_i - X_i \boldsymbol{\widehat\beta} \Big)^2 + \text{df}\dfrac{\log n}{n}.
    \]
    \item If $\sigma^2$ is unknown, using $\widehat{\sigma}^2_{\text{MLE}} = n^{-1} \sum_{i=1}^{n} \left(y_i - X_i \boldsymbol{\widehat \beta} \right)^2 $ leads to
    \[
    \text{\BIC} = n\log \Big\{n^{-1}\sum_{i=1}^{n} \Big(y_i - X_i \boldsymbol{\widehat \beta} \Big)^2 \Big\} + \text{df} \log n.
    \]
\end{enumerate}

\cite{Wilms:2015wna} use criterion 2 for canonical correlation analysis by substituting $\| X_1 \widetilde w_1 - X_2 w_2\|_2^2 /n$ instead of $\sum_{i=1}^{n} (y_i - X_i \boldsymbol{\widehat \beta})^2 /n$ for centered $X_1$ and $X_2$. Since $\| X_1 \widetilde w_1 - X_2 w_2\|_2^2 /n = \widetilde w_1^{\top}S_1 \widetilde w_1 - 2 \widetilde w_1^{\top}S_{12}w_2 + w_2^{\top}S_2w_2$, and we use $\widetilde R$ instead of the sample covariance matrix $S$, we substitute
\[
f(\widetilde{w}_1) = \widetilde{w}_1^\top \widetilde R_{1} \widetilde{w}_1 - 2 \widetilde{w}_1^\top \widetilde R_{12} w_2 + w_2 \widetilde R_{2} w_2
\]
instead of residual sum of squares. Furthermore, motivated by the performance of the adjusted degrees of freedom variance estimator in \citet{reid2016study}, we also adjust $f(\widetilde w_1)$ for the 2nd criterion leading to
\begin{equation}
    \begin{split}\nonumber
        \text{\BIC}_1  = f(\widetilde{w}_1) + \text{df}_{\widetilde{w}_1} \dfrac{\log n}{n};\quad
        \text{\BIC}_2  = \log \Big\{ \dfrac{n}{n-\text{df}_{\widetilde{w}_1}} f(\widetilde{w}_1) \Big\}  + \text{df}_{\widetilde{w}_1} \dfrac{\log n}{n}.
    \end{split}
\end{equation}
Here df$_{\widetilde w_1}$ coincides with the size of the support of $\widetilde w_1$ \citep{Tibshirani:2012jk}. The \BIC~criteria for $w_2$ are defined analogously to those for $w_1$.

We use both criteria in evaluating our approach. Given the selected criterion (either $\text{\BIC}_1$ or 
$\text{\BIC}_2$), we apply it sequentially at each step of the biconvex optimization algorithm of Section~\ref{sec:optim}, and each time select the tuning parameter corresponding to the smallest value of the criterion. Due to alternating minimization, the solution will in general depend on the choice of the initial starting point. By default, we initialize the algorithm with the unpenalized solution to \eqref{eq:pCCA} obtained using $\widetilde R + 0.25I$, which corresponds to canonical ridge solution with fixed amount of regularization \citep{RCCA:2008}. We find that this initialization works well compared to a random initialization, more details are provided in Section~S3$\cdot$2 of the Supplementary Material.

\begin{remark}
    A sequence of $\lambda$ values for $w_1$ and $w_2$ are separately generated for the algorithm if there is no specification. For example, a sequence for $\lambda_1$ is generated as follows. We first calculate $\lambda_{\rm max} = \widetilde R_{12} \widehat w^{(0)}_{2}$ and $\lambda_{\rm min} = \epsilon \lambda_{\rm max}$, where $\widehat w^{(0)}_{2}$ is the initial starting point for $w_2$. Then, from $\lambda_{\rm min}$ to $\lambda_{\rm max}$, the sequence is generated to be equally spaced on a logarithmic scale. As a default, we use $20$ lambda values for each side with $\epsilon = 0.01$. The sequence for $\lambda_2$ is analogously defined.
\end{remark}


\section{Simulation studies}\label{sec:sim}


In this section we evaluate the performance of the following methods: (i) Classical canonical correlation analysis based on the sample covariance matrix; (ii) Canonical ridge available in the R package \texttt{CCA} \citep{RCCA:2008}; (iii) Sparse canonical correlation analysis of \cite{Witten:2009PMD} available in the R package \texttt{PMA}; (iv) Sparse canonical correlation analysis of \citet{Gao:2017SCCA} available in the Matlab package \texttt{SCCALab}; (v) Sparse canonical correlation analysis via Kendall's $\tau$ proposed in this paper. For our method, we evaluate both types of \BIC~criteria as described in Section~\ref{sec:tuning}. We also consider using the Pearson sample correlation instead of $\widetilde R$ within our optimization framework with the same \BIC-criteria for parameter selection. For fair comparison with $\widetilde R$, we also apply shrinkage to the Pearson correlation matrix as in~\eqref{d:Rtilde}. Direct comparison of estimation performance between our rank-based estimator and Pearson sample correlation as a function of sample size and level of truncation can be found in the Supplementary Material Section S3.1.

We generate $n=100$ independent pairs $(\mathbf{Z}_1, \mathbf{Z}_2) \in \R^{p_1+p_2}$ following
\[
\bpm \mathbf{Z}_1\\\mathbf{Z}_2 \epm \sim \N \left\{ \bpm 0\\0 \epm, \bpm \Sigma_1 & \rho \Sigma_1 w_1 w_2^{\top} \Sigma_2\\
\rho \Sigma_2 w_2 w_1^{\top} \Sigma_1 & \Sigma_2 \\
\epm
\right\}.
\]
We consider two settings for the number of variables: low-dimensional ($p_1=p_2=25$) and high-dimensional ($p_1=p_2=100$).
Each canonical vector $w_g$ ($g=1, 2$) is defined by taking a vector of ones at the coordinates $(1, 6, 11, 16, 21)$ and zeros elsewhere, and normalizing it such that $w_g^\top \Sigma_g w_g =1$; a similar model is used in \citet{Chen:2013}. We use an autoregressive structure for $\Sigma_{1} = \{\gamma^{|j-k|}\}_{j,k=1}^{p_1}$ and a block-diagonal structure for $\Sigma_2=$block-diag$(\Sigma_\gamma, \ldots, \Sigma_\gamma)$, where $\Sigma_\gamma\in R^{d\times d}$ is an equicorrelated matrix with value $1$ on the diagonal and $\gamma$ off the diagonal. We use five blocks of size $d\in\{6, 6, 3, 7, 3\}$ for low-dimensional, and $d\in\{14, 21, 12, 25, 28\}$ for high-dimensional setting.
We set $\gamma = \text{0$\cdot$7}$ for both $\Sigma_1$ and $\Sigma_2$. We further randomly permute the order of variables in each $\mathbf{Z}_g$ to remove the covariance-induced ordering.  The value of the canonical correlation is set at $\rho = \text{0$\cdot$9}$.

We consider transformations $\mathbf{U}_g = f_g(\mathbf{Z}_g +\mathbf{B}_g)$ where the elements of vector $\mathbf{B}_g$ are 0 or 1 with equal probability. The variation in the shift of $\mathbf{Z}_g$ across $p_g$ variables due to $\mathbf{B}_g$ leads to the variation in the proportion of zeros across the variables in the 5--80\% range for the same choice of truncation constant $C$. We consider three choices for $f_g$: (copula~0) no transformation, $f_g(z) = z$ for $g=1, 2$; (copula 1) exponential transformation for $\mathbf{U}_1$, $f_1(z) = \exp(z)$, and no transformation for $\mathbf{U}_2$, $f_2(z) = z$; (copula 2) exponential transformation for $\mathbf{U}_1$, $f_1(z) = \exp(z)$, and cubic transformation for $\mathbf{U}_2$, $f_2(z) = z^3$. Finally, we set $\mathbf{X}_g$ to be equal to $\mathbf{U}_g$ for continuous variable type, and dichotomize/truncate $\mathbf{U}_g$ at the same value $C$ for all variables to form binary/truncated $\mathbf{X}_g$. We set $C=\text{1$\cdot$5}$ for exponentially transformed variables, and $C=0$ for the others. For each case, we consider three combinations of variable types for $\mathbf{X}_1$/$\mathbf{X}_2$: truncated/truncated, truncated/continuous and truncated/binary.

To compare the methods' performance, we evaluate expected out-of-sample correlation
\begin{equation}\label{def:rhohat}
    \widehat{\rho} = \left| \dfrac{ \widehat{w}_1^\top \Sigma_{12} \widehat{w}_2 }{ (\widehat{w}_1^\top \Sigma_{1} \widehat{w}_1)^{1/2} (\widehat{w}_2^\top \Sigma_{2} \widehat{w}_2)^{1/2} } \right|,
\end{equation}
and predictive loss
\begin{equation}\label{def:predloss}
 L(w_g, \widehat w_g) = 1-\dfrac{|\widehat{w}_g^\top \Sigma_{g} w_g|}{(\widehat{w}_g^\top \Sigma_{g} \widehat{w}_g)^{1/2} } \quad (g=1,2);
\end{equation}
 a similar loss is used in \cite{Gao:2017SCCA}. By definition of the true canonical correlation $\rho$, for any $\widehat w_1$ and $\widehat w_2$ it holds that $\widehat \rho \leq \rho$, with equality when $\widehat w_1 = w_1$ and $\widehat w_2 = w_2$. Since $w_g^{\top}\Sigma_g w_g=1$, $L(w_g, \widehat w_g)\in[0,1]$ with $L(w_g, \widehat w_g)=0$ if $\widehat w_g = w_g$.
 We also evaluate the variable selection performance using the selected model size, true-positive rate and true-negative rate defined as
$$
\text{TPR}_{g} = \dfrac{\#\{j: \widehat{w}_{gj} \neq 0 \text{ and } w_{gj} \neq 0\}}{\#\{j: w_{gj} \neq 0\}}, \quad \text{TNR}_{g} = \dfrac{\#\{j: \widehat{w}_{gj} = 0 \text{ and } w_{gj} = 0\}}{\#\{j: w_{gj} = 0\}} \quad (g=1,2).
$$

\begin{figure}[!t]
\centering
    \includegraphics[scale = 0.55]{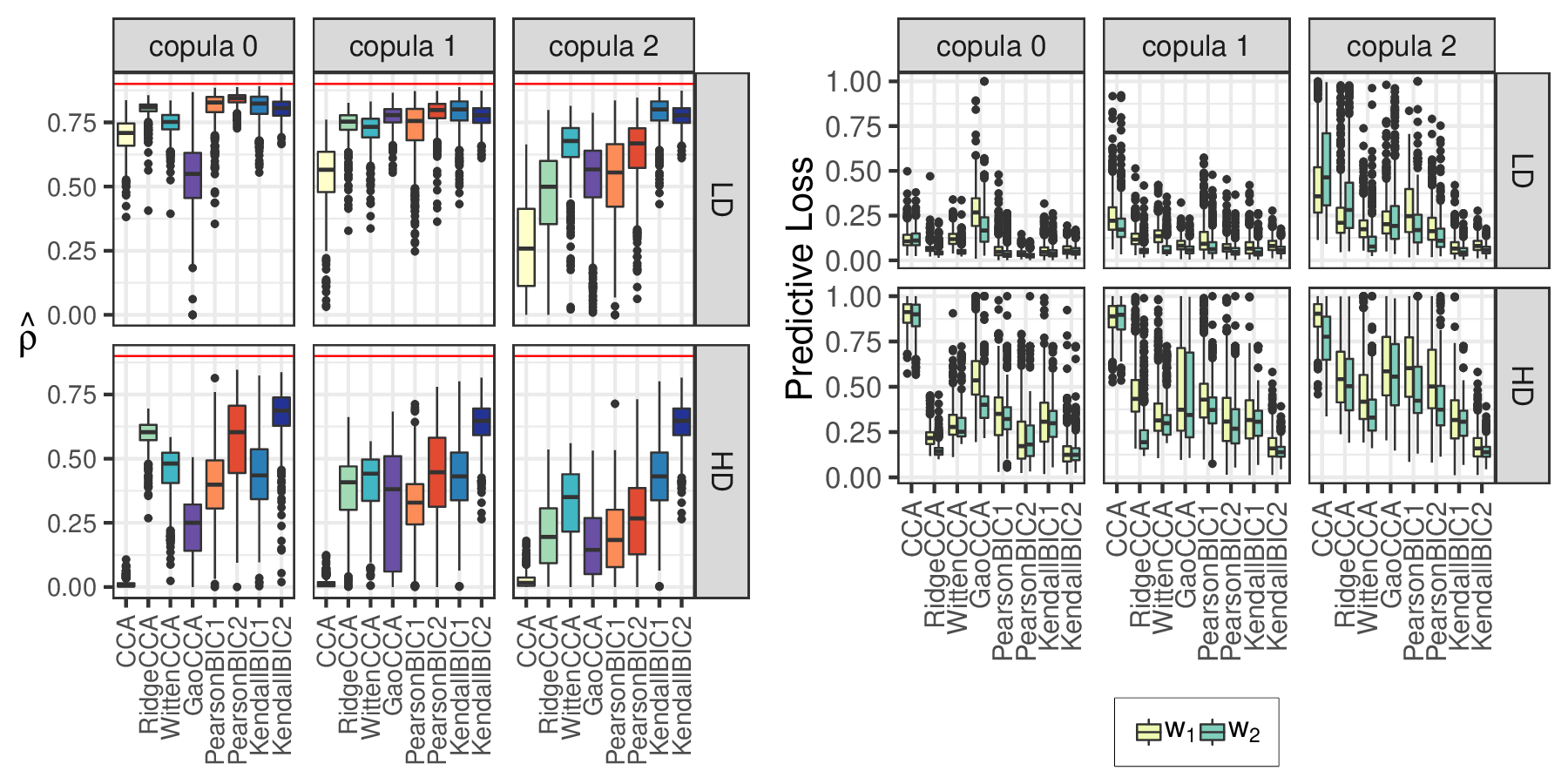}
        \caption{Truncated/truncated case. \textbf{Left:} The value of $\widehat{\rho}$ from \eqref{def:rhohat}. The horizontal lines indicate the true canonical correlation value $\rho = $ \text{0$\cdot$9}. \textbf{Right:} The value of predictive loss~\eqref{def:predloss}. Results over 500 replications. CCA:~Sample canonical correlation analysis; RidgeCCA: Canonical ridge of~\citet{RCCA:2008}; WittenCCA: method of~\cite{Witten:2009PMD}; GaoCCA:~method of ~\cite{Gao:2017SCCA}; PearsonBIC1, PearsonBIC2: proposed algorithm with Pearson sample correlation matrix; KendallBIC1, KendallBIC2: proposed algorithm with rank-based estimator $\widetilde R$; \BIC$_1$ or \BIC$_2$ refer to tuning parameter selection criteria; LD: low-dimensional setting ($p_1 = p_2 = 25$); HD: high-dimensional setting ($p_1 = p_2 = 100$).}\label{fig:TT_rhohatPredloss}
\end{figure}

\begin{figure}[!t]
\centering
    \includegraphics[scale = 0.55]{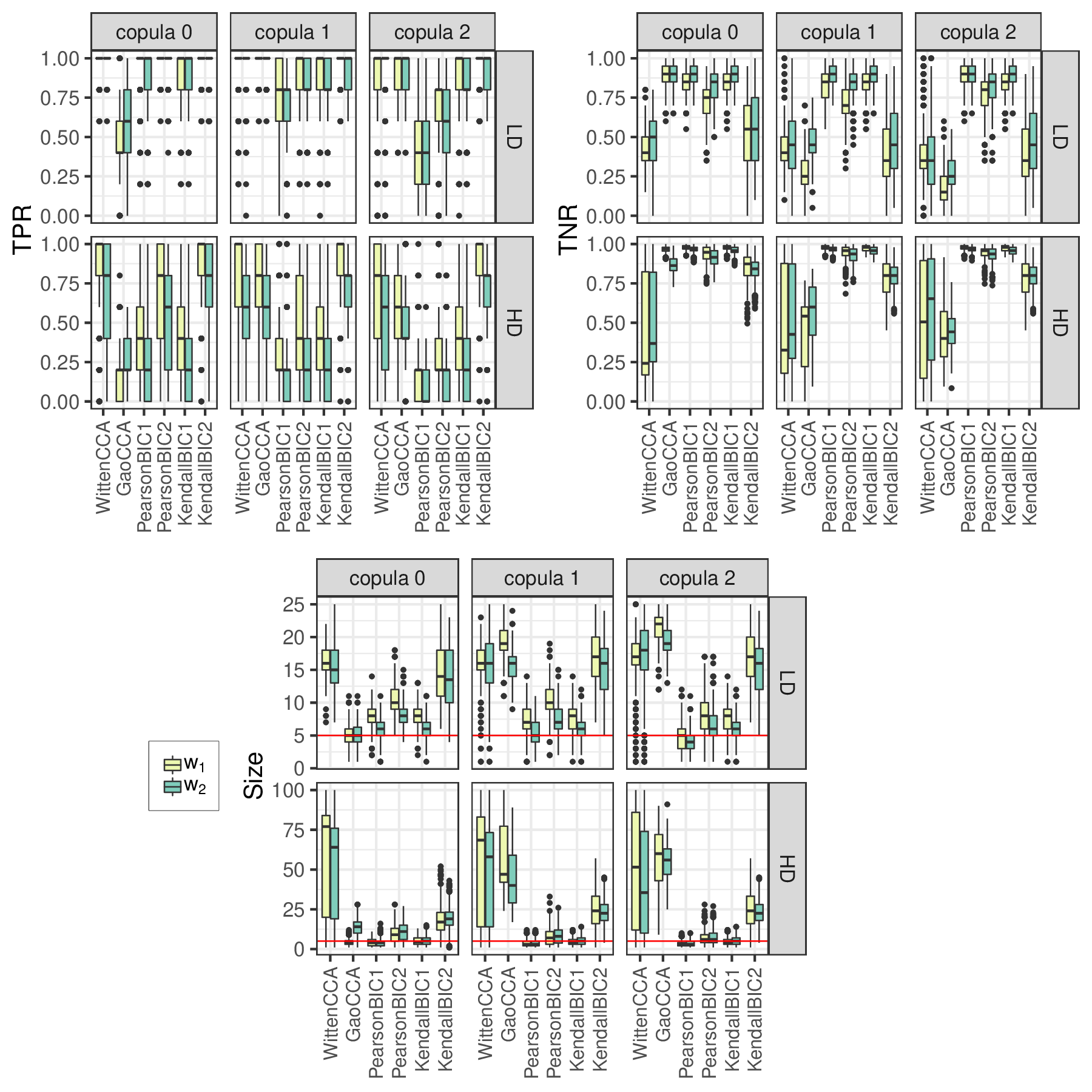}
    \caption{Truncated/truncated case. \textbf{TopLeft:} True positive rate (TPR); \textbf{TopRight:} True negative rate (TNR); \textbf{BottomMiddle:} Selected model size. Results over 500 replications. WittenCCA: method of~\cite{Witten:2009PMD}; GaoCCA:~method of ~\cite{Gao:2017SCCA}; PearsonBIC1, PearsonBIC2: proposed algorithm with Pearson sample correlation matrix; KendallBIC1, KendallBIC2: proposed algorithm with rank-based estimator $\widetilde R$; \BIC$_1$ or \BIC$_2$ refer to tuning parameter selection criteria; LD: low-dimensional setting ($p_1 = p_2 = 25$); HD: high-dimensional setting ($p_1 = p_2 = 100$).}\label{fig:TT_selection}
\end{figure}

The results for the truncated/truncated case over 500 replications are presented in Figures \ref{fig:TT_rhohatPredloss}--\ref{fig:TT_selection}. From Figure~\ref{fig:TT_rhohatPredloss}, the majority of methods achieve higher values of $\widehat \rho$ in the absence of data transformation (copula 0) compared to cases where transformation is applied (copula 1 and 2). The only exception is our approach based on Kendall's $\tau$, which as expected has comparable performance across the copula types. The performance of all methods deteriorates with increased dimension leading to smaller values of $\widehat\rho$ and larger predictive losses. The classical canonical correlation analysis performs especially poorly in high-dimensional settings with $\widehat \rho$ being almost 0 and predictive loss being close to 1 for both $w_1$ and $w_2$. Canonical ridge works well in the copula 0 setting, however its performance is strongly affected in the presence of transformations (copula 1 and 2). Surprisingly to us, Gao's method, as implemented in \texttt{SCCALab}, performs poorly compared to other approaches. Since Gao's method is designed for Gaussian data, the poor performance is likely due to its sensitivity to the presence of copulas and zero truncation (in the copula 0 case, proportions of zero values for each variable range from $5\%$ to $70\%$). We also use the default values in \texttt{SCCALab} for all of the parameters, so better performance could possibly be achieved by adjusting those values. Sparse canonical correlation analysis based on Pearson's correlation outperforms all other methods in low dimensional setting when no data transformation is applied (copula 0), however its performance deteriorates when the monotone transformations are applied to the data (copulas 1 and 2). It also performs worse than our rank-based approach in high-dimensional setting. This is likely due to the increase in variables with zero inflation due to truncation, which Pearson's correlation doesn't take into account. In low-dimensional settings, \BIC$_1$ and \BIC$_2$ criteria lead to similar values of $\widehat \rho$, with larger variance in \BIC$_1$ performance. In high-dimensional settings, \BIC$_2$ is clearly better than \BIC$_1$ in predictive performance, and this better performance is irrespective of the choice of the estimator for the latent correlation matrix (Pearson's correlation matrix or proposed rank-based correlation matrix). Overall, our method based on Kendall's $\tau$ with \BIC$_2$ criterion leads to highest values of $\widehat \rho$ and smallest values of predictive loss across dimensions and different copula types.

Figure~\ref{fig:TT_selection} illustrates variable selection performance of each method. The classical canonical correlation analysis and canonical ridge are excluded as they do not perform variable selection. To ensure the results are consistent with numerical precision of optimization algorithm, we treat variable as nonzero if its loading is above $10^{-6}$ threshold in absolute value. Unexpected to us, the number of selected variables varies significantly across replications for Witten's method (bottom figure in Figure~\ref{fig:TT_selection}), leading to significant variations in true positive and true negative rates. We suspect this is due to the use of a permutation approach for selection of tuning parameters. Our approach based on Kendall's $\tau$ leads to a more favorable combination of true positive and true negative rates compared to competing methods, especially when data transformations are applied. Furthermore, this advantage is maintained independently of tuning parameter selection scheme. In Section~S3$\cdot$3 of the Supplementary Material, we compare the true positive versus false positive curves obtained by each method over the range of tuning parameters, and find that our rank-based estimator leads to highest area under the curve in the copula settings.
Comparing \BIC$_1$ with \BIC$_2$ performance in Figure~\ref{fig:TT_selection}, \BIC$_1$ leads to the sparsest model and the highest true negative rate for both Pearson correlation and our rank-based correlation , at the expense of missing some true variables in the high-dimensional settings. Given the comparison in predictive performance between the two selection criteria, we conclude that \BIC$_1$ is better suited for variable selection, especially when it is desired to have a high true negative rate, whereas \BIC$_2$ works better for prediction.

In addition to the truncated/truncated case, we also consider truncated/continuous and truncated/binary cases in Section~S3$\cdot$4 of the Supplementary Material. The conclusions of methods' comparison are similar to the truncated/truncated case. Overall, all the methods perform best in the truncated/continuous case and worst in the truncated/binary case, which is not surprising, since dichotomization of continuous variable leads to a loss of information, thus reducing the effective sample size.


\section{Application to TCGA data}\label{sec:data}


The Cancer Genome Atlas (TCGA) project collects data from multiple platforms using high-throughput sequencing technologies. We consider gene expression data ($p_1=891$) and micro RNA data ($p_2=431$) for $n=500$ matched subjects from the TCGA breast cancer database. We treat gene expression data as continuous and micro RNA data as truncated continuous. The range of proportions of zero values contained in each variable in micro RNA data is $0 - \text{49$\cdot$8}\%$. The subjects belong to one of the 5 breast cancer subtypes: Normal, Basal, Her2, LumA and LumB, with 37 subjects having missing subtype information (denoted as NA). The goal of the analysis is to characterize the association between gene expression and micro RNA data, and investigate whether this association is related to breast cancer subtypes.

To investigate the performance of our method relative to other approaches, we randomly split the data 500 times. Each time $400$ samples are used for training, and the remaining $100$ test samples are used to assess the association via
\[
\widehat{\rho}_{\text{test}} = \left| \dfrac{\widehat{w}_{1, \text{train}}^\top \Sigma_{12, \text{test}} \widehat{w}_{2, \text{train}}}{(\widehat{w}_{1, \text{train}}^\top \Sigma_{1, \text{test}} \widehat{w}_{1, \text{train}})^{1/2}  (\widehat{w}_{2, \text{train}}^\top \Sigma_{2, \text{test}} \widehat{w}_{2, \text{train}})^{1/2}} \right|.
\]
Here $\Sigma_{\text{test}}$ is evaluated based on the test samples, and is either the rank-based estimator $\widetilde R$ for our method, or the sample covariance matrix for other methods. We also compare the number of selected genes and selected micro RNAs, and the results are presented in Table~\ref{tab:TCGA_cv}. We have not considered the method of \citet{Gao:2017SCCA} in this section due to its poor performance in Section~\ref{sec:sim} and high computational cost (it takes around 40 minutes per replication on these data on a Windows 3$\cdot$60GHz Intel Core i7 CPU machine).

\begin{table}[!b]
	\centering
  \begin{tabular}{lrrrrrrrrrrl}
  \hline
    Method & \multicolumn{4}{r}{Selected Genes} & \multicolumn{4}{r}{Selected micro RNAs} & \multicolumn{3}{c}{$\widehat{\rho}_{\text{test}}$}\\
    \hline
  CCA & & & 891$\cdot$00 & (0$\cdot$00) & & & 431$\cdot$00 & (0$\cdot$00) & & 0$\cdot$004 & (0$\cdot$109) \\ 
  RidgeCCA & & & 891$\cdot$00 & (0$\cdot$00) & & & 431$\cdot$00 & (0$\cdot$00) & & 0$\cdot$712 & (0$\cdot$126)\\ 
  WittenCCA & & &  338$\cdot$36 & (194$\cdot$58)& & & 165$\cdot$53 & (100$\cdot$30) & & 0$\cdot$789 & (0$\cdot$041)  \\ 
  PearsonBIC1 & & & 9$\cdot$92 & (2$\cdot$62) & & & 16$\cdot$08 & (3$\cdot$18) & & 0$\cdot$813 & (0$\cdot$044) \\ 
  PearsonBIC2 & & & 27$\cdot$68 & (6$\cdot$20)  & & & 40$\cdot$50 & (12$\cdot$54) & & 0$\cdot$857 & (0$\cdot$034) \\ 
  KendallBIC1 & & & 18$\cdot$24 & (3$\cdot$51) & & & 9$\cdot$68 & (3$\cdot$15) & &  $\mathbf{0}$\mbox{$\cdot$}$\mathbf{880}$ & (0$\cdot$030) \\ 
  KendallBIC2 & & & 38$\cdot$18  & (8$\cdot$47) & & & 31$\cdot$01 & (6$\cdot$74) & & $\mathbf{0}$\mbox{$\cdot$}$\mathbf{913}$ & (0$\cdot$029)\\
  \hline
  \end{tabular}
\caption{Mean support sizes and values of $\widehat{\rho}_{\text{test}}$'s over 500 random splits of breast cancer data. The standard deviation is given in parentheses}\label{tab:TCGA_cv}%
\end{table}

Of course, neither the sample canonical correlation analysis nor the canonical ridge method performs variable selection. In addition, $\widehat{\rho}_{\text{test}}$ is very close to $0$ for the sample canonical correlation, confirming poor performance of the method. Canonical ridge leads to significantly higher values of $\widehat \rho_{\text{test}}$ demonstrating the advantage of added regularization, however it still has smaller correlation values compared to other approaches. The method of~\cite{Witten:2009PMD} leads to higher correlation values compared to both sample canonical correlation analysis and canonical ridge, however it still selects a significant number of variables, with highly variable model sizes across replications. We suspect this is due to the use of a permutation-based algorithm for tuning parameter selection: similar behaviour is also observed in Section~\ref{sec:sim}. Sparse canonical correlation analysis based on Pearson's correlation selects a much smaller number of genes and micro RNAs but achieves higher values of $\widehat \rho_{\text{test}}$ than the method of \citet{Witten:2009PMD}. This is consistent with results in Section~\ref{sec:sim}. The highest values of $\widehat \rho_{\text{test}}$ are achieved by our approach based on Kendall's $\tau$ with smaller number of selected variables, confirming that found association is not due to over-fitting as it generalizes well to out-of-sample data. \BIC$_1$ criterion leads to the sparser model than \BIC$_2$ consistently for both Pearson and Kendall-based correlation estimates, with \BIC$_2$ criterion having the larger out-of-sample correlation value. In light of these results and results of Section~\ref{sec:sim}, we conclude that \BIC$_1$ is advantageous for variable selection due to its selection of sparser model and higher true negative rate observed in simulations, whereas \BIC$_2$ is advantageous for prediction.

\begin{figure}[!t]
\centering
\includegraphics[scale = 0.35]{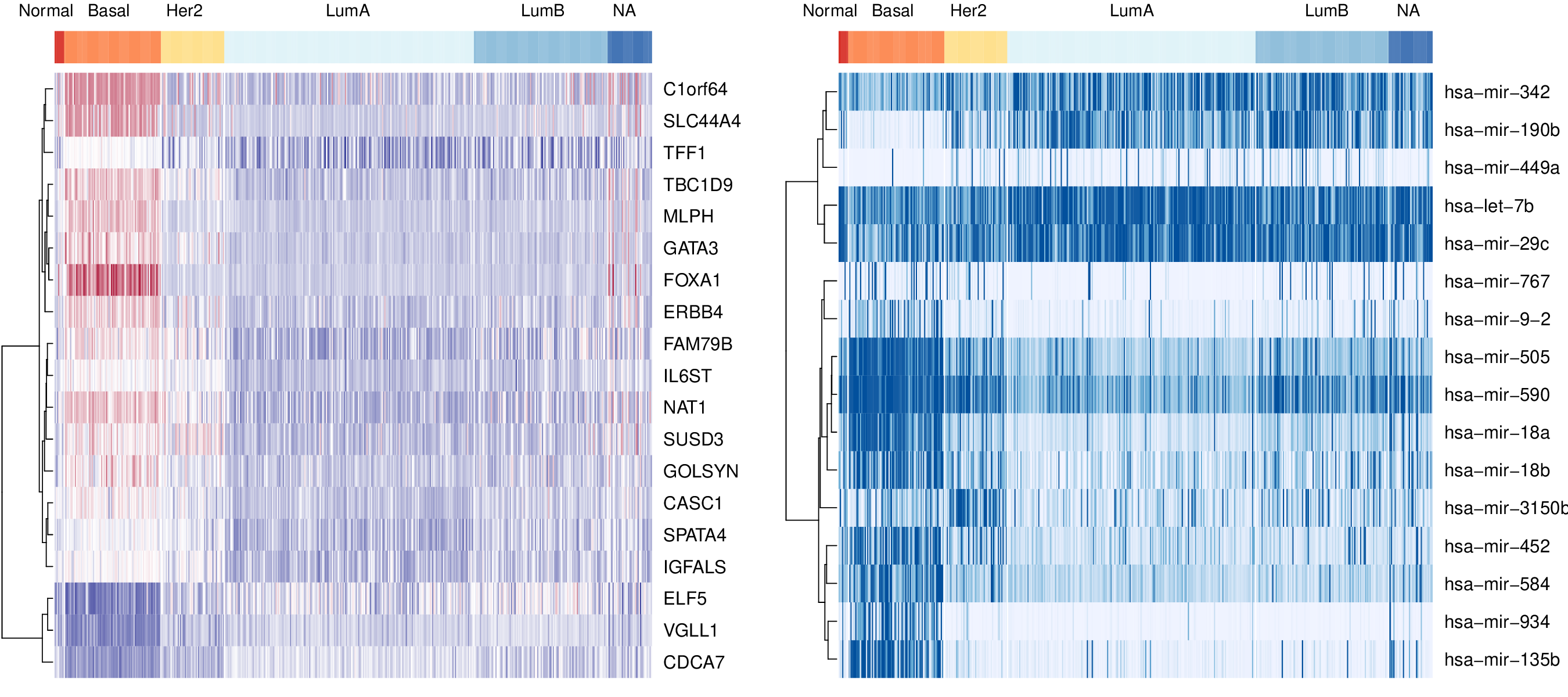}
        \caption{\baselineskip=12pt
        Genes and micro RNAs selected often more than 80\% of 500 repetitions by our approach with the~\BIC$_2$ criterion are used for heatmap. \textbf{Left:} A heatmap of 19 genes. The blue indicates positive expression level, and red for negative expression level. The white means zero expression level. \textbf{Right:} A heatmap of 16 micro RNAs. The saturation level of colors are assigned based on variable-specific quantiles. For both figures, the dissimilarity measure is set as $1-\widetilde{R}$ with our rank-based correlation $\widetilde{R}$, and Ward linkage is used.}\label{fig:TCGA_heatmapboth}
\end{figure}

We next investigate possible relationships between selected variables and breast cancer subtypes. Since the selected variables may change across the random data splits, we consider the selection frequency of each gene and micro RNA across all 500 replications of our method with \BIC$_2$ criterion, and choose the variables that are selected at least 80\% of the times. Figure~\ref{fig:TCGA_heatmapboth} shows heatmaps of expression levels of resulting 19 genes and 16 micro RNAs, with samples ordered by their respective cancer subtype. The heatmaps show clear separation between Basal and other subtypes, suggesting that the found association is relevant to cancer biology.

Many of the selected genes and micro RNAs can be found in recent literature which supports their association with breast cancer. \citet{ERBB4:kim2016prognostic} indicates that ERBB4 is a prognostic marker for triple negative breast cancer, which is often used interchangeably with Basal-like breast cancer. In agreement with our results, \citet{VGLL1} identifies that VGLL1 and miR-934 are highly correlated with each other, and that both are overexpressed in the Basal-like subtype. They also find that selected FOXA1 and GATA3 genes, as well as ESR1 gene (not selected at 80\% frequency threshold, but still has a 73.4\% frequency), have strong negative correlation with both VGLL1 and miR-934. The expression level of selected ELF5 is shown to play a key role in determining breast cancer molecular subtype in \citet{ELF5_kalyuga} and \citet{ELF5:piggin2016}. Furthermore, \citet{miRNA18a_505} validate that selected hsa-miR-18a and hsa-miR-505 miRNAs are significantly correlated with prognostic breast cancer biomarkers, and high expression of hsa-miR-18a is strongly associated with Basal-like breast cancer features. Finally, the selected hsa-miR-135b is reported to be related to breast cancer cell growth in \citet{miRNA135b_ref1} and \citet{miRNA135b_ref2}.


\section{Discussion}


One of the main contributions of this work is a truncated Gaussian copula model for the zero-inflated data, and corresponding development of a rank-based estimator for the latent correlation matrix. While our focus is on canonical correlation analysis, our estimator can be used in conjunction with other covariance-based approaches. For example it can be used for constructing graphical models as in~\citet{Fan:2016um} in cases where some or all of the variables have an excess of zeros. Micro RNA data is one example that we have explored in this work, however another prominent example is microbiome abundance data. It would be of interest to further explore the potential of our modeling approach in different application areas. The R package \texttt{mixedCCA} with our method's implementation is available from the authors github page \url{https://github.com/irinagain/mixedCCA}.

\section*{Acknowledgements}
Yoon's research was funded by a grant from the National Cancer Institute (T32-CA090301). Carroll's research was supported by a grant from the National Cancer Institute (U01-CA057030). Carroll is also Distinguished Professor, School of Mathematical and Physical Sciences, University of Technology Sydney, Broadway NSW 2007, Australia. Gaynanova's research was supported by National Science Foundation grant DMS-1712943.

\bibliographystyle{biometrika}
\bibliography{KendallCCAReferences}

\end{document}


\maketitle

\renewcommand{\theequation}{S.\arabic{equation}}
\renewcommand{\thesection}{S.\arabic{section}}
\renewcommand{\thesubsection}{S\arabic{section}.\arabic{subsection}}
\renewcommand{\thetheorem}{S.\arabic{theorem}}
\renewcommand{\thefigure}{S\arabic{figure}}

The Supplement Material is organized as follows. Section~\ref{appen2:thmproofs} includes the proofs of all results in the paper. Section~\ref{app:lemmas} includes the proofs of supporting lemmas. Section~\ref{app:extrasim} includes additional simulation results.

\section{Proofs of main results in the paper}\label{appen2:thmproofs}
\begin{proof}[Proof of Theorem 2]
		Without loss of generality, let $j=1$ and $k=2$.
		By the definition of Kendall's $\tau$,
		\[
		\tau_{12}  = \E(\widehat{\tau}_{12})
		= \E \Big[ \dfrac{2}{n(n-1)} \sum_{1\le i<i'\le n} \sign \left\{(X_{i1}-X'_{i1}) (X_{i2}-X'_{i2}) \right\} \Big].
		\]
		Since $X_2$ is binary,
		\begin{equation}
			\begin{split}\nonumber
				\sign\left(X_{2}-X'_{2}\right) & = I\left(U_{2} >C_2, U'_2\le C_2 \right)-I\left( U_2 \le C_2, U'_2 >C_2 \right)\\
				& = I\left(U_2 >C_2 \right) - I\left(U_2 >C_2, U'_2>C_2 \right)
				- I\left( U'_2 >C_2 \right) + I\left( U_2 > C_2, U'_2 >C_2 \right)\\
				& = I\left( U_2 >C_2 \right)
				- I\left( U'_2 >C_2 \right),
			\end{split}
		\end{equation}
		thus
		\begin{equation}
			\begin{split}\label{eq:binarysplit}
				\tau_{12}  & = \E \left[ \sign\left( X_{1}-X'_{1} \right) I(U_{2} >C_2)  \right] - \E \left[ \sign\left( X_{1}-X'_{1} \right) I(U'_{2} >C_2) \right].
			\end{split}
		\end{equation}
		Since $X_1$ is truncated, $C_1 >0$ and
		\begin{equation}\label{eq:lem1-truncsign}
			\begin{split}
				\sign\left( X_1-X'_1 \right)  &= -I(X_{1}=0, X'_{1}>0)
				+I(X_{1}>0, X'_{1}=0)
				+I(X_{1}>0, X'_{1}>0)\sign(X_1-X'_1)\\
				&= -I(X_{1}=0)
				+I(X'_{1}=0)
				+I(X_{1}>0, X'_{1}>0)\sign(X_1-X'_1).
			\end{split}
		\end{equation}
		Since we assume that $\mathbf{U} = (U_1,\ldots, U_p)^{\top} \sim \NPN(0, \Sigma, f)$, let $Z=f(U)$ where $Z \sim N(0, \Sigma)$ for the rest of the proofs. Since $f$ is monotonically increasing, $\sign(X_1-X'_1) = \sign(Z_1-Z'_1)$. Combining \eqref{eq:binarysplit} and \eqref{eq:lem1-truncsign} yields
		\begin{equation}
			\begin{split}\nonumber
				\tau_{12} & = - 2 \E \left\{ I( X_1=0) I( U_2>C_2) \right\} + 2 \E \left\{ I( X'_1=0) I( U_2>C_2) \right\}\\
				& \quad + \E \left\{ I( X_1>0, X'_1>0) \sign\left(Z_1-Z'_1\right) I( U_2>C_2) \right\}\\
				& \quad - \E \left\{ I( X_1>0, X'_1>0) \sign\left(Z_1-Z'_1\right) I( U'_2>C_2) \right\}.
			\end{split}
		\end{equation}
		
		From the definition of $U$, let $Z_j=f_j(U_j)$ and $\Delta_j=f_j(C_j)$ for $j=1,2$. Using $\sign\left(x\right) = 2 I(x>0) -1$, we obtain
		\begin{equation}
			\begin{split}\nonumber
				\tau_{12}	& = -2\E \left\{I\left( Z_1\le\Delta_1, Z_2>\Delta_2\right) \right\} + 2\E \left\{I\left( Z'_1 \le\Delta_1, Z_2>\Delta_2\right) \right\}\\
				&
				\quad
				+ 2\E \left\{ I{\left( Z_1>\Delta_1, Z'_1>\Delta_1 , Z_1-Z'_1>0 \right)}  I{\left( Z_2>\Delta_2 \right)} \right\}\\
				&
				\quad - 2\E \left\{ I{\left( Z_1>\Delta_1, Z'_1>\Delta_1 , Z_1-Z'_1>0 \right)}  I{\left( Z'_2>\Delta_2 \right)} \right\}.
			\end{split}
		\end{equation}
		Since $Z'_1>\Delta_1 , Z_1-Z'_1>0$ implies $Z_1>\Delta_1$, so $\tau_{12}$ can be further simplified as
		\begin{equation}
			\begin{split}\nonumber
				\tau_{12}	& = -2\E \left\{I\left( Z_1\le\Delta_1, Z_2>\Delta_2\right) \right\} + 2\E \left\{I\left( Z'_1 \le\Delta_1, Z_2>\Delta_2\right) \right\}\\
				&
				\quad
				+ 2\E \left\{ I{\left( Z'_1>\Delta_1 , Z_2>\Delta_2,  Z_1-Z'_1>0 \right)} \right\} - 2\E \left\{ I{\left( Z'_1>\Delta_1 , Z'_2>\Delta_2 , Z_1-Z'_1>0 \right)} \right\}.
			\end{split}
		\end{equation}
		Since $\left\{(Z'_1-Z_1)/\surd{2}, -Z'_1\right\}$, $\left\{(Z'_1-Z_1)/\surd{2}, -Z_2\right\}$ and $\left\{(Z'_1-Z_1)/\surd{2}, ~ -Z'_2\right\}$ are standard bivariate normally distributed with correlation $-1/\surd{2}$, $\Sigma_{12}/\surd{2}$ and $-\Sigma_{12}/\surd{2}$ respectively, by the definition of $\Phi(\cdot)$ and $\Phi_2(\cdot, \cdot; r)$, we have that
		\begin{equation}\label{eq:lem1-1}
			\begin{split}
				\tau_{12} & =  2\Phi_2(\Delta_1, -\Delta_2; -\Sigma_{12})  - 2\Phi(\Delta_1)\Phi(-\Delta_2)\\
				& \quad
				+ 2 \Phi_3 \left\{-\Delta_1, -\Delta_2, 0; \bpm
				1 & 0 & -1/\surd{2} \\
				0 & 1 & \Sigma_{12}/\surd{2}\\
				-1/\surd{2} & \Sigma_{12}/\surd{2} & 1
				\epm \right\}\\
				& \quad
				- 2 \Phi_3 \left\{ -\Delta_1, -\Delta_2, 0; \bpm
				1 & \Sigma_{12} & -1/\surd{2} \\
				\Sigma_{12} & 1 & -\Sigma_{12}/\surd{2}\\
				-1/\surd{2} & -\Sigma_{12}/\surd{2} & 1
				\epm \right\}.
			\end{split}
		\end{equation}
		
		Using that $\Phi(\Delta_1) + \Phi(-\Delta_1) = 1$, $\Phi(\Delta_1)  = \Phi_2(\Delta_1, \Delta_2; \Sigma_{12}) + \Phi_2(\Delta_1, -\Delta_2; -\Sigma_{12})$ and
		\begin{equation}
			\begin{split}\label{eq:lem1-prop}
				\Phi_2(\Delta_1, \Delta_2; \Sigma_{12}) & = \Phi_3  \left\{ \Delta_1, \Delta_2, \Delta_3; \bpm
				1 & \Sigma_{12} & \Sigma_{13} \\
				\Sigma_{12} & 1 & \Sigma_{23}\\
				\Sigma_{13} & \Sigma_{23} & 1
				\epm \right\} \\ & \quad + \Phi_3 \left\{ \Delta_1, \Delta_2, -\Delta_3; \bpm
				1 & \Sigma_{12} & -\Sigma_{13} \\
				\Sigma_{12} & 1 & -\Sigma_{23}\\
				-\Sigma_{13} & -\Sigma_{23} & 1
				\epm \right\}
			\end{split}
		\end{equation}
		we further simplify as
		\begin{equation}
			\begin{split}\nonumber
				F_{\rm TB}(\Sigma_{12};\Delta_1, \Delta_2) & =
				2 \left\{1-\Phi(\Delta_1)\right\}\Phi(\Delta_2)\\
				& \quad
				-2 \Phi_3 \left\{ -\Delta_1, \Delta_2, 0; \bpm
				1 & -\Sigma_{12} & 1/\surd{2} \\
				-\Sigma_{12} & 1 & -\Sigma_{12}/\surd{2}\\
				1/\surd{2} & -\Sigma_{12}/\surd{2} & 1
				\epm \right\}\\
				& \quad
				- 2 \Phi_3 \left\{ -\Delta_1, \Delta_2, 0; \bpm
				1 & 0 & -1/\surd{2} \\
				0 & 1 & -\Sigma_{12}/\surd{2}\\
				-1/\surd{2} & -\Sigma_{12}/\surd{2} & 1
				\epm \right\}.
			\end{split}
		\end{equation}
		
	\end{proof}

	\begin{proof}[Proof of Theorem~3]
		Without loss of generality, we set $j=1$ and $k=2$.
		Plugging \eqref{eq:lem1-truncsign} into the definition of the population Kendall's $\tau_{12}$ and using $\sign(X_1-X'_1) = \sign(Z_1-Z'_1)$, we find
		\begin{equation}
			\begin{split}\nonumber
				\tau_{12}  & = -\E \left\{ I\left(Z_1\le\Delta_1\right) \sign(Z_2-Z'_2) \right\}
				+\E \left\{ I\left(Z'_1\le\Delta_1\right) \sign(Z_2-Z'_2) \right\}\\
				&
				\quad
				+\E \left\{ I\left( Z_1>\Delta_1, Z'_1>\Delta_1 \right) \sign(Z_1-Z'_1)
				\sign\left(Z_2-Z'_2\right) \right\}.
			\end{split}
		\end{equation}
		
		Using $\sign\left(x\right) = 2 I(x>0) -1$, it holds that
		\begin{equation}
			\begin{split}\nonumber
				\tau_{12}  & = -2 \E \left\{ I\left(Z_1\le\Delta_1, Z'_2-Z_2 <0\right) \right\} +2\E \left\{ I\left(Z'_1\le\Delta_1, Z'_2-Z_2 <0\right) \right\} \\
				& \quad  + \E \left\{ I\left( Z_1>\Delta_1, Z'_1>\Delta_1 \right) \sign(Z_{1}-Z'_{1})
				\sign\left(Z_2-Z'_2\right) \right\}.
			\end{split}
		\end{equation}

		It remains to show that the last term can be rewritten using cumulative normal distribution functions. The last term consists of four terms,
		\begin{equation}
			\begin{split}\nonumber
				& \E \left\{ I\left( Z_1>\Delta_1, Z'_1>\Delta_1 \right) \sign(Z_1-Z'_1)
				\sign\left(Z_2-Z'_2\right) \right\}\\
				& = \P \left( Z_1 >\Delta_1, Z'_1 >\Delta_1, Z_1-Z'_1 >0, Z_2-Z'_2 >0 \right)\\
				& \quad + \P \left( Z_1 >\Delta_1, Z'_1 >\Delta_1, Z_1-Z'_1 <0, Z_2-Z'_2 <0 \right)\\
				&  \quad -\P \left( Z_1 >\Delta_1, Z'_1 >\Delta_1, Z_1-Z'_1 >0, Z_2-Z'_2 <0 \right)\\
				& \quad -\P \left( Z_1 >\Delta_1, Z'_1 >\Delta_1, Z_1-Z'_1 <0, Z_2-Z'_2 >0 \right).
			\end{split}
		\end{equation}
		Consider the first term
		\begin{equation}\label{lem2:trick1}
			\begin{split}
				& \P \left( Z_1 >\Delta_1, Z'_1 >\Delta_1, Z_1-Z'_1 >0, Z_2-Z'_2 >0 \right) \\
				& = \P \left( Z'_1 >\Delta_1, Z_1-Z'_1 >0, Z_2-Z'_2 >0 \right)- \P \left( Z_1 \le \Delta_1, Z'_1 >\Delta_1, Z_1-Z'_1 >0, Z_2-Z'_2 >0 \right) \\
				& = \P \left( Z'_1 >\Delta_1, Z_1-Z'_1 >0, Z_2-Z'_2 >0 \right).
			\end{split}
		\end{equation}
		
		The last equality comes from the fact that $Z_1-Z'_1 >0$ cannot hold when $Z_1 \le \Delta_1$ and  $Z'_1 >\Delta_1$. Applying this argument to all the four terms leads to
		\begin{equation}
			\begin{split}\nonumber
				\E &\left\{ I\left( Z_1>\Delta_1, Z'_1>\Delta_1 \right) \sign(Z_1-Z'_1)
				\sign\left(Z_2-Z'_2\right) \right\}\\
				& = \P \left( -Z'_1 <\Delta_1, Z'_1-Z_1 <0, Z'_2-Z_2 <0 \right) + \P \left( -Z_1 <\Delta_1, Z_1-Z'_1 <0, Z_2-Z'_2 <0 \right)\\
				& \quad  - \P \left( -Z'_1 <\Delta_1, Z'_1-Z_1 <0, Z_2-Z'_2 <0 \right)  - \P \left( -Z_1 <\Delta_1, Z_1-Z'_1 <0, Z'_2-Z_2 <0 \right).
			\end{split}
		\end{equation}

		Using the definition of $\Phi(\cdot)$ and $\Phi_2(\cdot, \cdot; r)$,
		\begin{equation}
			\begin{split}\nonumber
				\tau_{12}  & = -2\Phi_2(\Delta_1, 0; -\Sigma_{12}/\surd{2}) +2\Phi_2(\Delta_1, 0; \Sigma_{12}/\surd{2}) \\
				& \quad  +
				2 \Phi_3 \left\{ -\Delta_1,0,0; \bpm
				1 & -1/\surd{2} & -\Sigma_{12}/\surd{2}\\
				-1/\surd{2} & 1 & \Sigma_{12}\\
				-\Sigma_{12}/\surd{2} & \Sigma_{12} & 1
				\epm \right\}\\
				& \quad - 2 \Phi_3 \left\{ -\Delta_1,0,0; \bpm
				1 & -1/\surd{2} & \Sigma_{12}/\surd{2} \\
				-1/\surd{2} & 1 & -\Sigma_{12}\\
				\Sigma_{12}/\surd{2} & -\Sigma_{12} & 1
				\epm
				\right\}
			\end{split}
		\end{equation}
		
		The second property in \eqref{eq:lem1-prop} yields
		\begin{equation}
			\begin{split}\nonumber
				-2\Phi_2(\Delta_1, 0; -\Sigma_{12}/\surd{2}) & = -2 \Phi(0) + 2\Phi_2(-\Delta_1, 0; \Sigma_{12}/\surd{2})\\
				2\Phi_2(\Delta_1, 0; \Sigma_{12}/\surd{2}) & = 2 \Phi(0) - 2\Phi_2(-\Delta_1, 0; -\Sigma_{12}/\surd{2}),
			\end{split}
		\end{equation}
		and we can simplify further using the third property in  \eqref{eq:lem1-prop}
		\begin{equation}
			\begin{split}\nonumber
				2\Phi_2(-\Delta_1, 0; \Sigma_{12}/\surd{2}) - 2\Phi_3 \Bigg\{ -\Delta_1,0,0; & \bpm
				1 & -1/\surd{2} & \Sigma_{12}/\surd{2}\\
				-1/\surd{2} & 1 & -\Sigma_{12}\\
				\Sigma_{12}/\surd{2} & -\Sigma_{12} & 1
				\epm \Bigg\} \\
				& = 2\Phi_3 \Bigg\{-\Delta_1,0,0; \bpm
				1 & 1/\surd{2} & \Sigma_{12}/\surd{2}\\
				1/\surd{2} & 1 & \Sigma_{12}\\
				\Sigma_{12}/\surd{2} & \Sigma_{12} & 1
				\epm\Bigg\}\\
				- 2\Phi_2(-\Delta_1, 0; -\Sigma_{12}/\surd{2}) + 2\Phi_3 \Bigg\{-\Delta_1,0,0; &  \bpm
				1 & -1/\surd{2} & -\Sigma_{12}/\surd{2}\\
				-1/\surd{2} & 1 & \Sigma_{12}\\
				-\Sigma_{12}/\surd{2} & \Sigma_{12} & 1
				\epm \Bigg\} \\
				& \hfill  = - 2\Phi_3 \Bigg\{-\Delta_1,0,0; \bpm
				1 & 1/\surd{2} & -\Sigma_{12}/\surd{2}\\
				1/\surd{2} & 1 & -\Sigma_{12}\\
				-\Sigma_{12}/\surd{2} & -\Sigma_{12} & 1
				\epm\Bigg\}.
			\end{split}
		\end{equation}

		Since
		\begin{equation}
			\begin{split}\nonumber
				& -2\Phi_3 \left\{-\Delta_1,0,0; \bpm
				1 & 1/\surd{2} & -\Sigma_{12}/\surd{2}\\
				1/\surd{2} & 1 & -\Sigma_{12}\\
				-\Sigma_{12}/\surd{2} & -\Sigma_{12} & 1
				\epm \right\}\\
				&  \quad = -2 \Phi_2 (-\Delta_1,0; 1/\surd{2} ) +2 \Phi_3 \left\{ -\Delta_1,0,0; \bpm
				1 & 1/\surd{2} & \Sigma_{12}/\surd{2}\\
				1/\surd{2} & 1 & \Sigma_{12}\\
				\Sigma_{12}/\surd{2} & \Sigma_{12} & 1
				\epm \right\},
			\end{split}
		\end{equation}
		
		we finally obtain
		\begin{equation}
			\begin{split}\nonumber
				\tau_{12}  & =  -2 \Phi_2 (-\Delta_1,0; 1/\surd{2} ) +4 \Phi_3 \left\{ -\Delta_1,0,0; \bpm
				1 & 1/\surd{2} & \Sigma_{12}/\surd{2}\\
				1/\surd{2} & 1 & \Sigma_{12}\\
				\Sigma_{12}/\surd{2} & \Sigma_{12} & 1
				\epm
				\right\}.
			\end{split}
		\end{equation}
	\end{proof}
	
	\begin{proof}[Proof of Theorem~4]
		Without loss of generality, we set $j=1$ and $k=2$.
		By the definition,
		\[
		\tau_{12}  = \E(\widehat{\tau}_{12})
		= \E \Big[ \dfrac{2}{n(n-1)} \sum_{1\le i<i'\le n} \sign \left\{(X_{i1}-X'_{i1}) (X_{i2}-X'_{i2}) \right\} \Big].
		\]
		Plugging \eqref{eq:lem1-truncsign} into the previous display and rearranging yields
		\begin{equation}
			\begin{split}\nonumber
				\tau_{12}  & = \E \left\{ 2 I\left(X_1=0, X_2=0\right) - 2I\left(X_1=0\right)I\left( X'_2=0\right) \right\}\\
				& \quad - \E \left\{ 2 I\left(X_1=0, X_2>0, X'_2>0 \right) \sign(X_2-X'_2) \right\}\\
				& \quad - \E \left\{ 2 I\left(X_2=0, X_1>0, X'_1>0 \right) \sign(X_1-X'_1) \right\}\\
				& \quad + \E \left\{ I\left(X_1>0, X'_1>0, X_2>0, X'_2>0 \right) \sign(X_1-X'_1) \sign(X_2-X'_2) \right\}.
			\end{split}
		\end{equation}
		Using the definition of variable $X$ in terms of variable $Z$, the above display can be rewritten as
		\begin{equation}
			\begin{split}\nonumber
				\tau_{12}  & = \E \left\{ 2 I\left(Z_1<\Delta_1, Z_2<\Delta_2 \right) - 2I\left( Z_1<\Delta_1 \right)I\left( Z'_2<\Delta_2 \right) \right\}\\
				& \quad - \E \left\{ 2 I\left( Z_1<\Delta_1, Z_2>\Delta_2, Z'_2>\Delta_2 \right) \sign(Z_2-Z'_2) \right\}\\
				& \quad - \E \left\{ 2 I\left(Z_2 <\Delta_2, Z_1>\Delta_1, Z'_1>\Delta_1 \right) \sign(Z_1-Z'_1) \right\}\\
				& \quad + \E \left\{ I\left(Z_1>\Delta_1, Z'_1>\Delta_1, Z_2 >\Delta_2, Z'_2>\Delta_2 \right) \sign(Z_1-Z'_1) \sign(Z_2-Z'_2) \right\}.
			\end{split}
		\end{equation}
		Using $\sign\left(x\right) = I(x>0) - I(x<0)$ and \eqref{lem2:trick1}, it holds that
		\begin{equation}
			\begin{split}\nonumber
				\tau_{12}  & = 2\Phi_2(\Delta_1,\Delta_2;\Sigma_{12}) - 2 \Phi(\Delta_1)\Phi(\Delta_2) \\
				& \quad - 2\E \left\{ I\left( Z_1<\Delta_1, Z'_2>\Delta_2, Z_2-Z'_2>0 \right) \right\} + 2 \E \left\{ I\left( Z_1<\Delta_1, Z_2>\Delta_2, Z_2-Z'_2<0 \right) \right\}\\
				& \quad - 2\E \left\{ I\left( Z_2<\Delta_2, Z'_1>\Delta_1, Z_1-Z'_1>0 \right) \right\} + 2 \E \left\{ I\left( Z_2<\Delta_2, Z_1>\Delta_1, Z_1-Z'_1<0 \right) \right\}\\
				& \quad + \E \left\{ 2 I\left(Z_2 >\Delta_2, Z_1>\Delta_1, Z_1-Z'_1<0, Z_2-Z'_2 <0 \right) \right.\\
				& \quad \quad - 2  I\left. \left(Z'_1>\Delta_1, Z_2 >\Delta_2, Z'_1-Z_1<0, Z_2-Z'_2 <0 \right) \right\}.
			\end{split}
		\end{equation}
		
		Using the definition of the normal cumulative distribution function, $\tau_{12}$ can be re-written as
		\begin{equation}\label{eq:lem3-start}
			\begin{split}
				\tau_{12}  = & ~ 2\Phi_2(\Delta_1,\Delta_2;\Sigma_{12}) - 2 \Phi(\Delta_1)\Phi(\Delta_2) \\
				& - 2\Phi_3 \left\{ \Delta_1, -\Delta_2, 0;  \bpm
				1 & 0 & -\Sigma_{12}/\surd{2}\\
				0 & 1 & -1/\surd{2}\\
				-\Sigma_{12}/\surd{2} & -1/\surd{2} & 1
				\epm\right\}\\
				& + 2\Phi_3 \left\{ \Delta_1, -\Delta_2, 0;  \bpm
				1 & -\Sigma_{12} & \Sigma_{12}/\surd{2}\\
				-\Sigma_{12} & 1 & -1/\surd{2}\\
				\Sigma_{12}/\surd{2} & -1/\surd{2} & 1
				\epm \right\}\\
				& - 2\Phi_3 \left\{  -\Delta_1, \Delta_2, 0;  \bpm
				1 & 0 & -1/\surd{2}\\
				0 & 1 & -\Sigma_{12}/\surd{2}\\
				-1/\surd{2} & -\Sigma_{12}/\surd{2} & 1
				\epm\right\}\\
				& + 2\Phi_3 \left\{  -\Delta_1, \Delta_2, 0;  \bpm
				1 & -\Sigma_{12} & -1/\surd{2}\\
				-\Sigma_{12} & 1 & \Sigma_{12}/\surd{2}\\
				-1/\surd{2} & \Sigma_{12}/\surd{2} & 1
				\epm\right\}\\
				& +2 \Phi_4 \left\{  -\Delta_1, -\Delta_2, 0,0; \bpm
				1 & \Sigma_{12} & -1/\surd{2} & -\Sigma_{12}/\surd{2}\\
				\Sigma_{12} & 1 & -\Sigma_{12}/\surd{2} & -1/\surd{2}\\
				-1/\surd{2} & -\Sigma_{12}/\surd{2} & 1 & \Sigma_{12}\\
				-\Sigma_{12}/\surd{2} & -1/\surd{2} & \Sigma_{12} & 1
				\epm \right\}\\
				& -2 \Phi_4 \left\{ -\Delta_1, -\Delta_2, 0,0;  \bpm
				1 & 0 & -1/\surd{2} & \Sigma_{12}/\surd{2}\\
				0 & 1 & \Sigma_{12}/\surd{2} & -1/\surd{2}\\
				-1/\surd{2} & \Sigma_{12}/\surd{2} & 1 & -\Sigma_{12}\\
				\Sigma_{12}/\surd{2} & -1/\surd{2} & -\Sigma_{12} & 1
				\epm \right\}.
			\end{split}
		\end{equation}
		
		Let the last six terms be denoted as $T_1$ to $T_6$ respectively. Then,
		\begin{equation}
			\begin{split}\nonumber
				\tau_{12} & = 2\Phi_2(\Delta_1,\Delta_2;\Sigma_{12}) - 2 \Phi(\Delta_1)\Phi(\Delta_2)
				+T_1 + T_2 + T_3 + T_4 + T_5 + T_6.
			\end{split}
		\end{equation}
		Next we consider each term separately, and apply the same technique as in \eqref{eq:lem1-prop} for normal cdfs $\Phi_3(\cdot, \cdot, \cdot; r)$ and $\Phi_4(\cdot, \cdot, \cdot, \cdot; r)$.
		
		\begin{equation}
			\begin{split}\label{eq:lem3-t1t6}
				T_3 +T_4  = &  2\Phi_3 \left\{-\Delta_1, -\Delta_2, 0;  \bpm
				1 & 0 & -1/\surd{2}\\
				0 & 1 & \Sigma_{12}/\surd{2}\\
				-1/\surd{2} & \Sigma_{12}/\surd{2} & 1
				\epm\right\}\\
				& - 2\Phi_3 \left\{-\Delta_1, -\Delta_2, 0;  \bpm
				1 & \Sigma_{12} & -1/\surd{2}\\
				\Sigma_{12} & 1 & -\Sigma_{12}/\surd{2}\\
				-1/\surd{2} & -\Sigma_{12}/\surd{2} & 1
				\epm \right\}\\
				= &  T'_3 + T'_4
			\end{split}
		\end{equation}
		
		\begin{equation}
			\begin{split}\nonumber
				T'_4 + T_5= & -2\Phi_4 \left\{-\Delta_1, -\Delta_2, 0,0; \bpm
				1 & \Sigma_{12} & -1/\surd{2} & \Sigma_{12}/\surd{2}\\
				\Sigma_{12} & 1 & -\Sigma_{12}/\surd{2} & 1/\surd{2}\\
				-1/\surd{2} & -\Sigma_{12}/\surd{2} & 1 & -\Sigma_{12}\\
				\Sigma_{12}/\surd{2} & 1/\surd{2} & -\Sigma_{12} & 1
				\epm \right\};\\
				T'_3 + T_6= & 2 \Phi_4 \left\{-\Delta_1, -\Delta_2, 0,0;  \bpm
				1 & 0 & -1/\surd{2} & -\Sigma_{12}/\surd{2}\\
				0 & 1 & \Sigma_{12}/\surd{2} & 1/\surd{2}\\
				-1/\surd{2} & \Sigma_{12}/\surd{2} & 1 & \Sigma_{12}\\
				-\Sigma_{12}/\surd{2} & 1/\surd{2} & \Sigma_{12} & 1
				\epm \right\}.
			\end{split}
		\end{equation}
		
		\begin{equation}
			\begin{split}\nonumber
				T_1 +T_2  = &  2\Phi_3 \left\{-\Delta_1, -\Delta_2, 0;  \bpm
				1 & 0 & \Sigma_{12}/\surd{2}\\
				0 & 1 & -1/\surd{2}\\
				\Sigma_{12}/\surd{2} & -1/\surd{2} & 1
				\epm\right\}\\
				& - 2\Phi_3 \left\{-\Delta_1, -\Delta_2, 0;  \bpm
				1 & \Sigma_{12} & -\Sigma_{12}/\surd{2}\\
				\Sigma_{12} & 1 & -1/\surd{2}\\
				-\Sigma_{12}/\surd{2} & -1/\surd{2} & 1
				\epm\right\}.
			\end{split}
		\end{equation}
		
		Applying the \eqref{eq:lem1-prop}-type result again yields
		\begin{equation}
			\begin{split}\nonumber
				T_1 +T_2  & = 2 \Phi_2 (-\Delta_1, -\Delta_2; 0) - 2 \Phi_3 \left\{ -\Delta_1, -\Delta_2, 0; \bpm
				1 & 0 & -\Sigma_{12}/\surd{2}\\
				0 & 1 & -1/\surd{2}\\
				-\Sigma_{12}/\surd{2} & -1/\surd{2} & 1
				\epm \right\}\\
				& \quad -2 \Phi_2 (-\Delta_1, -\Delta_2; \Sigma_{12}) + 2 \Phi_3 \left\{ -\Delta_1, -\Delta_2, 0; \bpm
				1 & \Sigma_{12} & \Sigma_{12}/\surd{2}\\
				\Sigma_{12} & 1 & 1/\surd{2}\\
				\Sigma_{12}/\surd{2} & 1/\surd{2} & 1
				\epm \right\}\\
				& = 2 \Phi_2 (-\Delta_1, -\Delta_2; 0) - 2 \Phi_2 (-\Delta_1, -\Delta_2; \Sigma_{12}) + T'_1 + T'_2
			\end{split}
		\end{equation}
		
		The first terms in the previous display are cancelled out with the first two terms in \eqref{eq:lem3-t1t6} because
		\begin{equation}
			\begin{split}\nonumber
				2 \Phi_2 (-\Delta_1, -\Delta_2; 0) & = 2 - 2 \Phi(\Delta_1) - 2\Phi(\Delta_2) + 2 \Phi(\Delta_1)\Phi(\Delta_2)\\
				- 2 \Phi_2 (-\Delta_1, -\Delta_2; \Sigma_{12}) & =  2 \Phi(\Delta_1) + 2\Phi(\Delta_2) -2 - 2\Phi_2(\Delta_1,\Delta_2;\Sigma_{12}).
			\end{split}
		\end{equation}
		Furthermore, $T'_1$ and $T'_3+T_6$, $T'_2$ and $T'_4+T_5$ can be combined into one term, respectively
		\begin{equation}
			\begin{split}\nonumber
				T'_1 + (T'_3+T_6) & = -2 \Phi_4 \left\{ -\Delta_1, -\Delta_2, 0,0;\bpm
				1 & 0 & 1/\surd{2} & -\Sigma_{12}/\surd{2}\\
				0& 1 & -\Sigma_{12}/\surd{2}& 1/\surd{2}\\
				1/\surd{2}& -\Sigma_{12}/\surd{2} & 1& -\Sigma_{12}\\
				-\Sigma_{12}/\surd{2}& 1/\surd{2}& -\Sigma_{12} & 1
				\epm \right\},\\
				T'_2 + (T'_4+T_5) & = 2 \Phi_4 \left\{ -\Delta_1, -\Delta_2, 0,0; \bpm
				1 & \Sigma_{12} & 1/\surd{2} & \Sigma_{12}/\surd{2}\\
				\Sigma_{12} & 1 & \Sigma_{12}/\surd{2}& 1/\surd{2}\\
				1/\surd{2}& \Sigma_{12}/\surd{2} & 1& \Sigma_{12}\\ \Sigma_{12}/\surd{2}& 1/\surd{2}& \Sigma_{12} & 1
				\epm \right\}.
			\end{split}
		\end{equation}
		
		This concludes the proof of Theorem ~4.
		
	\end{proof}

	\begin{proof}[Proof of Theorem~5]
		Let $\Sigma_{jk}=r$. We consider separately each of the three cases.
		
		1) For the truncated/binary case, the bridge function (Theorem~2) has the form
		\[
		F_{\rm TB}(r;\Delta_j, \Delta_k) =
		2 \{1-\Phi(\Delta_j)\}\Phi(\Delta_k)
		-2 \Phi_3\left\{-\Delta_j, \Delta_k, 0; \Sigma_{3a}(r) \right\}
		-2 \Phi_3\left\{-\Delta_j, \Delta_k, 0; \Sigma_{3b}(r) \right\},
		\]
		where
		\[
		\Sigma_{3a}(r)=\bpm
		1 & -r & 1/\surd{2} \\
		-r & 1 & -r/\surd{2}\\
		1/\surd{2} & -r/\surd{2} & 1
		\epm, \quad
		\Sigma_{3b}(r)=\bpm
		1 & 0 & -1/\surd{2} \\
		0 & 1 & -r/\surd{2}\\
		-1/\surd{2} & -r/\surd{2} & 1
		\epm.
		\]
		From Lemma~\ref{l:partial} in Section~\ref{app:lemmas},
		\begin{align*}
			\dfrac{\partial \Phi_3\left(-\Delta_j, \Delta_k, 0; \Sigma_{3a}(r) \right)}{\partial r} &= \sum_{i=1}^{2}\sum_{i'=i+1}^3h_{ii'}(r)\dfrac{\partial \rho_{ii'} (r)}{\partial r} \\
			&= h_{12}(r)(-1) +h_{13}(r)(0)+ h_{23}(r)(-1/\surd{2}) <0,
		\end{align*}
		and similarly
		\begin{align*}
			\dfrac{\partial \Phi_3\left(-\Delta_j, \Delta_k, 0; \Sigma_{3b}(r) \right)}{\partial r}  = h_{23}(r)(-1/\surd{2})<0.
		\end{align*}
		Therefore,
		$$
		\dfrac{\partial F_{\rm TB}(r;\Delta_j, \Delta_k)}{\partial r} = -2 \dfrac{\partial \Phi_3\left(-\Delta_j, \Delta_k, 0; \Sigma_{3a}(r) \right)}{\partial r} - 2\dfrac{\partial \Phi_3\left(-\Delta_j, \Delta_k, 0; \Sigma_{3b}(r) \right)}{\partial r} >0.
		$$
		It follows that $F_{\rm TB}(r;\Delta_j, \Delta_k)$ is increasing in $r$.

		2) For the truncated/continuous case, the bridge function (Theorem~3) has the form
		\[
		F_{\rm TC}(r;\Delta_j) = -2 \Phi_2 (-\Delta_j,0; 1/\surd{2} ) +4\Phi_3 \left(-\Delta_j,0,0; \Sigma_3(r)\right),
		\]
		with
		\[
		\Sigma_3(r) = \bpm
		1 & 1/\surd{2} & r/\surd{2}\\
		1/\surd{2} & 1 & r\\
		r/\surd{2} & r & 1
		\epm.
		\]
		Using Lemma~\ref{l:partial} in Section~\ref{app:lemmas},
		\begin{align*}
			\dfrac{\partial \Phi_3 \left(-\Delta_j,0,0; \Sigma_3(r)\right)}{\partial r} = \sum_{i=1}^{2}\sum_{i'=i+1}^3h_{ii'}(r)\dfrac{\partial \rho_{ii'} (r)}{\partial r} = h_{12}(r)(0) +h_{13}(r)(1/\surd{2})+ h_{23}(r)(1) >0.
		\end{align*}
		Thus,
		$$
		\dfrac{\partial F_{\rm TC}(r;\Delta_j)}{\partial r} = 4 \dfrac{\partial \Phi_3\left(-\Delta_j,0,0; \Sigma_3(r)\right)}{\partial r}  >0,
		$$
		which implies that $F_{\rm TC}(r;\Delta_j)$ is increasing in $r$.
		
		3) For the truncated/truncated case, the bridge function (Theorem~4) has the form
		\[
		F_{\rm TT}(r;\Delta_j, \Delta_k)  = ~-2 \Phi_4 (-\Delta_j, -\Delta_k, 0,0; \Sigma_{4a}(r)) + 2 \Phi_4 (-\Delta_j, -\Delta_k, 0,0; \Sigma_{4b}(r)),
		\]
		\[
		\Sigma_{4a}(r) = \bpm
		1 & 0 & 1/\surd{2} & -r/\surd{2}\\
		0& 1 & -r/\surd{2}& 1/\surd{2}\\
		1/\surd{2}& -r/\surd{2} & 1& -r\\
		-r/\surd{2}& 1/\surd{2}& -r & 1
		\epm
		, \quad
		\Sigma_{4b}(r) = \bpm
		1 & r & 1/\surd{2} & r/\surd{2}\\
		r & 1 & r/\surd{2}& 1/\surd{2}\\
		1/\surd{2}& r/\surd{2} & 1& r\\ r/\surd{2}& 1/\surd{2}& r & 1
		\epm.
		\]
		From Lemma~\ref{l:partial} in Section~\ref{app:lemmas},
		\begin{align*}
			\dfrac{\partial \Phi_4 (-\Delta_j, -\Delta_k, 0,0; \Sigma_{4a}(r)) }{\partial r} &= \sum_{i=1}^{3}\sum_{i'=i+1}^4 h_{ii'}(r)\dfrac{\partial \rho_{ii'} (r)}{\partial r}\\
			& = h_{14}(r)(-1/\surd{2}) +h_{23}(r)(-1/\surd{2})+ h_{34}(r)(-1) <0,
		\end{align*}
		and similarly
		\begin{align*}
			\dfrac{\partial \Phi_4 (-\Delta_j, -\Delta_k, 0,0; \Sigma_{4b}(r)) }{\partial r}  = h_{12}(r)(1) + h_{14}(r)(1/\surd{2}) +h_{23}(r)(1/\surd{2})+ h_{34}(r)(1) >0.
		\end{align*}
		Therefore,
		$$
		\dfrac{\partial F_{\rm TT}(r;\Delta_j, \Delta_k)}{\partial r} = -2 \dfrac{\partial  \Phi_4 (-\Delta_j, -\Delta_k, 0,0; \Sigma_{4a}(r)) }{\partial r} + 2\dfrac{\partial \Phi_4 (-\Delta_j, -\Delta_k, 0,0; \Sigma_{4b}(r)) }{\partial r} >0.
		$$
		
		Thus, all the bridge functions are strictly increasing with $r$.
	\end{proof}

	\begin{proof}[Proof of Theorem~6]
		It is sufficient to show that $F^{-1}(\cdot)$ has bounded first derivative, $\left|\partial F^{-1}(\tau)/\partial \tau\right|\leq L.$ Since the derivative of the bridge function is strictly positive, it is equivalent to show $\partial F(r)/\partial r \geq L^{-1}$.
		We consider separately each of the three cases using the form of the derivative from Theorem~5, and set $L^{-1} = \min(L_1^{-1}, L_2^{-1}, L_3^{-1})$ from below.
		
		1) For the truncated/binary case, we need to prove that there exists constant $L_1>0$ such that
		\begin{align*}
			\dfrac{\partial F_{\rm TB}(r)}{\partial r} = 2h_{12a}(r) + \surd{2}h_{23a}(r)+\surd{2}h_{23b}(r)  \geq L_1^{-1}.
		\end{align*}
		Since $h_{12a}(r)$, $h_{23a}(r)$ and $h_{23b}(r)$ are all strictly positive, it is sufficient to show $h_{23b}(r)\geq L_1^{-1}$. 
		
		Using the Assumption~2, $\Delta_j \leq M $, therefore 
		\begin{align*}
			h_{23b}(r) &=\int_{-\infty}^{-\Delta_j}\phi_3\{x, \Delta_k, 0; \Sigma_{3b}(r)\}dx \geq \int_{-\infty}^{-M}\phi_3\{x, \Delta_k, 0; \Sigma_{3b}(r)\}dx.
		\end{align*}
		Consider $\phi_3\{x, \Delta_k, 0; \Sigma_{3b}(r)\} = f(X_1 = x, X_2 = \Delta_k|X_3 = 0)\phi(0)$, where
		$$
		f(X_1 = x, X_2 = \Delta_k|X_3 = 0) = \frac1{2\pi|V|^{1/2}}\exp\Big\{-\dfrac{1}{2} (x, \Delta_k) V^{-1} \bpm x\\ \Delta_k \epm \Big\}
		$$
		with
		$$
		V = \cov\{X_1, X_2 |X_3=0; \Sigma_{3b}(r)\}=\bpm
		1 & 0\\
		0 & 1
		\epm - \bpm -1/\surd{2}\\ -r/\surd{2}\epm(-1/\surd{2}\ -r/\surd{2}) = \dfrac{1}{2}\bpm
		1 & -r\\
		-r & 2-r^2
		\epm.
		$$
		Since $|V|=(1-r^2)/2 \leq 1/2$, and
		$$
		V^{-1} = \dfrac{1}{1-r^2}\bpm
		2-r^2 & r\\
		r & 1
		\epm,
		$$
		from above displays
		\begin{align*}
			f(X_1 = x, X_2=\Delta_k|X_3 = 0) &\geq \dfrac1{\surd{2}\pi}\exp\left\{-\frac12\left(\dfrac{x^2 (2-r^2)}{(1-r^2)}+ \dfrac{\Delta_k^2}{(1-r^2)}+\dfrac{2r\Delta_k x}{1-r^2}\right)\right\}.
		\end{align*}
		By Assumption~1,  $1-r^2 \ge 2\delta - \delta^2 > 0$, hence
		\begin{align*}
			f(X_1 = x, X_2=\Delta_k|X_3 = 0) &\geq \dfrac1{\surd{2}\pi}\exp\left\{-\frac12\left(\dfrac{2x^2}{2\delta - \delta^2}+ \dfrac{\Delta_k^2}{2\delta-\delta^2}+\dfrac{2r\Delta_k x}{1-r^2}\right)\right\}.
		\end{align*}
		By Assumption~2, $|\Delta_k|\leq M$, hence
		\begin{align*}
			f(X_1 = x, X_2=\Delta_k|X_3 = 0) &\geq \dfrac1{\surd{2}\pi}\exp\left\{-\frac12\left(\dfrac{2x^2}{2\delta - \delta^2}+ \dfrac{M^2}{2\delta-\delta^2}+\dfrac{2r\Delta_k x}{1-r^2}\right)\right\}\\
			&\geq \dfrac1{\surd{2}\pi}\exp\left\{-\frac12\left(\dfrac{2x^2}{2\delta - \delta^2}+ \dfrac{M^2}{2\delta-\delta^2}+2(\delta-1)Mx\right)\right\},
		\end{align*}
		where the last inequality follows since $r\Delta_k x/(1-r^2)\leq (\delta-1)Mx/(1-r^2)\leq (\delta-1)Mx$ by Assumptions~1--2 with $x<0$. Combining the above displays
		\begin{align*}
			h_{23b}(r) &\geq \int_{-\infty}^{-M}\dfrac1{\surd{2}\pi}\exp\left\{-\frac12\left(\dfrac{2x^2}{2\delta - \delta^2}+ \dfrac{M^2}{2\delta-\delta^2}+2(\delta-1)Mx\right)\right\}\phi(0)dx = L_{1}^{-1},
		\end{align*}
		where $L_{1}$ is independent of $r$, $\Delta_j$ and $\Delta_k$.
		
		2) For the truncated/continuous case, we need to prove that there exists $L_2>0$ such that
		$$
		\frac{\partial F_{\rm TC}(r)}{\partial r} = 2\surd{2}h_{13}(r) + 4h_{23}(r) \geq L_2^{-1}.
		$$
		Since both $h_{13}(r)$ and $h_{23}(r)$ are strictly positive, it is sufficient to provide lower bound on $h_{23}(r)$. By Assumption~2, $\Delta_j \leq M$, hence
		\begin{align*}
			h_{23}(r) &= \int_{-\infty}^{-\Delta_j}\phi_3\{x,0,0;\Sigma_3(r)\}dx \geq \int_{-\infty}^{-M}\phi_3\{x,0,0;\Sigma_3(r)\}dx\\
			& = \int_{-\infty}^{-M}\phi_2\{0,0; \Sigma_{-1,-1}(r)\}f\{X_1 = x|X_2=0;X_3=0; \Sigma_3(r)\}dx\\
			&= \phi_2\{0,0; \Sigma_{-1,-1}(r)\}\int_{-\infty}^{-M}f\{X_1 = x|X_2=0;X_3=0; \Sigma_3(r)\}dx,
		\end{align*}
		where $\Sigma_{-1,-1}(r)$ is the upper left 2 by 2 submatrix of $\Sigma_3(r)$.
		Since
		$$
		\phi_2\{0,0; \Sigma_{-1,-1}(r)\} = \frac1{2\pi|\Sigma_{-1,-1}(r)|^{1/2}}\exp(0)=\frac1{2\pi(1-r^2)^{1/2}}>\frac1{2\pi},
		$$
		and $X_1|X_2=0, X_3 = 0 \sim N(0, 1/2)$, that is $f\{X_1 = x|X_2=0;X_3=0; \Sigma_3(r)\}$ does not depend on $r$, it follows that there exists constant $L_2$ independent of $r$ and $\Delta_j$ such that
		$
		h_{23}(r) \geq L_2^{-1}.
		$
		
		3) For the truncated/truncated case, we need to prove that there exists constant $L_3>0$ such that
		\begin{equation}
			\begin{split}\nonumber
				\dfrac{\partial F_{\rm TT}(r)}{\partial r} & = \surd{2}h_{14a}(r) + \surd{2}h_{23a}(r) + 2h_{34a}(r)
				+  2h_{12b}(r) + \surd{2}h_{14b}(r) + \surd{2}h_{23b}(r) +  2h_{34b}(r)\\ & \geq L_3^{-1}.
			\end{split}
		\end{equation}
		It is sufficient to show $h_{34a}(r) \geq L_3^{-1}$.
		Since $\Delta_j \leq M$ and $\Delta_k \leq M$ by Assumption~2, 
		\begin{align*}
			h_{34a}(r) &= \int_{-\infty}^{-\Delta_j} \int_{-\infty}^{-\Delta_k} \phi_4\{x_1, x_2,0,0;\Sigma_{4a}(r)\}dx_2 dx_1 \\
			& \ge \int_{-\infty}^{-M} \int_{-\infty}^{-M}\phi_4\{x_1, x_2,0,0;\Sigma_{4a}(r)\}dx_2 dx_1.
		\end{align*}
		Consider conditional representation
		\begin{align*}
			\phi_4\{x_1, x_2,0,0;\Sigma_{4a}(r)\}& = f\left\{x_1, x_2 \left| X_3=0,X_4=0 \right.; V \right\} \phi_2(0, 0; -r),
		\end{align*}
		where
		\begin{align*}
			V = \bpm1& 0\\ 0&1\epm - \bpm1/\surd{2}&-r/\surd{2}\\-r/\surd{2}&1/\surd{2}\epm \frac1{1-r^2}\bpm1&r\\r&1\epm \bpm1/\surd{2}&-r/\surd{2}\\-r/\surd{2}&1/\surd{2}\epm=\frac12\bpm1 & r\\
			r& 1\epm.
		\end{align*}
		By Assumption~1, $1-r^2\ge 2\delta-\delta^2$ and $-r/(1-r^2)\leq (1-\delta)/(2\delta - \delta^2)$. Therefore, since $x_1x_2>0$ when $x_1\leq -M$, $x_2\leq -M$,
		\begin{align}\label{eq:tt_f2dim}
			f \left\{ x_1, x_2  | X_3=0, X_4=0 ; V\right\}  &= \dfrac{1}{2\pi \{(1-r^2)/4\}^{1/2}} \exp \left\{-\dfrac{1}{2} (x_1, x_2) \dfrac{2}{1-r^2} \bpm 1&-r\\-r&1 \epm \bpm x_1\\ x_2 \epm \right\} \nonumber\\
			& =\dfrac{1}{\pi (1-r^2)^{1/2}} \exp \left\{ -\Big(\dfrac{2x_1^2 + 2x_2^2}{1-r^2} - \dfrac{4rx_1 x_2}{1-r^2}\Big)\right\} \nonumber\\
			& \ge \dfrac{1}{\pi} \exp \left\{ -\Big(\dfrac{2x_1^2 + 2x_2^2}{2\delta - \delta^2} + 4\frac{1-\delta}{2\delta - \delta^2}x_1 x_2\Big)\right\}.
		\end{align}
		Since
		$
		\phi_2(0,0; -r) = \exp(0)/\{2\pi(1-r^2)^{1/2}\}>1/(2\pi),
		$
		and the bound in \eqref{eq:tt_f2dim} does not depend on $r$, there exists constant $L_3$ such that
		combining the above displays $
		h_{34a}(r) \geq L_3^{-1}.
		$
	\end{proof}
	
	\begin{proof}[Proof of Theorem~7]
		Corollary~1 in \cite{Fan:2016um} covers continuous and binary cases, therefore it remains to prove consistency for truncated-continuous, truncated-binary and truncated-truncated cases. For clarity, we separate the three cases into  Theorems~\ref{thm:consistencyTC}--\ref{thm:consistencyTT}. Combining these results together with the union bound leads to the desired rate.
	\end{proof}
	
	\begin{theorem}[Truncated-continuous case]\label{thm:consistencyTC} Let variable $j$ be truncated and variable $k$ be continuous. Under Assumptions~1 and~2, for any $t>0$ and for constants $L>0$ from Theorem~6, $L_{\Phi}>0$ from Lemma~\ref{lem:Fan} and $M' = \Phi(2M) - \Phi(M)$,
		\[
		\P\left(  |\widehat R_{jk} - \Sigma_{jk} | > t \right)  \le 2 \left[\exp{\left\{ -2n(M')^2\right\} } + \exp \left( {-\dfrac{nt^2}{2 L^2 }} \right) + \exp \left( -\dfrac{nt^2}{36L^2L_{\Phi}^2}\right) \right].
		\]
	\end{theorem}
	
	\begin{proof}
		Let the event $A_j = \{ |\widehat \Delta_j | \le 2M \}$. For any $t>0$, 
		\begin{equation}
			\begin{split}\nonumber
				\P \left\{ |\widehat R_{jk} - \Sigma_{jk} | >t \right\}=\P \left\{ |F^{-1}_{\text{TC}} (\widehat \tau_{jk} ; \widehat \Delta_j ) - \Sigma_{jk} | >t \right\} \le \P \left[ \left\{ |F^{-1}_{\text{TC}} (\widehat \tau_{jk} ; \widehat \Delta_j ) - \Sigma_{jk} | >t \right\} \cap A_j \right] + \P \left( A_j^c\right).
			\end{split}
		\end{equation}
		Consider the second term. Using Assumption~2, $\Phi(\Delta_j)\in [\Phi(-M), \Phi(M)]$, therefore
		\begin{equation}
			\begin{split}\label{eq:TC_complement}
				\P(A_j^c) & = \P \left( | \widehat \Delta_j| > 2M \right) = \P \left( \widehat \Delta_j > 2M \right) + \P \left( \widehat \Delta_j < -2M \right)\\
				& = \P \left\{ \Phi(\widehat \Delta_j) > \Phi(2M) \right\} + \P \left\{ \Phi(\widehat \Delta_j) < \Phi(-2M) \right\}\\
				&\leq\P \left\{ \Phi(\widehat \Delta_j)-\Phi(\Delta_j) > \Phi(2M)-\Phi(\Delta_j) \right\} + \P \left\{ \Phi(\widehat \Delta_j)-\Phi(\Delta_j) < \Phi(-2M) - \Phi(\Delta_j) \right\}\\
				& \le \P \left\{ \left| \Phi(\widehat \Delta_j) - \Phi( \Delta_j) \right|  > \Phi(2M) - \Phi(M) \right\}\\
				& =\P \left\{ \left| \dfrac{\sum_{i=1}^{n} I(X_{ij} = 0)}{n} - \Phi(\Delta_j) \right| > M' \right\} \text{ where } M' = \Phi(2M) - \Phi(M)\\
				& \le 2 \exp\left\{ -2n(M')^2  \right\}\text{ by Hoeffding's inequality}.
			\end{split}
		\end{equation}
		Consider the first term
		\begin{equation}
			\begin{split}\nonumber
				& \P \left[ \left\{ |F^{-1}_{\text{TC}} (\widehat \tau_{jk} ; \widehat \Delta_j ) - \Sigma_{jk} | >t \right\} \cap A_j \right]\\
				& = \P \left[ \left\{ |F^{-1}_{\text{TC}} (\widehat \tau_{jk} ; \widehat \Delta_j ) -  F^{-1}_{\text{TC}} (F_{\text{TC}} (\Sigma_{jk} ; \widehat \Delta_j ) ; \widehat \Delta_j )| >t \right\} \cap A_j \right]\\
				& \le \P \left[ \left\{ L|\widehat \tau_{jk} - F_{\text{TC}} (\Sigma_{jk} ; \widehat \Delta_j ) | >t \right\} \cap A_j \right] \quad \mbox{ by Theorem~ 6}\\
				& \le \P \left[ \left\{ |\widehat \tau_{jk} - F_{\text{TC}} (\Sigma_{jk} ; \Delta_j ) | > \dfrac{t}{2L} \right\} \cap A_j \right] + \P \left[ \left\{ | F_{\text{TC}} (\Sigma_{jk} ; \Delta_j ) - F_{\text{TC}} (\Sigma_{jk} ; \widehat \Delta_j ) | > \dfrac{t}{2L} \right\} \cap A_j \right]\\
				& \le \P \left[ \left\{ |\widehat \tau_{jk} - F_{\text{TC}} (\Sigma_{jk} ; \Delta_j ) | > \dfrac{t}{2L} \right\} \right] + \P \left[ \left\{ | F_{\text{TC}} (\Sigma_{jk} ; \Delta_j ) - F_{\text{TC}} (\Sigma_{jk} ; \widehat \Delta_j ) | > \dfrac{t}{2L} \right\} \cap A_j \right]\\
				& \equiv I_1 + I_2.
			\end{split}
		\end{equation}
		
		Consider $I_1$. Since $\widehat \tau_{jk}$ is a U-statistic with bounded kernel and $F_{\text{TC}} (\Sigma_{jk} ; \Delta_j )$ is the expected value of the kernel, by Hoeffding's inequality
		\begin{equation}\label{eq:TC_i1}
			I_1 \le 2 \exp \left( {-\dfrac{nt^2}{2 L^2 }} \right).
		\end{equation}
		
		Consider $I_2$. Using Lemma~\ref{lem:Fan} and Assumption~2, on the event $A_j = \{ |\widehat \Delta_j | \le 2M \}$
		\begin{equation*}
			\begin{split}
				| \widehat \Delta_j- \Delta_j| = \left| \Phi^{-1} \left( \dfrac{\sum_{i=1}^{n} I(X_{ij} = 0)}{n} \right) - \Phi^{-1} \left\{ \Phi(\Delta_j)\right\} \right| \le L_{\Phi} \left| \dfrac{\sum_{i=1}^{n} I(X_{ij} = 0)}{n} - \Phi(\Delta_j) \right|.
			\end{split}
		\end{equation*}
		Using the above display with Lemma~\ref{lem:BridgeTCdelta}, by Hoeffding's inequality
		\begin{equation}
			\begin{split}\label{eq:TC_i2}
				I_2 & =  \P \left[ \left\{ | F_{\text{TC}} (\Sigma_{jk} ; \Delta_j ) - F_{\text{TC}} (\Sigma_{jk} ; \widehat \Delta_j ) | > \dfrac{t}{2L} \right\} \cap A_j \right]\\
				& \le \P \left[ \left\{ \dfrac{6L_{\Phi}}{(2\pi)^{1/2}} \left| \dfrac{\sum_{i=1}^{n} I(X_{ij} = 0)}{n} - \Phi(\Delta_j) \right| > \dfrac{t}{2L} \right\} \cap A_j \right]\\
				& \le \P \left[ \left\{ \dfrac{6L_{\Phi}}{(2\pi)^{1/2}} \left| \dfrac{\sum_{i=1}^{n} I(X_{ij} = 0)}{n} - \Phi(\Delta_j) \right| > \dfrac{t}{2L} \right\} \right]\\
				& \le 2 \exp \left( -\dfrac{nt^2\pi}{36L^2L_{\Phi}^2}\right).
			\end{split}
		\end{equation}
		Combining \eqref{eq:TC_complement}, \eqref{eq:TC_i1} and \eqref{eq:TC_i2} yields the desired result.
	\end{proof}

	\begin{theorem}[Truncated-binary case]\label{thm:consistencyTB} Let variable $j$ be truncated and variable $k$ be binary. Under Assumptions~1 and~2, for any $t>0$ and for constants $L>0$ from Theorem~6, $L_{\Phi}>0$ from Lemma~\ref{lem:Fan} and $M' = \Phi(2M) - \Phi(M)$,
		\begin{equation}
			\begin{split}
				\P\left(  | \widehat R_{jk} - \Sigma_{jk} | >t \right)   \le 4 \exp\left\{-n(M')^2\right\} + 2 \exp \left( {-\dfrac{nt^2}{2 L^2 }} \right)  + 4 \exp \left( -\dfrac{nt^2 \pi}{36L^2L_{\Phi}^2}\right).
			\end{split}
		\end{equation}
	\end{theorem}

	\begin{proof}
		Let $A_j = \{ |\widehat \Delta_j | \le 2M \}$ and $A_k = \{ |\widehat \Delta_k | \le 2M \}$ with $A = A_j\cap A_k$. The proof follows the proof of Theorem~\ref{thm:consistencyTC} by replacing event $A_j$ with event $A$.
		Using~\eqref{eq:TC_complement}
		$$
		\P(A^c) = \P\{(A_j\cap A_k)^c\} = \P(A_j^c\cup A_k^c)\leq P(A_j^c) + P(A_k^c)\leq 4\exp\{-2n(M')^2\}.
		$$
		Using decomposition into $I_1$ and $I_2$ as in the proof of Theorem~\ref{thm:consistencyTC}, \eqref{eq:TC_i1} holds for $I_1$. To control $I_2$, from Lemma~\ref{lem:Fan} and Assumption~2, on the event $A$
		\begin{equation*}
			\begin{split}
				| \widehat \Delta_j- \Delta_j|  \le L_{\Phi} \left| \dfrac{\sum_{i=1}^{n} I(X_{ij} = 0)}{n} - \Phi(\Delta_j) \right|,
				| \widehat \Delta_k- \Delta_k|  \le L_{\Phi} \left| \dfrac{\sum_{i=1}^{n} I(X_{ik} = 0)}{n} - \Phi(\Delta_k) \right|.
			\end{split}
		\end{equation*}
		Combining the above display with Lemma~\ref{lem:BridgeTBdelta} gives
		\begin{equation}
			\begin{split}\label{eq:TB_i2}
				I_2 & =  \P \left[ \left\{ | F_{\text{TB}} (\Sigma_{jk} ; \Delta_j, \Delta_k ) - F_{\text{TB}} (\Sigma_{jk} ; \widehat \Delta_j, \widehat \Delta_k ) | > \dfrac{t}{2L} \right\} \cap A \right]\\
				& \le \P \left[ \left\{ \dfrac{4 L_{\Phi}}{(2\pi)^{1/2}} \left| \dfrac{\sum_{i=1}^{n} I(X_{ij} = 0)}{n} - \Phi(\Delta_j) \right| > \dfrac{t}{2L} \right\} \right] \\
				& \quad + \P \left[ \left\{ \dfrac{6 L_{\Phi}}{(2\pi)^{1/2}} \left| \dfrac{\sum_{i=1}^{n} I(X_{ik} = 0)}{n} - \Phi(\Delta_k) \right| > \dfrac{t}{2L} \right\} \right]\\
				& \le 4 \exp \left( -\dfrac{nt^2\pi}{36L^2L_{\Phi}^2}\right).
			\end{split}
		\end{equation}
		Combining bound on $\P(A^c)$ with \eqref{eq:TC_i1} and \eqref{eq:TB_i2} completes the proof.
	\end{proof}

	\begin{theorem}[Truncated-truncated case]\label{thm:consistencyTT} Let both variables $j$ and $k$ be truncated. Under Assumptions~1 and~2, for any $t>0$ and for constants $L>0$ from Theorem~6, $L_{\Phi}>0$ from Lemma~\ref{lem:Fan} and $M' = \Phi(2M) - \Phi(M)$,
		\[
		\P\left( | \widehat R_{jk} - \Sigma_{jk} | >t \right)  \le 4 \exp\{-n(M')^2\} + 2 \exp \left( {-\dfrac{nt^2}{2 L^2 }} \right) + 4\exp\left( -\dfrac{nt^2 \pi}{16L^2L_{\Phi}^2}\right).
		\]
	\end{theorem}
	
	\begin{proof}
		The proof follows the proof of Theorem~\ref{thm:consistencyTB} by invoking Lemma~\ref{lem:BridgeTTdelta} instead of Lemma~\ref{lem:BridgeTBdelta} to control term $I_2$ leading to 
		$$
		I_2  =  \P \left[ \left\{ | F_{\text{TT}} (\Sigma_{jk} ; \Delta_j, \Delta_k ) - F_{\text{TT}} (\Sigma_{jk} \widehat \Delta_j, \widehat \Delta_k ) | > \dfrac{t}{2L} \right\} \cap A \right]\le 4 \exp \left( -\dfrac{nt^2\pi}{16L^2L_{\Phi}^2}\right)
		$$
		under the conditions of the theorem, thus completing the proof.
	\end{proof}

	\begin{proof}[Proof of Proposition~1]
		This proof follows the proof of Proposition 2 in \cite{Witten:2011JRSSB}. Consider the Karush-Kuhn-Tucker (KKT) conditions for problem~(5):
		\begin{align*}
			\mbox{Lagrangian condition: }&-\widetilde R_{12}w_2 + \lambda_1 z + 2\mu \widetilde R_1 w_1 = 0;\\
			\mbox{Complementary slackness: }& \mu (w_1^{\top} \widetilde R_1 w_1 -1) = 0;\\
			\mbox{Primal/dual feasibility: }& \mu \ge 0, ~~~ w_1^{\top} \widetilde R_1 w_1 -1 \le 0;
		\end{align*}
		where $z$ is the subgradient of $\|w_1\|_1$, i.e. $z_j=\sign(w_{1j})$ if $w_{1j}\neq0$ and $z_j\in [-1,1]$ if $w_{1j}=0$.
		
		First, if $w_1=0$, then we must have $\mu=0$ and $-(\widetilde R_{12}w_2)_j + \lambda_1 z_j = 0$ should hold for all $j$. This is only possible when $z_j={(\widetilde R_{12}w_2)_j}/{\lambda_1} \in [-1,1]$ for all $j$, that is $\|\widetilde R_{12}w_2 \|_\infty \le \lambda_{1}$. Therefore, if $\|\widetilde R_{12}w_2 \|_\infty \le \lambda_{1}$, then $w_1=0$ solves~(5). For problem (6), since $-\widetilde R_{12}w_2 + \lambda_1 z + \widetilde R_1 w_1 = 0$ where $z$ is the subgradient vector of $\|w_1\|_1$, $w_1=0$ solves problem (6) and if $\widehat{w}_1=0$, then $w_1=0$.
		
		Second, if we suppose instead that $\| \widetilde R_{12}w_2 \|_\infty > \lambda_{1}$, then $w_1 \neq 0$ and $w_1^{\top} \widetilde R_1 w_1 =1$ should be the case, which now simplifies conditions to
		\[
		-\widetilde R_{12}w_2 + \lambda_1 z + 2\mu \widetilde R_1 w_1 = 0; \quad w_1^{\top} \widetilde R_1 w_1 = 1; \quad \mu > 0.
		\]
		
		If we let $\widetilde{w}_1=2\mu w_1$, then this is equivalent to solving problem (6) and then dividing the solution by $({\widetilde w}_1^{\top} \widetilde R_1 \widetilde{w}_1)^{1/2}$.
	\end{proof}

	\section{Supporting lemmas}\label{app:lemmas}
	
	\begin{lemma}\label{l:partial} For any constants $a_1, \dots, a_d$, let $\Phi_d(a_1, \ldots, a_d; \Sigma_d (r))$ be the cumulative distribution function of $d$-dimensional central normal distribution with covariance matrix
		$$
		\Sigma_d (r) = \bpm 1 & \rho_{12} (r) & \rho_{13} (r) & \cdots & \rho_{1d} (r)\\ \rho_{21} (r)  & 1 & \rho_{23} (r) & \cdots & \rho_{2d} (r)\\
		& & 1 & &  \\
		\vdots & & & \ddots & \vdots \\
		\rho_{d1} (r) & & \cdots & & 1
		\epm.
		$$
		Then there exist $h_{ij}(r)>0$ for all $r\in (-1,1)$ such that
		$$
		\dfrac{\partial \Phi_d(a_1, \ldots, a_d; \Sigma_d (r))}{\partial r} =  \sum_{i=1}^{d-1}\sum_{j=i+1}^d h_{ij}(r) \dfrac{\partial \rho_{ij} (r)}{\partial r}.
		$$
	\end{lemma}
	\begin{proof}[Proof of Lemma~\ref{l:partial}]
		Using the multivariate chain rule,
		\[
		\dfrac{\partial \Phi_d(a_1, \ldots, a_d; \Sigma_d (r))}{\partial r} = \sum_{i<j} \Big\{ \dfrac{\partial \Phi_d(a_1, \ldots, a_d; \Sigma_d (r))}{\partial \rho_{ij} (r)} \dfrac{\partial \rho_{ij} (r)}{\partial r} \Big\} := \sum_{i<j} h_{ij}(r)\dfrac{\partial \rho_{ij} (r)}{\partial r}.
		\]
		Without loss of generality, let $i=1$, $j=2$ and consider $h_{12}(r)$. By dominated convergence theorem, we can interchange integration and differentiation to get
		
		\begin{align*}
			\dfrac{\partial \Phi_d(a_1, \ldots, a_d; \Sigma_d (r))}{\partial \rho_{12} (r)} & = \int_{-\infty}^{a_1} \cdots \int_{-\infty}^{a_d}  \dfrac{\partial \phi_d(x_1, x_2, x_3 \ldots, x_d; \Sigma_d (r))}{\partial \rho_{12}(r)} dx_d \cdots dx_3 dx_2 dx_1\\
			& = \int_{-\infty}^{a_3} \cdots \int_{-\infty}^{a_d} \int_{-\infty}^{a_1}\int_{-\infty}^{a_2} \dfrac{\partial^2 \phi_d(x_1, x_2, x_3 \ldots, x_d; \Sigma_d (r))}{\partial x_1 \partial x_2}dx_2 dx_1 dx_d \cdots dx_3 \\
			& = \int_{-\infty}^{a_3} \cdots \int_{-\infty}^{a_d}  \phi_d(a_1, a_2, x_3 \ldots, x_d; \Sigma_d (r)) dx_d \cdots dx_3,
		\end{align*}
		where in the 2nd equality we used $\partial\phi_d / \partial \rho_{ij} = \partial^2\phi_d /(\partial x_i\partial x_j)$ \citep{Plackett:1954}.
		Since $\phi$ is multivariate density function, $h_{12}(r)=\partial \Phi_d(a_1, \ldots, a_d; \Sigma_d (r))/\partial \rho_{12} (r)$ is positive. The proof for other $i$, $j$ is analogous.
	\end{proof}
	
	
	\begin{lemma}[Lemma A.2 in \citet{Fan:2016um}]\label{lem:Fan} For any $y_1, y_2\in [\Phi(-2M), \Phi(2M)]$, there exists a Lipschitz constant $L_{\Phi}>0$ such that
		$$
		|\Phi^{-1}(y_1) - \Phi^{-1}(y_2)|\leq L_{\Phi}|y_1 - y_2|.
		$$
	\end{lemma}
	
	\begin{lemma}\label{lem:BridgeTCdelta} Let $F_{\text{TC}}(\Sigma_{jk}; \Delta_j)$ be the bridge function from Theorem~3 for the truncated/continuous case, and let $\widehat \Delta_j = \Phi^{-1} \left(n_{\text{zero}}/{n}\right)$. Then
		$$
		| F_{\text{TC}}(\Sigma_{jk}; \Delta_j) - F_{\text{TC}}(\Sigma_{jk};  \widehat \Delta_j)| \le \dfrac{6}{\surd{(2\pi)}} | \widehat \Delta_j - \Delta_j |.
		$$
	\end{lemma}
	\begin{proof}[Proof of Lemma~\ref{lem:BridgeTCdelta}] Using the bridge function formula from Theorem~3 leads to
		\begin{equation}
			\begin{split}\nonumber
				| & F_{\text{TC}}(\Sigma_{jk}; \Delta_j) - F_{\text{TC}}(\Sigma_{jk};  \widehat \Delta_j)|\\
				& = | -2\Phi_2(-\Delta_j, 0; 1/\surd{2}) + 4\Phi_3(-\Delta_j, 0, 0; \Sigma_3(r))
				+2\Phi_2(-\widehat \Delta_j, 0; 1/\surd{2}) - 4\Phi_3(-\widehat \Delta_j, 0, 0; \Sigma_3(r))| \\
				& \le 2| \Phi_2(-\Delta_j, 0; 1/\surd{2}) - \Phi_2(-\widehat \Delta_j, 0; 1/\surd{2})| + 4 | \Phi_3(-\Delta_j, 0, 0; \Sigma_3(r)) - \Phi_3(-\widehat \Delta_j, 0, 0; \Sigma_3(r)) |.
			\end{split}
		\end{equation}
		By the mean value theorem
		\begin{equation}\label{eq:lem2-meanvalue}
			| F_{\text{TC}}(\Sigma_{jk}; \Delta_j) - F_{\text{TC}}(\Sigma_{jk};  \widehat \Delta_j)|
			\le 2 \Phi_{21} (\xi_1) | \Delta_j-\widehat \Delta_j| + 4 \Phi_{31} (\xi_2) |  \Delta_j-\widehat \Delta_j|,
		\end{equation}
		where $\xi_1$ and $\xi_2$ are the intermediate values and $\Phi_{21} (x, y; r) = \partial \Phi_2(x, y; r)/\partial x$, $\Phi_{31} (x, y, z; \Sigma_3(r)) = \partial \Phi_3(x, y, z; \Sigma_3(r))/\partial x$.
		Since a bivariate random variable $(X_1, X_2)$ with distribution $\Phi_2(\cdot, \cdot; r)$ has a conditional distribution $X_2|X_1=x_1 \sim \N(rx_1, 1-r^2)$, 
		\begin{align}
			\Phi_{21} (x, y; r) & = \dfrac{\partial \Phi_2(x, y; r)}{\partial x}
			= \dfrac{\partial}{\partial x} \int_{-\infty}^{x} \int_{-\infty}^{y} \phi_2 (x_1, x_2; r) dx_2 dx_1 \nonumber\\
			& = \dfrac{\partial}{\partial x} \int_{-\infty}^{x} \int_{-\infty}^{y} \phi(x_2|x_1) \phi(x_1) dx_2 dx_1
			= \dfrac{\partial}{\partial x} \int_{-\infty}^{x} \Phi \left( \dfrac{y-rx_1}{(1-r^2)^{1/2}}\right) \phi(x_1) dx_1 \nonumber\\
			& = \Phi \left(\dfrac{y-rx}{(1-r^2)^{1/2}}\right) \phi(x) \le (2\pi)^{-1/2}. \label{eq:2dim-upperbound}
		\end{align}
		Let the density function and the distribution function of $X_2, X_3 | X_1 = x_1$ be $\phi_2 (x_2, x_3 | x_1)$ and $\Phi_2 (x_2, x_3 | x_1)$, respectively. Then
		\begin{equation}
			\begin{split}\label{eq:3dim-upperbound}
				\Phi_{31} (x, y, z; \Sigma_3(r)) & = \dfrac{\partial \Phi_3(x, y, z; \Sigma_3(r))}{\partial x}\\
				& = \dfrac{\partial}{\partial x} \int_{-\infty}^{x}\Phi_2(y, z|x_1)\phi(x_1) dx_1
				= \Phi_2(y, z|x)\phi(x)
				\le (2\pi)^{-1/2}.
			\end{split}
		\end{equation}
		Plugging \eqref{eq:2dim-upperbound} and \eqref{eq:3dim-upperbound} into \eqref{eq:lem2-meanvalue} completes the proof.
		
	\end{proof}

	\begin{lemma}\label{lem:BridgeTBdelta}Let $F_{\text{TB}}(\Sigma_{jk}; \Delta_j, \Delta_k)$ be the bridge function from Theorem~2 for the truncated/binary case, and let $\widehat \Delta_j$, $\widehat \Delta_k$ be the method of moments estimators for $\Delta_j$, $\Delta_k$, respectively. Then
		$$
		| F_{TB}(\Sigma_{jk}; \Delta_j , \Delta_k) - F_{TB}(\Sigma_{jk};  \widehat \Delta_j ,  \widehat \Delta_k)| \le \dfrac{6}{\surd{(2\pi)}}| \widehat \Delta_j - \Delta_j | + \dfrac{8}{\surd{(2\pi)}}| \widehat \Delta_k - \Delta_k |.
		$$
	\end{lemma}
	\begin{proof}[Proof of Lemma~\ref{lem:BridgeTBdelta}]
		Using the bridge function formula from Theorem~2 leads to
		\begin{equation}
			\begin{split}\label{eq:Bridgedelta_TB_main}
				| & F_{TB}(\Sigma_{jk}; \Delta_j , \Delta_k) - F_{TB}(\Sigma_{jk};  \widehat \Delta_j ,  \widehat \Delta_k)| \\
				& = | 2 \{1-\Phi(\Delta_j)\}\Phi(\Delta_k) -2 \Phi_3\left\{-\Delta_j, \Delta_k, 0; \Sigma_{3a}(r) \right\} -2 \Phi_3\left\{-\Delta_j, \Delta_k, 0; \Sigma_{3b}(r) \right\}\\
				& \quad - 2 \{1-\Phi(\widehat \Delta_j)\}\Phi(\widehat \Delta_k) +2 \Phi_3 \{-\widehat \Delta_j, \widehat \Delta_k, 0; \Sigma_{3a}(r)  \} +2 \Phi_3 \{-\widehat \Delta_j, \widehat \Delta_k, 0; \Sigma_{3b}(r) \} |\\
				& \le 2 | \Phi(\Delta_k) - \Phi(\widehat \Delta_k)| + 2| \Phi(\Delta_j)\Phi(\Delta_k) - \Phi(\widehat \Delta_j)\Phi(\widehat \Delta_k) |\\
				& \quad +2 | \Phi_3 \{-\Delta_j, \Delta_k, 0; \Sigma_{3a}(r)  \} - \Phi_3 \{-\widehat \Delta_j, \widehat \Delta_k, 0; \Sigma_{3a}(r)  \} |\\
				& \quad +2 |\Phi_3 \{- \Delta_j, \Delta_k, 0; \Sigma_{3b}(r) \} - \Phi_3 \{-\widehat \Delta_j, \widehat \Delta_k, 0; \Sigma_{3b}(r) \}|
			\end{split}
		\end{equation}
		From the mean value theorem, there exists intermediate values $\xi$'s such that
		\begin{equation}
			\begin{split}\label{eq:meanvalueTB-simplify1}
				| \Phi(\Delta_k) - \Phi(\widehat \Delta_k)| & = \phi(\xi_1)|\Delta_k - \widehat\Delta_k|\\
				| \Phi(\Delta_j)\Phi(\Delta_k) - \Phi(\widehat \Delta_j)\Phi(\widehat \Delta_k) | & \le \Phi (\Delta_j) \phi (\xi_2) | \Delta_k - \widehat \Delta_k | + \Phi (\widehat \Delta_k) \phi (\xi_3) | \Delta_j - \widehat \Delta_j |
			\end{split}
		\end{equation}
		
		The mean value theorem also can be applied for two dimension functions using chain rule. Thus we also have intermediate values $\xi_3, \ldots, \xi_7$ such that
		\begin{equation}
			\begin{split}\label{eq:meanvalueTB-simplify2}
				| \Phi_3 \{-\Delta_j, \Delta_k, 0; \Sigma_{3a}(r)  \} - \Phi_3 \{-\widehat \Delta_j, \widehat \Delta_k, 0 & ; \Sigma_{3a}(r)  \} |\\
				& \le \Phi_{3a1}(\xi_4) | \Delta_j - \widehat\Delta_j | + \Phi_{3a2}(\xi_5) | \Delta_k - \widehat\Delta_k |\\
				|\Phi_3 \{- \Delta_j, \Delta_k, 0; \Sigma_{3b}(r) \} - \Phi_3 \{-\widehat \Delta_j, \widehat \Delta_k, 0 & ; \Sigma_{3b}(r) \}|\\
				& \le \Phi_{3b1}(\xi_6) | \Delta_j - \widehat\Delta_j | + \Phi_{3b2}(\xi_7) | \Delta_k - \widehat\Delta_k |.
			\end{split}
		\end{equation}
		
		Similar to \eqref{eq:3dim-upperbound}, $\Phi_{3a1}$, $\Phi_{3a2}$, $\Phi_{3b1}$ and $\Phi_{3b2}$ are all bounded by $(2\pi)^{-1/2}$. Therefore, plugging \eqref{eq:meanvalueTB-simplify1} and \eqref{eq:meanvalueTB-simplify2} into \eqref{eq:Bridgedelta_TB_main} concludes the proof of the lemma.

	\end{proof}

	\begin{lemma}\label{lem:BridgeTTdelta} Let $F_{\text{TT}}(\Sigma_{jk}; \Delta_j, \Delta_k)$ be the bridge function from Theorem~4 for the truncated/truncated case, and let $\widehat \Delta_j$, $\widehat \Delta_k$ be the method of moments estimators for $\Delta_j$, $\Delta_k$, respectively. Then
		$$
		| F_{TT}(\Sigma_{jk}; \Delta_j , \Delta_k) - F_{TT}(\Sigma_{jk};  \widehat \Delta_j ,  \widehat \Delta_k)| \le 
		\dfrac{4}{\surd{(2\pi)}} \{ | \widehat \Delta_j - \Delta_j | + | \widehat \Delta_k - \Delta_k |\}.
		$$
	\end{lemma}
	
	\begin{proof}[Proof of Lemma~\ref{lem:BridgeTTdelta}]
		Using the bridge function formula from Theorem~4, and the mean value theorem gives
		\begin{equation}
			\begin{split}\label{eq:Bridgedelta_TT_main}
				| & F_{TT}(\Sigma_{jk}; \Delta_j , \Delta_k) - F_{TT}(\Sigma_{jk};  \widehat \Delta_j , \widehat  \Delta_k)|\\
				& = | -2 \Phi_4 (-\Delta_j, -\Delta_k, 0,0; \Sigma_{4a}(r)) + 2 \Phi_4 (-\Delta_j, -\Delta_k, 0,0; \Sigma_{4b}(r))\\
				& \quad + 2 \Phi_4 (- \widehat\Delta_j, -\widehat\Delta_k, 0,0; \Sigma_{4a}(r)) - 2 \Phi_4 (-\widehat\Delta_j, -\widehat\Delta_k, 0,0; \Sigma_{4b}(r)) |\\
				& \le 2 | \Phi_4 (-\Delta_j, -\Delta_k, 0,0; \Sigma_{4a}(r)) - \Phi_4 (-\widehat \Delta_j, -\widehat\Delta_k, 0,0; \Sigma_{4a}(r))|\\
				& \quad + 2 |\Phi_4 (- \Delta_j, -\Delta_k, 0,0; \Sigma_{4b}(r)) - \Phi_4 (-\widehat \Delta_j, -\widehat\Delta_k, 0,0; \Sigma_{4b}(r)) |\\
				&\leq 2\Phi_{4a1}(\xi_1) | \Delta_j - \widehat\Delta_j | + 2\Phi_{4a2}(\xi_2) | \Delta_k - \widehat\Delta_k | + 2\Phi_{4b1}(\xi_3) | \Delta_j - \widehat\Delta_j | + 2\Phi_{4b2}(\xi_4) | \Delta_k - \widehat\Delta_k|,
			\end{split}
		\end{equation}
		where all $\xi$'s are the intermediate values.
		
		By definition
		\begin{equation}\label{eq:4dim-a}
			\begin{split}
				\Phi_{4a1} (x, y, & z, w; \Sigma_{4a}(r))\\
				& = \dfrac{\partial \Phi_4(x, y, z, w; \Sigma_{4a}(r))}{\partial x}\\
				& = \dfrac{\partial}{\partial x} \int_{-\infty}^{x} \int_{-\infty}^{y} \int_{-\infty}^{z} \int_{-\infty}^{w} \phi(x_3, x_4|x_1, x_2)\phi(x_1, x_2) dx_4 dx_3 dx_2 dx_1\\
				& = \dfrac{\partial}{\partial x} \int_{-\infty}^{x} \int_{-\infty}^{y} \Phi_2\left(z, w | x_1, x_2\right)\phi(x_1, x_2) dx_2 dx_1\\
				& = \int_{-\infty}^{y}\Phi_2\left(z, w | x, x_2\right)\phi(x, x_2) dx_2\\
				&\leq \int_{-\infty}^{y}\phi(x, x_2) dx_2\\
				& = \int_{-\infty}^{y}\phi(x)\phi(x_2) dx_2\\
				& = \phi(x)\Phi(y)\\
				&\leq (2\pi)^{-1},
			\end{split}
		\end{equation}
		where $\phi(x, x_2) = \phi(x)\phi(x_2)$ holds because the $\{\Sigma_{4a}(t)\}_{12} = 0$.
		
		Similarly, $\Phi_{4a2} (x, y, z, w; \Sigma_{4a}(r))$ is bounded by $(2\pi)^{-1/2}$.
		For $\Phi_{4b1}(\xi_3)$, this leads to
		\begin{align*}
			\Phi_{4b1} (x, y, z, w; \Sigma_{4b}(r)) \leq \int_{-\infty}^{y}\phi_2(x, x_2; r)dx_2 = \phi(x)\int_{-\infty}^{y} \phi(x_2| x_1 = x; r)dx_2 \leq \phi(x) \leq (2\pi)^{-1}.
		\end{align*}
		Similarly for $\Phi_{4b2}(\xi_3)$. Plugging these upper bounds in \eqref{eq:Bridgedelta_TT_main} completes the proof.
	\end{proof}

	\section{Additional simulation results}\label{app:extrasim}

	\subsection{Comparison of rank-based correlation estimator with Pearson sample correlation}\label{app:Pearson-vs-Kendall}

	In this section, we compare our rank-based estimator of latent correlation matrix with Pearson sample correlation. For clarity, we focus on the $p=2$ case with true latent correlation value $\Sigma_{12} = 0.8$, and both variables being of truncated type. The corresponding rank-based estimator is calculated as $\widehat R_{12} = \argmin_{r}\{F_{TT}(r)-\widehat \tau_{12}\}^2$, where $F_{TT}(\cdot)$ is the bridge function from Theorem~4 and $\widehat \tau_{12}$ is the sample Kendall's $\tau$.  We investigate the performance by varying the sample size $n\in \{100, 500, 1000\}$ as well as the truncation rate from 20\% to 80\% of the sample size for each of the two variables.

	\begin{figure}[!t]
		\centering
		\includegraphics[scale = 0.8]{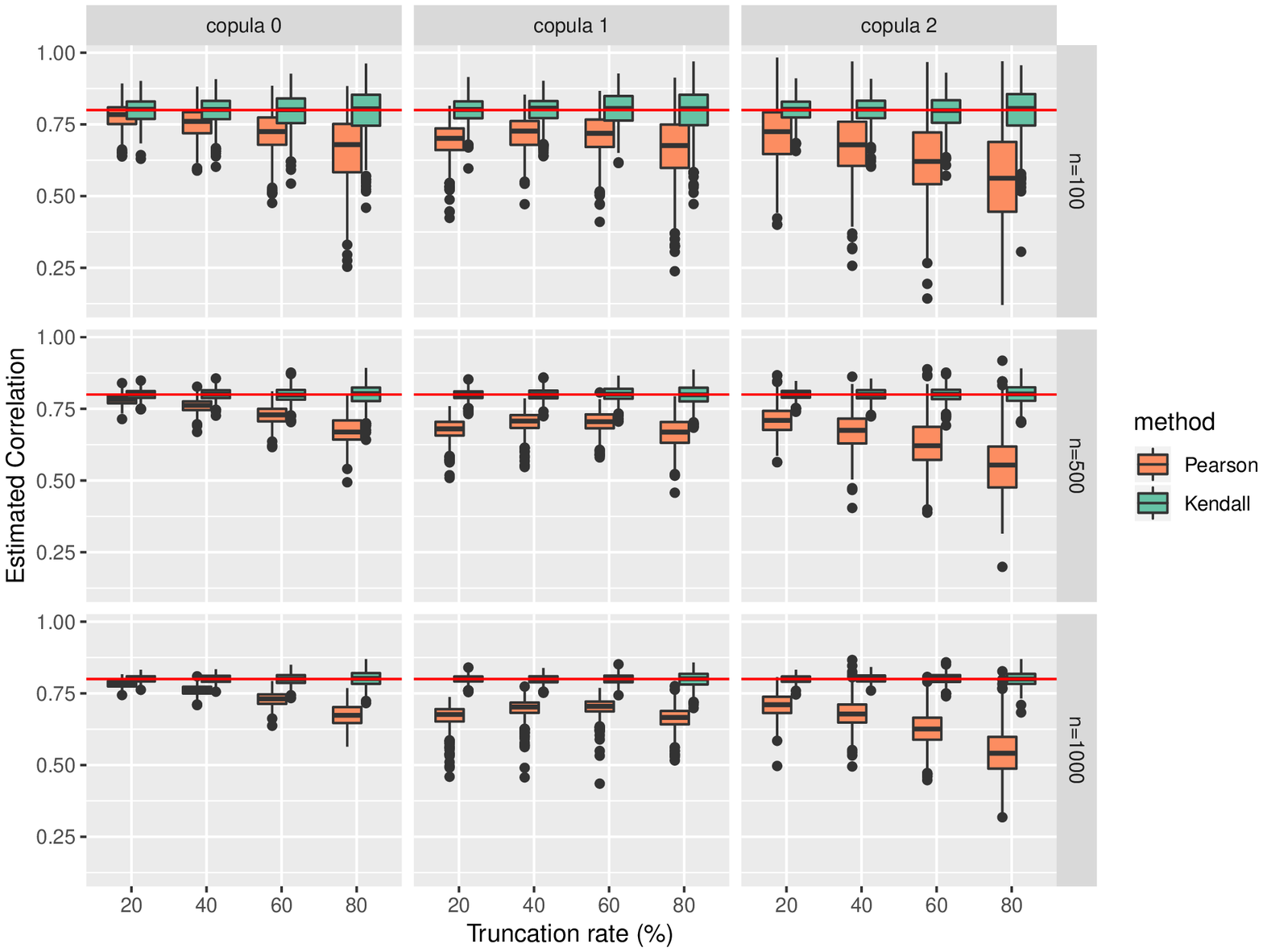}
		\caption{Truncated/truncated case with $p=2$ and $\Sigma_{12} = 0.8$ (red horizontal line). Comparison between Pearson sample correlation and the proposed rank-based correlation estimator based on Kendall's $\tau$ over 500 replications as a function of sample size $n$ and truncation rate.}
		\label{fig:Pearson-vs-Kendall}
	\end{figure}
	
	Figure~\ref{fig:Pearson-vs-Kendall} displays the values of Pearson sample correlation between the two variables as well as the values of the proposed $\widehat R_{12}$ over 500 replications for each combination of sample size $n$, truncation rate, and copula types as described in Section~4. Overall, both Pearson and rank-based correlation estimates have lower variance as the sample size increases, and larger variance as the truncation level increases. The Pearson sample correlation values are biased downwards even when no monotone transformations are applied (copula 0) but the truncation level is large ($\geq 40\%$), and the bias increases both with the truncation level and with the use of copula transformations. On the other hand, our rank-based estimator of latent correlation matrix is approximately unbiased in all cases, and has smaller variance compared to the Pearson sample correlation. We conclude that the proposed rank-based estimator has excellent performance for a wide range of sample size and truncation levels, although larger sample sizes are required for larger truncation levels to minimize the variance.

	\subsection{Sensitivity to the choice of initial optimization values}\label{app:initval}
	
	In this section, we investigate the sensitivity of the proposed method and tuning parameter selection scheme to the choice of starting values $w_1^{(0)}$ and $w_2^{(0)}$ in the optimization algorithm. By default, we initialize the algorithm with the unpenalized solution ($\lambda_1 = \lambda_2 =0$) obtained using $\widetilde R + 0.25I$, which corresponds to canonical ridge solution with fixed amount of regularization \citep{RCCA:2008}. We compare this default initialization with 50 random staring points generated as follows: (1) we generate a random $M \in \R^{p\times p}$, $p = p_1+p_2$ with independent elements $m_{ij}\sim N(0,1)$; (2) we use the first 50 right singular vectors of $M$ to get starting $w_{1}^{(0)}$ (first $p_1$ elements of a selected singular vector) and $w_2^{(0)}$ (last $p_2$ elements of a selected singular vector). This generation approach ensures that the starting points are in orthogonal directions. We further standardize all starting points to satisfy $w_1^{\top}\widetilde R_1 w_1 = 1$, $w_2^{\top}\widetilde R_2 w_2 = 1$.

	\begin{figure}[!t]
		\centering
		\includegraphics[scale = 0.6]{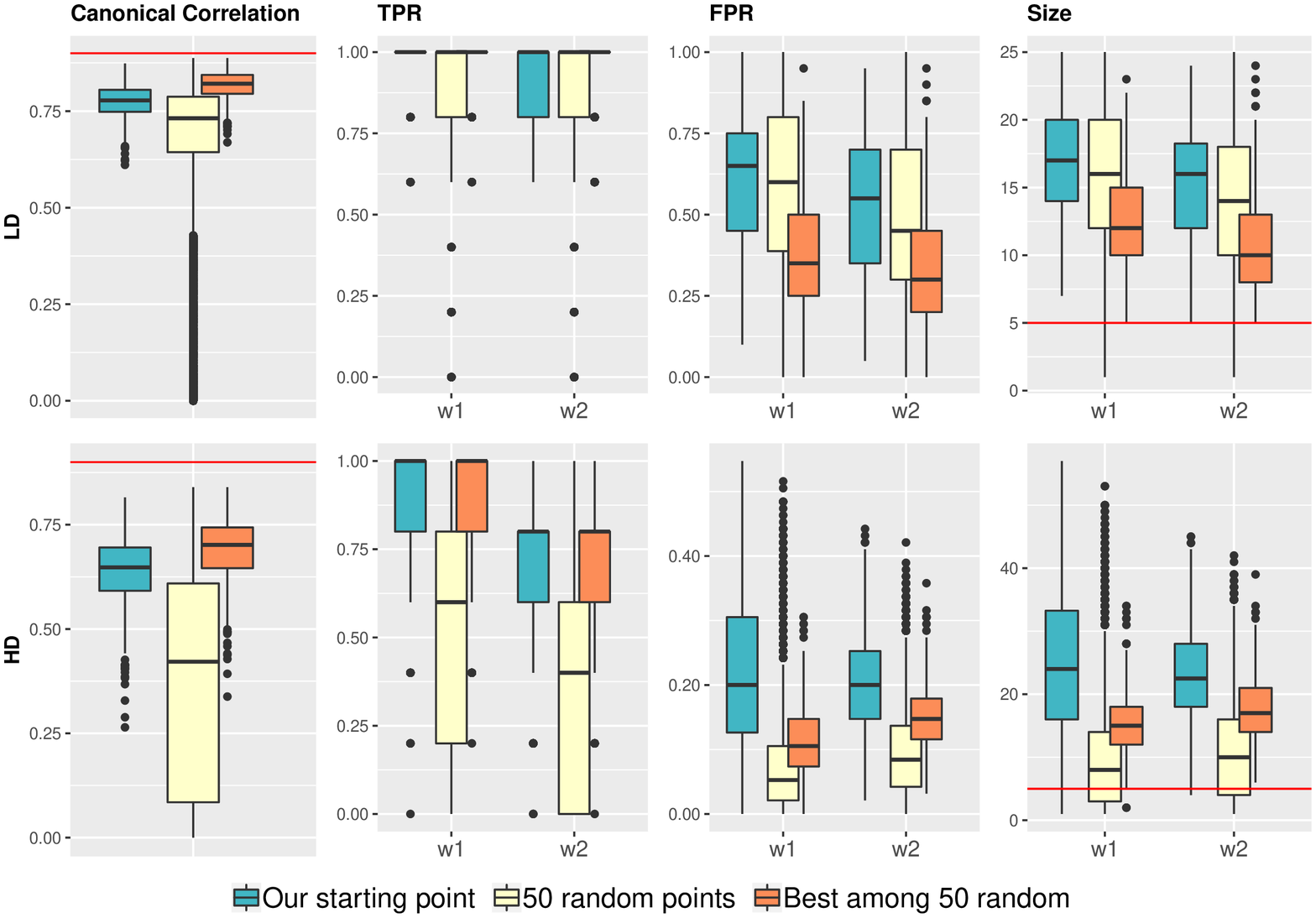}\label{fig:initSensitivity}
		\caption{Truncated/truncated case with copula 2 and \BIC$_2$ selection criterion. Value for out-of-sample canonical correlation $\widehat \rho$, true positive rate (TPR), false positive rate (FPR) and the size of selected model across 500 replications depending on the initial starting points. ``Our starting point" indicates our default initialization based on canonical ridge solution. ``50 random points" correspond to 50 random initializations across total 500 replications, and ``Best among 50 random" corresponds to best initialization out of 50 random for each replication in terms of highest value of $\widehat \rho$. \textbf{Top row}: Low dimensional (LD) case ($p_1 = p_2 = 25$). \textbf{Bottom row}: High dimensional (HD) case ($p_1 = p_2 = 100$).}
	\end{figure}

	We consider the truncated/truncated case with sample size $n=100$ and copula 2 model as described in Section~4 with \BIC$_2$ tuning parameter selection criterion. Figure~S2 displays the achieved out-of-sample correlation $\widehat \rho$, true positive rate, false positive rate and the selected model size over 500 replications. The results are separated by the proposed default initialization (blue), combined results from 50 random initializations (yellow, total $50 \text{ initial values} \times500 \text{ replications} = 25000$ estimates), and the best initialization out of 50 random starting points (orange), where we define the best initialization as the one with the highest value of $\widehat \rho$ in~(7). By definition, the best initialization always leads to better performance than a random initialization, and these differences are more pronounced in the high-dimensional case. Nevertheless, the median values of $\widehat{\rho}$ obtained by the proposed approach across the initializations and replications are still higher than the values of $\widehat{\rho}$ obtained by competing methods on copula 2 as seen in Figure~1 of the main manuscript. Although it is possible to have a ``bad" random initialization (see the outliers in low-dimensional case for the values of $\widehat \rho$ in Figure~S2), these outliers are not present for the proposed initial starting point. Specifically, the proposed default initialization leads to better performance than an average random initialization, and is comparable to the performance of the ``best" initial starting point. Furthermore, all the empirical analyses of Sections~4 and~5 have been performed using the proposed default initialization, thus demonstrating the excellent performance of the method as implemented in practice.

	\subsection{TPR versus FPR curves for truncated/truncated cases}\label{app:tprfpr}
	
	\begin{figure}
		\centering
		\includegraphics[scale = 0.73]{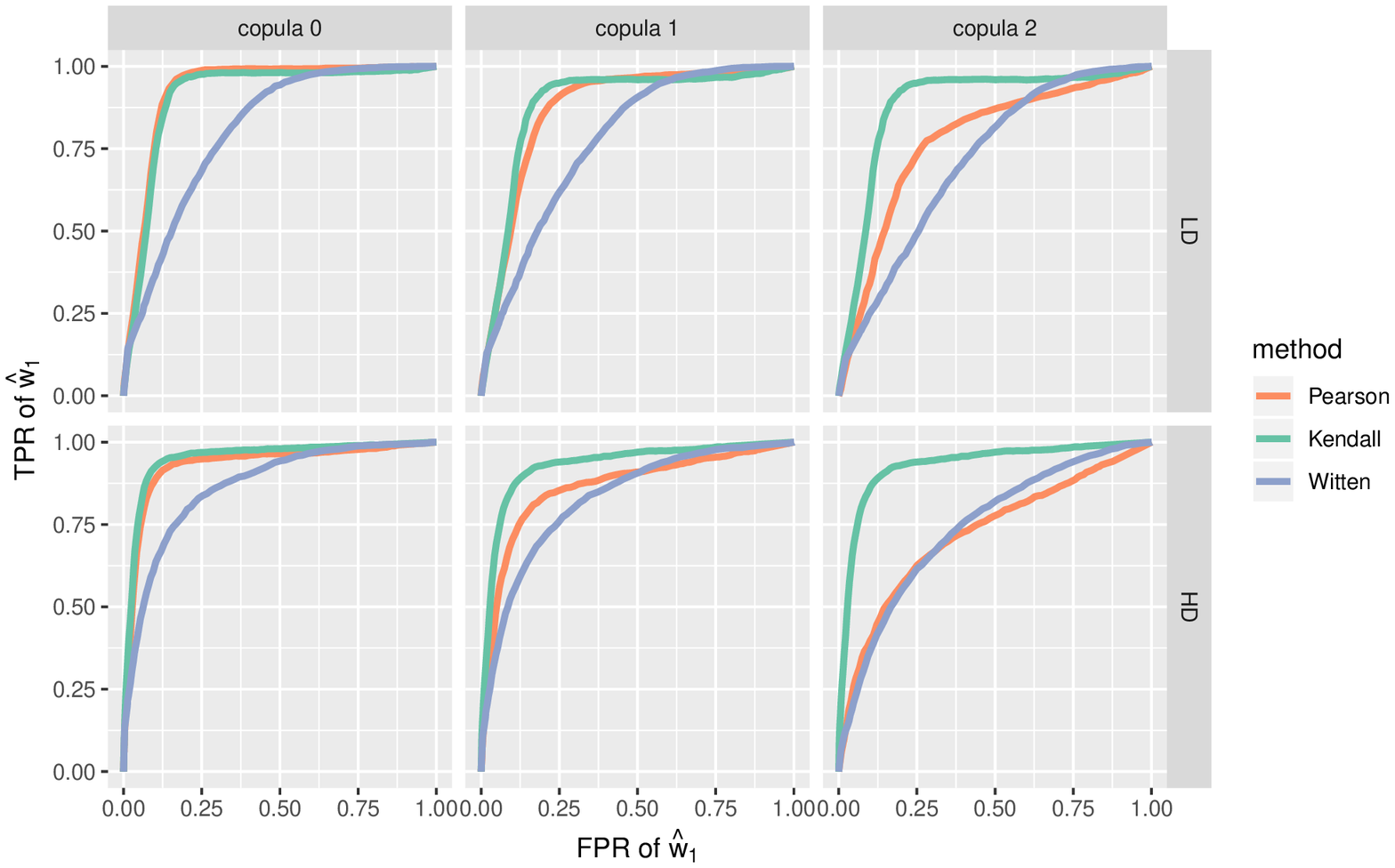}\\
		\includegraphics[scale = 0.73]{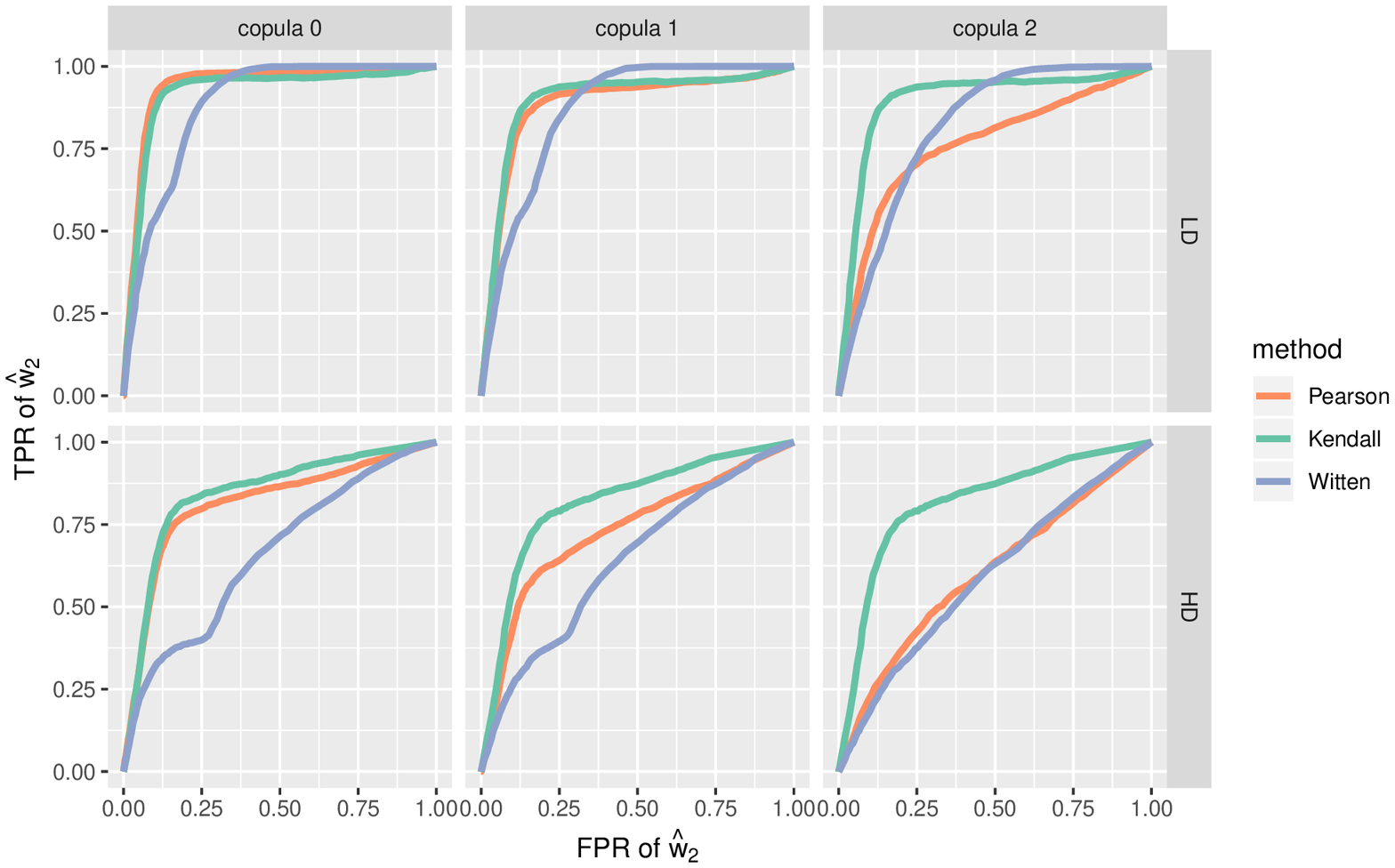}
		\caption{Truncated/truncated case. LD: low-dimensional setting ($p_1 = p_2 = 25$); HD: high-dimensional setting ($p_1=p_2=100$). Pearson (orange): our optimization framework with Pearson sample correlation. Kendall (green): our optimization framework with our rank-based estimator using Kendall's $\tau$. Witten (blue): method of \citet{Witten:2009PMD}. As described in Section~4, $X_1$ has an autoregressive correlation structure and $X_2$ has a block-diagonal correlation structure. \textbf{Top}: average TPR versus FPR value curve for $\widehat{w}_1$; \textbf{Bottom}: average TPR versus FPR value curve for $\widehat{w}_2$.}
		\label{fig:TPRFPRcurve}
	\end{figure}
	
	In this section we investigate the variable selection performance of different methods by comparing true positive rate (TPR) versus false positive rate (FPR) curves for $w_1$ and $w_2$. Since these curves are obtained by considering the range of corresponding tuning parameters, this comparison is invariant to the chosen tuning parameter selection scheme.  Specifically, given a value of $\lambda$, we define $\text{TPR}_{g, \lambda}$ and $\text{FPR}_{g, \lambda}$ for each $w_g$, $g\in\{1,2\}$, as 
	$$
	\text{TPR}_{g, \lambda} = \dfrac{\#\{(g,j): \widehat{w}_{gj} \neq 0 \text{ and } w_{gj} \neq 0\}}{\#\{(g,j): w_{gj} \neq 0\}}, \quad
	\text{FPR}_{g, \lambda} = \dfrac{\#\{(g,j): \widehat{w}_{gj} \neq 0 \text{ and } w_{gj} = 0\}}{\#\{(g,j): w_{gj} = 0\}}.
	$$
	For simplicity, we set $\lambda_1  = \lambda_2 = \lambda$, and consider 50 values of $\lambda$, using logarithmic grid from 0$\cdot$01 to 0$\cdot$7 for sparse canonical correlation analysis based on Pearson correlation and our rank-based correlation, and equally spaced grid from 0$\cdot$01 to 0$\cdot$9 for the Witten's method (the grid is different due to the use of $\ell_1$ constraint rather than $\ell_1$ penalty in \citet{Witten:2009PMD}). 
	
	We consider the truncated/truncated case with sample size $n=100$ in low-dimensional and high-dimensional settings as described in Section~4. The average TPR and FPR values for each value of $\lambda$ over 500 replications are plotted in Figure~\ref{fig:TPRFPRcurve}. In the copula 0 case (no data transformation), the method based on Pearson's correlation performs as well or better than our approach, however the performance deteriorates when data transformation is applied (copulas 1 and 2). The Witten's method has worse performance than Pearson-based method, especially in the copula 0 case, which is likely due to the fact that the Witten's method uses the diagonal approximation of correlation structure within each dataset. We suspect that the visual dent observed for $\widehat{w}_2$ in the Witten's TPR versus FPR curve (see the bottom figure in Figure~\ref{fig:TPRFPRcurve}) is due to the block-diagonal correlation structure of $X_2$ (see Section 4). Since the Witten's method ignores this structure, but the variables within the same block are highly correlated, the Witten's method includes all the variables from the same block first before adding other variables. This leads to increase in FPR while keeping the same level of TPR. In contrast, the other methods take the block-diagonal structure into account, and therefore do not exhibit this behavior.
	
	In conclusion, when no data transformation is applied (copula 0), our method based on Kendall's $\tau$ performs as well as Pearson correlation. In the copula settings (copulas 1 and 2), our method has the highest area under the TPR versus FPR curve, confirming its excellent variable selection performance independently of tuning parameter selection scheme.

	\subsection{Truncated/continuous and truncated/binary cases}\label{app:extrasim-tbtc}
	
\begin{figure}[t]
		\centering
		\hspace{1.2cm}\includegraphics[scale = 0.44]{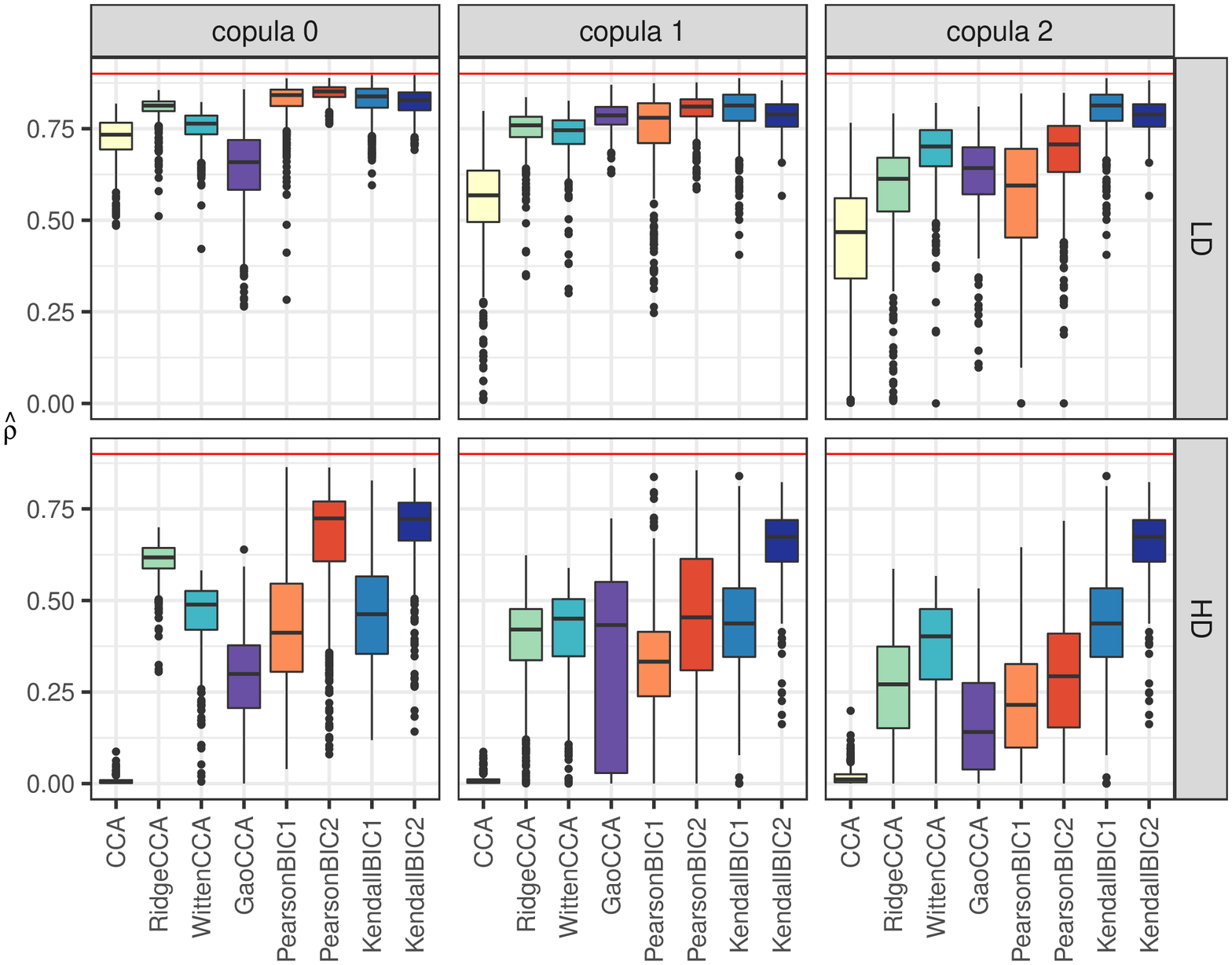}
		\includegraphics[scale = 0.44]{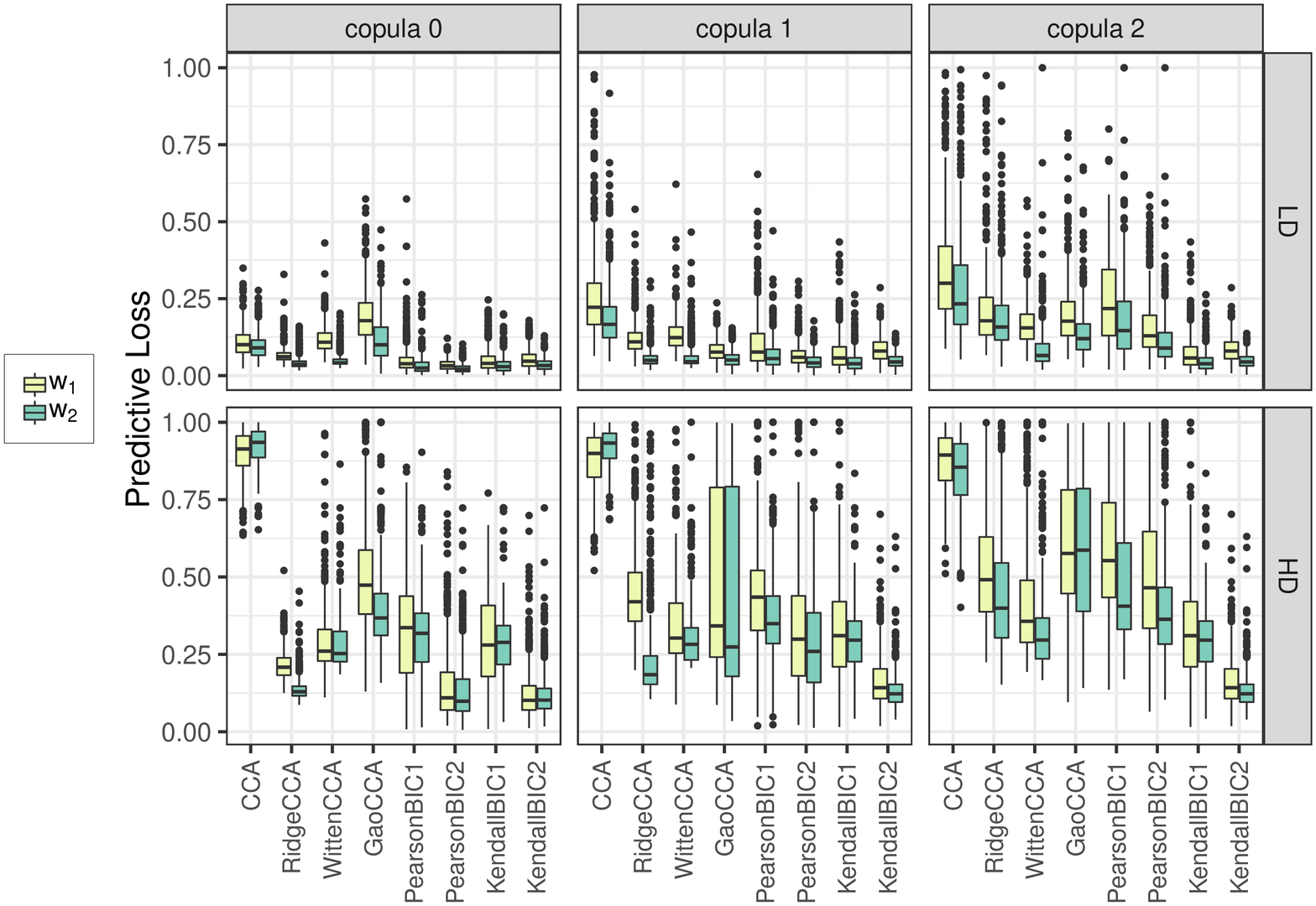}
		\caption{Truncated/continuous case. \textbf{Top:} The value of $\widehat{\rho}$ from (7). The horizontal lines indicate the true canonical correlation value $\rho = 0.9$. \textbf{Bottom:} The value of predictive loss~(8). Results over 500 replications. CCA:~Sample canonical correlation analysis; RidgeCCA: Canonical ridge of~\citet{RCCA:2008}; WittenCCA: method of~\cite{Witten:2009PMD}; GaoCCA:~method of ~\cite{Gao:2017SCCA}; PearsonBIC1, PearsonBIC2: proposed algorithm with Pearson sample correlation matrix;  KendallBIC1, KendallBIC2: proposed method with tuning parameter selected using either \BIC$_1$ or \BIC$_2$ criterion; LD: low-dimensional setting ($p_1 = p_2 = 25$); HD: high-dimensional setting ($p_1 = p_2 = 100$).}
		\label{fig:TC_rhohatPredloss}
	\end{figure}

\begin{figure}[t]
		\centering
		\includegraphics[scale = 0.44]{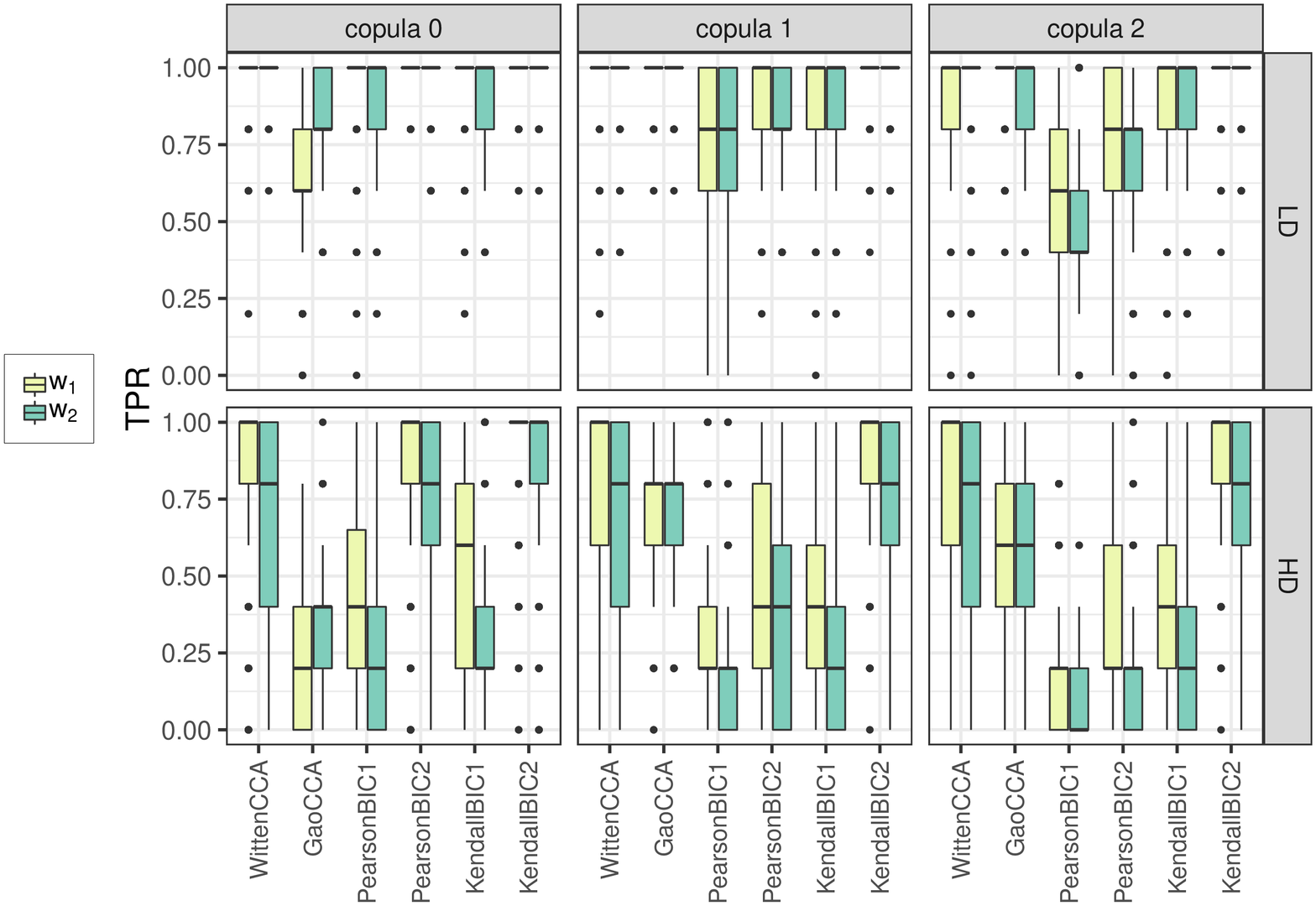}
		
		\includegraphics[scale = 0.44]{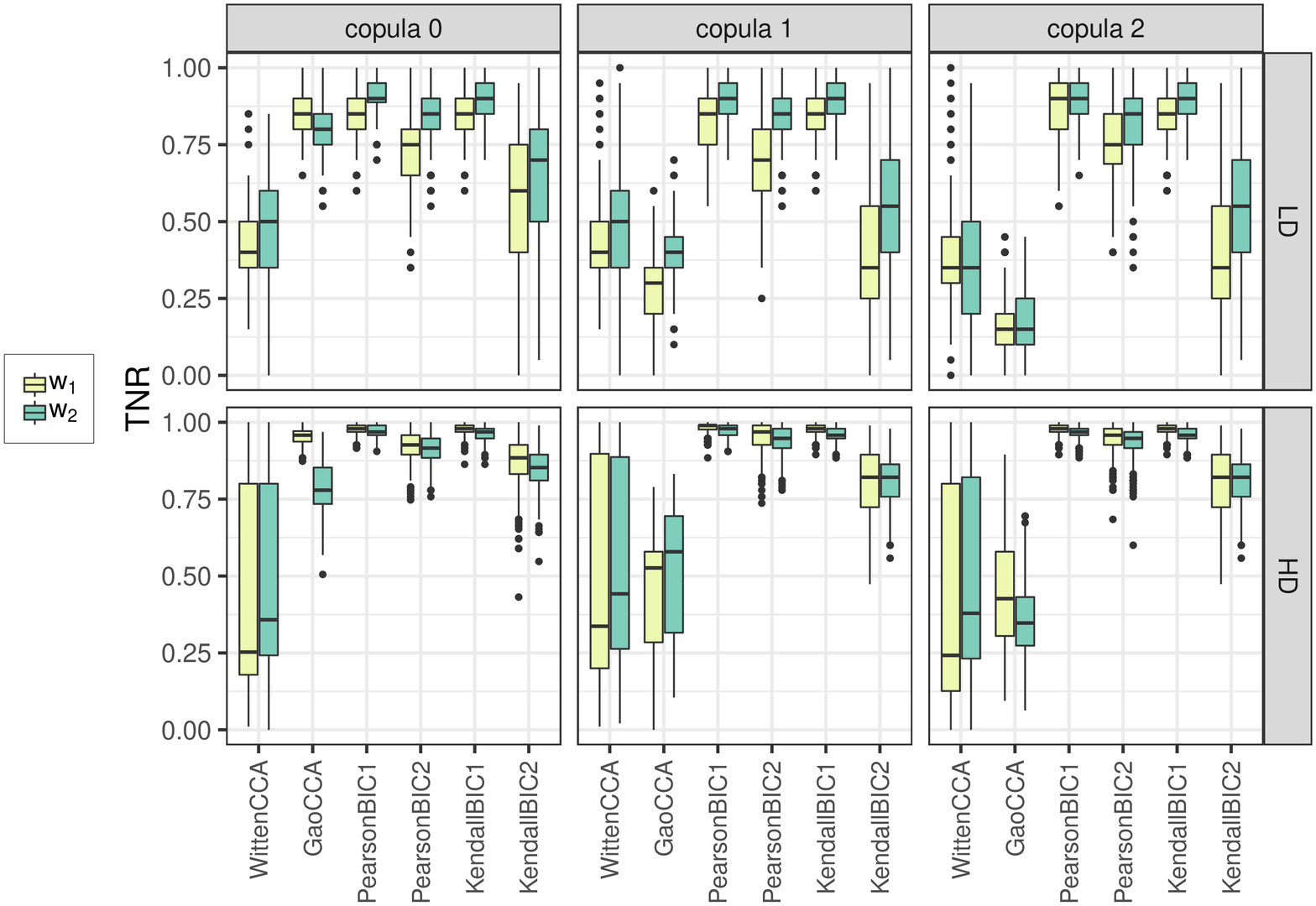}
		\caption{\baselineskip=12pt Truncated/continuous case. \textbf{Top:} True positive rate (TPR); \textbf{Bottom:} True negative rate (TNR). Results over 500 replications. WittenCCA: method of~\cite{Witten:2009PMD}; GaoCCA:~method of ~\cite{Gao:2017SCCA}; PearsonBIC1, PearsonBIC2: proposed algorithm with Pearson sample correlation matrix; KendallBIC1, KendallBIC2: proposed method with tuning parameter selected using either \BIC$_1$ or \BIC$_2$ criterion; LD: low-dimensional setting ($p_1 = p_2 = 25$); HD: high-dimensional setting ($p_1 = p_2 = 100$).}
		\label{fig:TC_TPRTNR}
	\end{figure}

\begin{figure}[t]
		\centering
		\includegraphics[scale = 0.44]{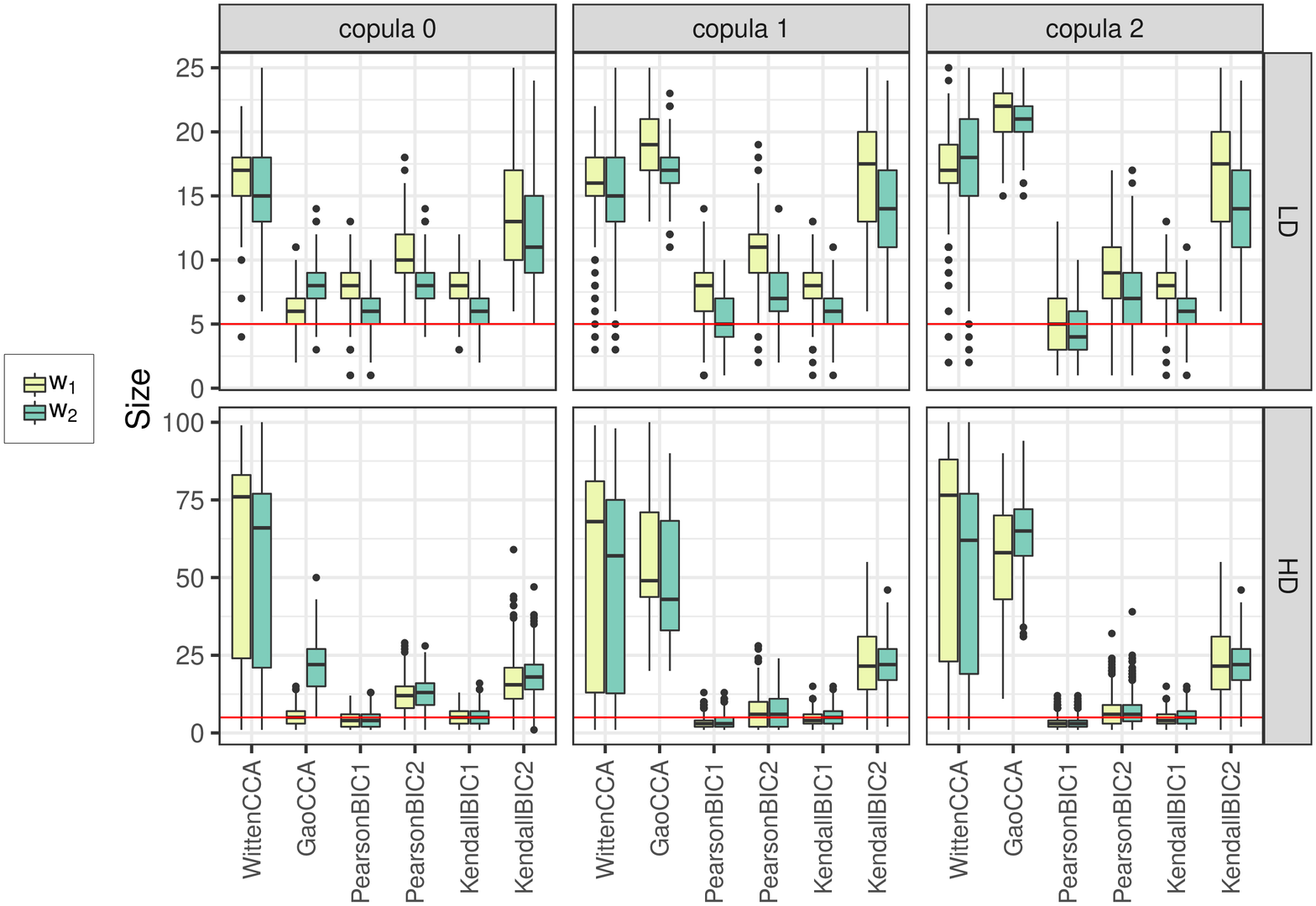}
		\caption{\baselineskip=12pt Truncated/continuous case. Selected model size over 500 replications. The horizontal lines indicate the true model size $5$. WittenCCA: method of~\cite{Witten:2009PMD}; GaoCCA:~method of ~\cite{Gao:2017SCCA}; PearsonBIC1, PearsonBIC2: proposed algorithm with Pearson sample correlation matrix; KendallBIC1, KendallBIC2: proposed method with tuning parameter selected using either \BIC$_1$ or \BIC$_2$ criterion; LD: low-dimensional setting ($p_1 = p_2 = 25$); HD: high-dimensional setting ($p_1 = p_2 = 100$).}
		\label{fig:TC_size}
	\end{figure}

\begin{figure}[t]
		\centering
		\hspace{1.2cm}\includegraphics[scale = 0.44]{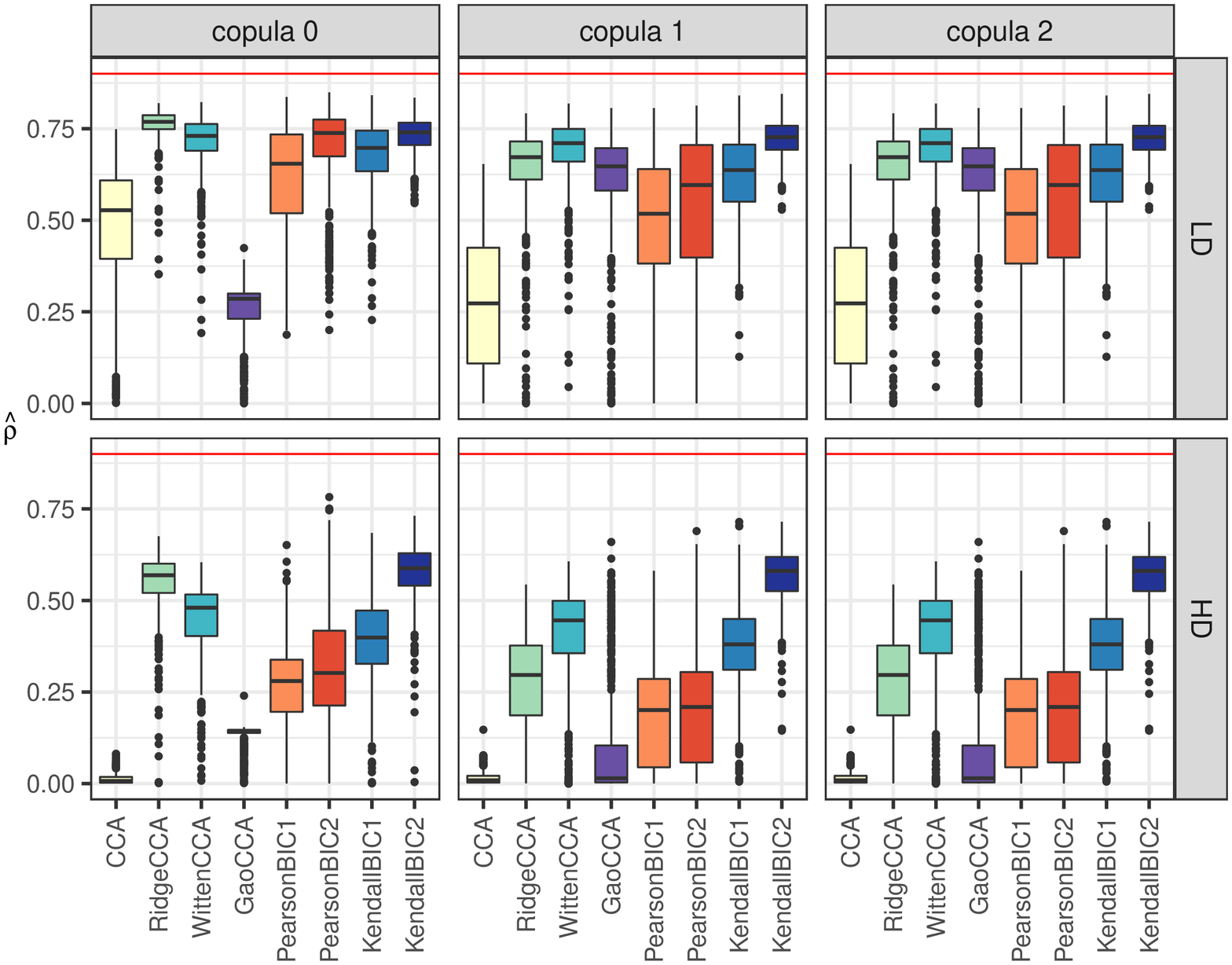}
		
		\includegraphics[scale = 0.45]{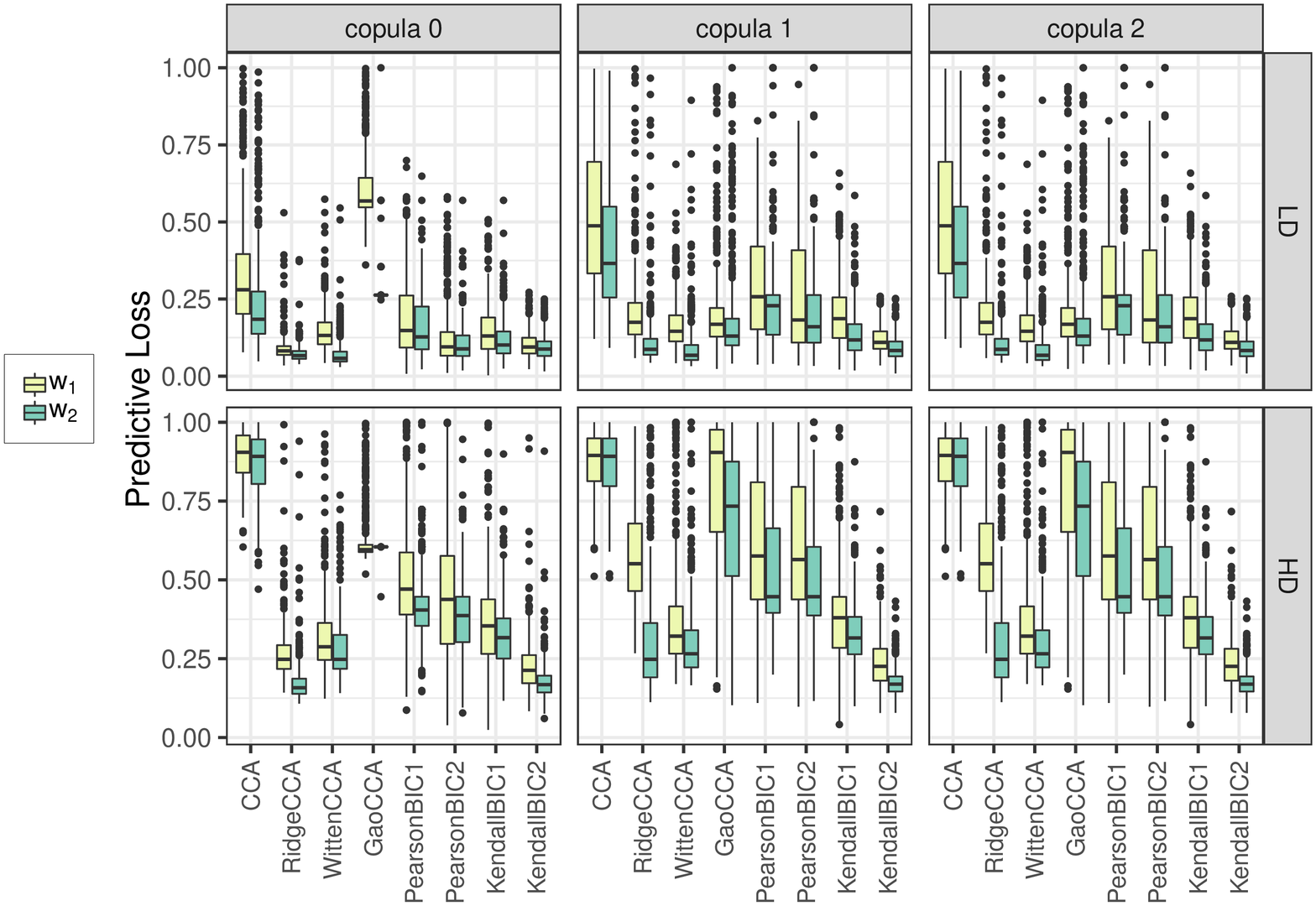}
		\caption{Truncated/binary case. \textbf{Top:} The value of $\widehat{\rho}$ from (7). The horizontal lines indicate the true canonical correlation value $\rho = 0.9$. \textbf{Bottom:} The value of predictive loss~(8). Results over 500 replications. CCA:~Sample canonical correlation analysis; RidgeCCA: Canonical ridge of~\citet{RCCA:2008}; WittenCCA: method of~\cite{Witten:2009PMD}; GaoCCA:~method of ~\cite{Gao:2017SCCA}; PearsonBIC1, PearsonBIC2: proposed algorithm with Pearson sample correlation matrix; KendallBIC1, KendallBIC2: proposed method with tuning parameter selected using either \BIC$_1$ or \BIC$_2$ criterion; LD: low-dimensional setting ($p_1 = p_2 = 25$); HD: high-dimensional setting ($p_1 = p_2 = 100$).}
		\label{fig:TB_rhohatPredloss}
	\end{figure}

\begin{figure}[t]
		\centering
		\includegraphics[scale = 0.44]{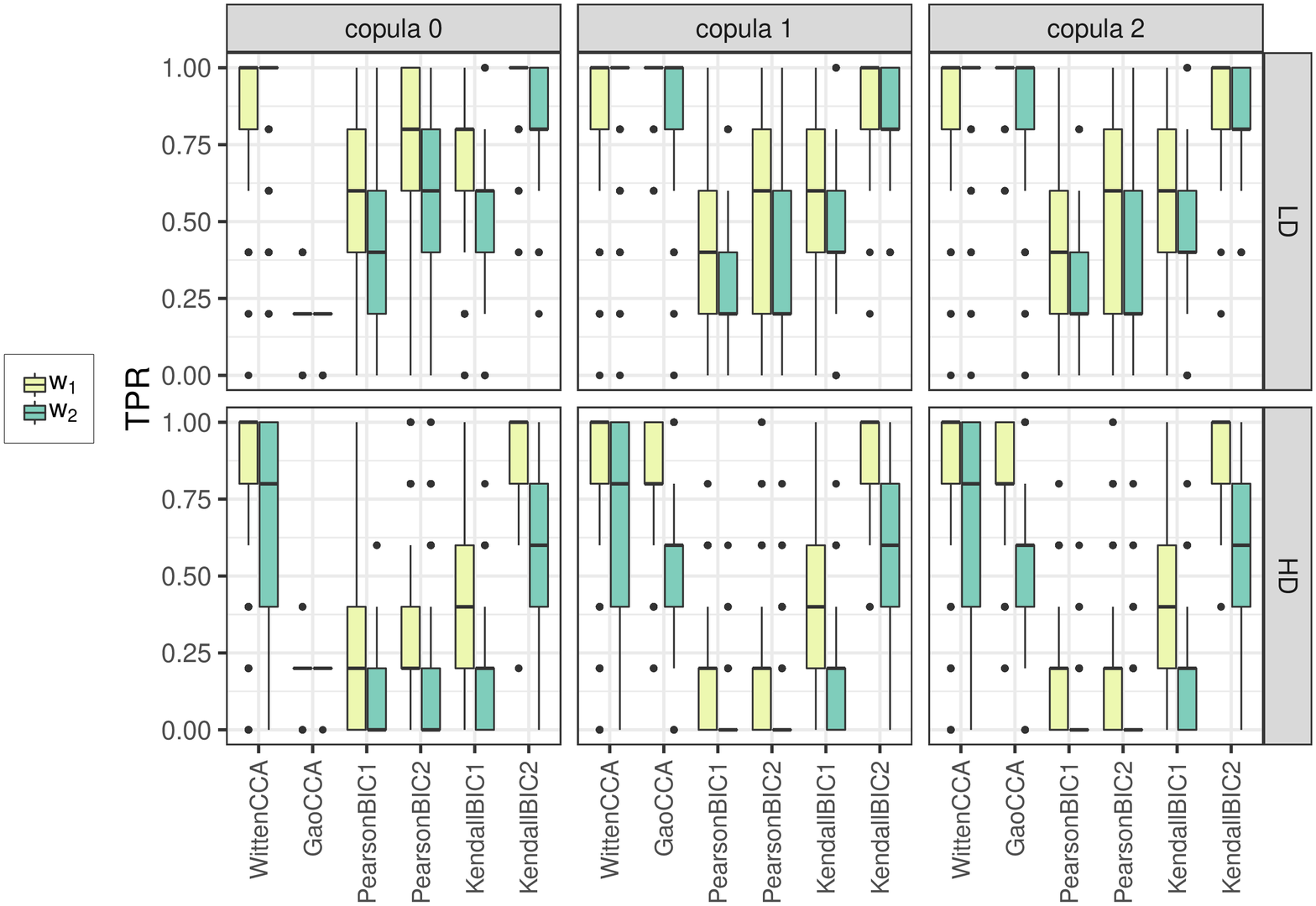}
		
		\includegraphics[scale = 0.44]{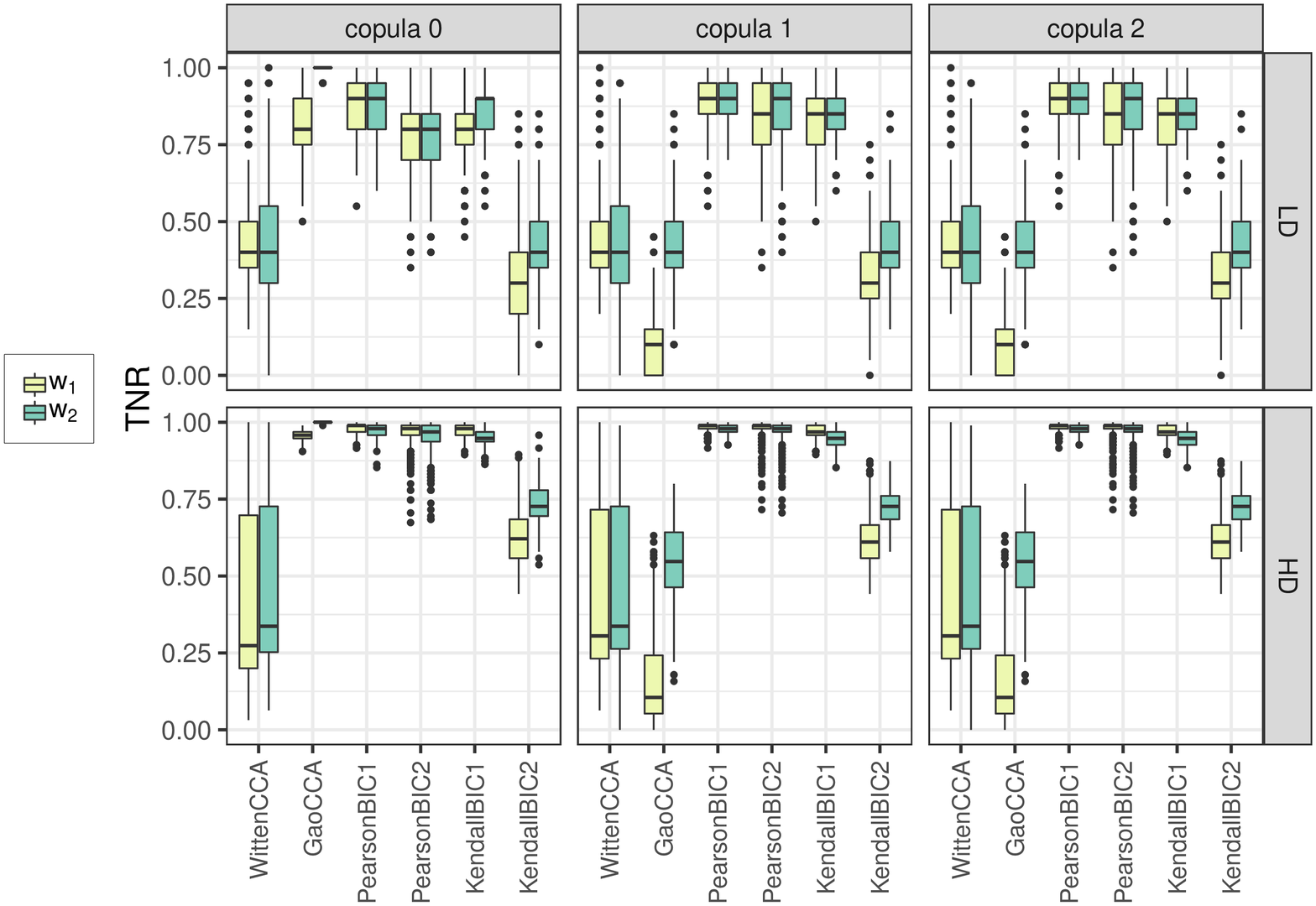}
		\caption{\baselineskip=12pt Truncated/binary case. \textbf{Top:} True positive rate (TPR); \textbf{Bottom:} True negative rate (TNR). Results over 500 replications. WittenCCA: method of~\cite{Witten:2009PMD}; GaoCCA:~method of ~\cite{Gao:2017SCCA}; PearsonBIC1, PearsonBIC2: proposed algorithm with Pearson sample correlation matrix; KendallBIC1, KendallBIC2: proposed method with tuning parameter selected using either \BIC$_1$ or \BIC$_2$ criterion; LD: low-dimensional setting ($p_1 = p_2 = 25$); HD: high-dimensional setting ($p_1 = p_2 = 100$).}
		\label{fig:TB_TPRTNR}
	\end{figure}

\begin{figure}[t]
		\centering
		\includegraphics[scale = 0.44]{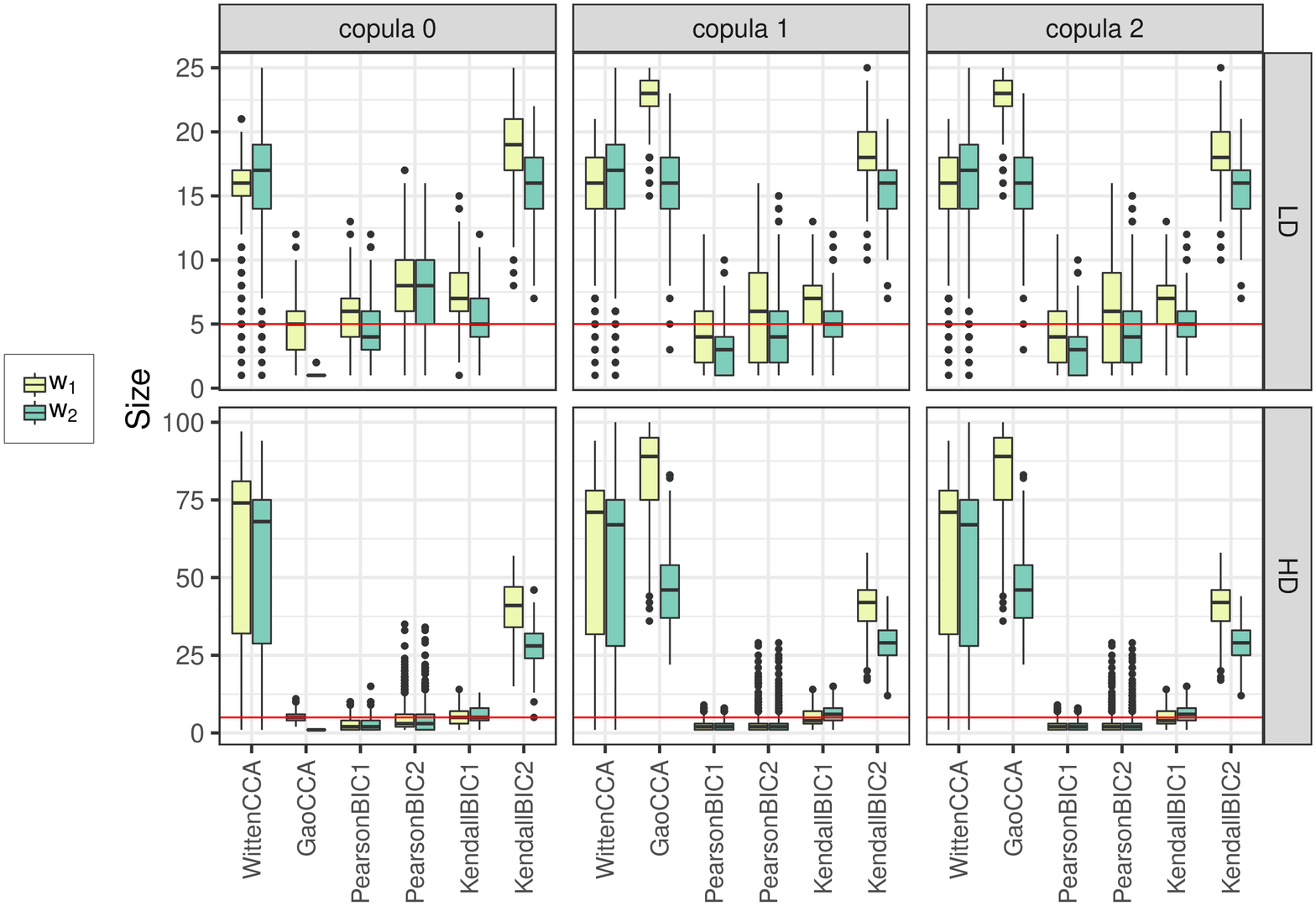}
		\caption{\baselineskip=12pt Truncated/binary case. Selected model size over 500 replications. The horizontal lines indicate the true model size $5$. WittenCCA: method of~\cite{Witten:2009PMD}; GaoCCA:~method of ~\cite{Gao:2017SCCA}; PearsonBIC1, PearsonBIC2: proposed algorithm with Pearson sample correlation matrix;  KendallBIC1, KendallBIC2: proposed method with tuning parameter selected using either \BIC$_1$ or \BIC$_2$ criterion; LD: low-dimensional setting ($p_1 = p_2 = 25$); HD: high-dimensional setting ($p_1 = p_2 = 100$).}
		\label{fig:TB_size}
	\end{figure}

	In this section, we complement the results of Section~4 with additional simulation results for the truncated/binary and truncated/continuous cases in Figures~\ref{fig:TC_rhohatPredloss}--\ref{fig:TB_size}. 
	For truncated/continuous case, the overall performance of all methods is slightly better or similar to the truncated/truncated case (Figures~ in the main manuscript). 
	For truncated/binary case, the method of \citet{Witten:2009PMD} has comparable prediction performance than both variations of our approach; however, it has worse variable selection performance due to significantly larger support sizes for both canonical vectors. Even when no data transformation is applied (copula 0), the method of \citet{Gao:2017SCCA} and the method based on Pearson correlation deteriorate compared to the truncated/continuous case. This is likely due to the binary case leading to the smaller effective sample size. For both truncated/continuous and truncated/binary cases, the conclusions on methods' comparison are qualitatively similar to Section~4 with our method having the best overall prediction performance with \BIC$_2$ criterion, and best variable selection performance with \BIC$_1$ criterion. 
	
\clearpage
\bibliographystyle{biometrika}
\bibliography{KendallCCAReferences}